\documentclass{emulateapj}
\usepackage{apjfonts}
\newcommand{\figexpand}{\epsscale{1.1}}
\newcommand{\ascaleup}{\epsscale{1.15}}
\newcommand{\plotter}{\plotone}

\newcommand{\tableset}{deluxetable}
\newcommand{\tableclear}{\clearpage}
\newcommand{\Mdot}{\dot{M}}

\newcommand{\etal}{et al.}
\newcommand{\mbh}{M_{\rm BH}}
\newcommand{\mstar}{M_{\ast}}
\newcommand{\msun}{M_{\sun}}
\newcommand{\lstar}{L_{\ast}}

\newcommand{\qeos}{q_{\rm eos}}
\newcommand{\fgas}{f_{\rm gas}}
\newcommand{\mdyn}{M_{\rm dyn}}
\newcommand{\re}{R_{\rm e}}
\newcommand{\rj}{RJ04}

\shorttitle{Extra Central Light in Merger Remnants}
\shortauthors{Hopkins \etal}
\slugcomment{Submitted to ApJ, October 24, 2007}
\begin{document}

\title{Dissipation and Extra Light in Galactic Nuclei: \textrm{I}.\ Gas-Rich Merger Remnants}
\author{Philip F. Hopkins\altaffilmark{1}, 
Lars Hernquist\altaffilmark{1}, 
Thomas J. Cox\altaffilmark{1,2}, 
Suvendra N. Dutta\altaffilmark{1}, 
\&\ Barry Rothberg\altaffilmark{3,4}
}
\altaffiltext{1}{Harvard-Smithsonian Center for Astrophysics, 
60 Garden Street, Cambridge, MA 02138, USA}
\altaffiltext{2}{W.~M.\ Keck Postdoctoral Fellow at the 
Harvard-Smithsonian Center for Astrophysics}
\altaffiltext{3}{Space Telescope Science Institute, 3700 
San Martin Drive, Baltimore, MD 21218, USA}
\altaffiltext{4}{Naval Research Laboratory, Remote Sensing 
Division, Code 7211, Washington, DC 20375, USA}

\begin{abstract}

We study the origin and properties of ``extra'' or ``excess'' central
light in the surface brightness profiles of remnants of gas-rich
mergers.  By combining a large set of hydrodynamical simulations with
data on observed mergers that span a broad range of profiles at
various masses and degrees of relaxation, we show how to robustly
separate the physically meaningful extra light -- i.e.\ the stellar
population formed in a compact central starburst during a gas-rich
merger -- from the outer profile established by violent relaxation
acting on stars already present in the progenitor galaxies prior to
the final stages of the merger.  This separation is sensitive to the
treatment of the profile, and we demonstrate that certain fitting
procedures can yield physically misleading results.  We show that our
method reliably recovers the younger starburst population, and examine
how the properties and mass of this component scale with e.g.\ the
mass, gas content, and other aspects of the progenitors.  We consider
the time evolution of the profiles in different bands, and estimate
the biases introduced by observational studies at different phases and
wavelengths.  We show that, when appropriately quantified, extra light
is ubiquitous in both observed and simulated gas-rich merger remnants,
with sufficient mass ($\sim3-30\%$ of the stellar mass) to explain the
apparent discrepancy in the maximum phase-space densities of
ellipticals and their progenitor spirals.  The nature of this central
component provides a powerful new constraint on the formation
histories of observed systems and can inform both our studies of their
progenitors and our understanding of the global kinematics and
structure of spheroids.

\end{abstract}

\keywords{quasars: general --- galaxies: nuclei --- galaxies: active --- 
galaxies: evolution --- cosmology: theory}

\section{Introduction}
\label{sec:intro}

A wide range of observed phenomena support the view that gas-rich
mergers play a central role in galaxy evolution.  The most intense
starbursts, ultraluminous infrared galaxies (ULIRGs), are always
associated with mergers \citep[e.g.][]{joseph85,sanders96:ulirgs.mergers}, with
dense gas in their centers providing material to feed black hole (BH)
growth and to boost the concentration and central phase space density
of merging spirals to match those of ellipticals
\citep{hernquist:phasespace,robertson:fp}.  Studies of ongoing mergers
and remnants from mergers of gas-rich systems
\citep[e.g.,][]{LakeDressler86,Doyon94,ShierFischer98,James99,
Genzel01,tacconi:ulirgs.sb.profiles,dasyra:mass.ratio.conditions,dasyra:pg.qso.dynamics,
rj:profiles,rothberg.joseph:kinematics}, as
well as post-starburst (E+A/K+A) galaxies
\citep{goto:e+a.merger.connection}, have shown that the kinematic
and photometric properties of these objects,
including velocity dispersions,
concentrations, stellar masses, light profiles, and phase space
densities, are consistent with their eventual evolution into typical
$\sim L_{\ast}$ elliptical galaxies. 

The correlations obeyed by these
mergers and remnants
\citep[e.g.,][]{Genzel01,tacconi:ulirgs.sb.profiles,
rothberg.joseph:kinematics,rothberg.joseph:rotation,dasyra:pg.qso.dynamics}
are similar to e.g.\ the observed fundamental plane and Kormendy
relations for relaxed ellipticals, and indicate that they will evolve
onto these relations as their stellar populations age.  This is
further supported by the ubiquitous presence of fine structures such
as shells, ripples, and tidal plumes in ordinary ellipticals
\citep[e.g.][]{schweizerseitzer92,schweizer96}, which are signatures
of mergers
\citep[e.g.][]{quinn.84,hernquist.quinn.87,hernquist.spergel.92}, and
the clustering and mass density of ellipticals, which
are consistent with
passive evolution after formation in mergers
\citep{hopkins:clustering}.  These various lines of evidence are in
accord with the ``merger hypothesis'' \citep{toomre72,toomre77}, that
elliptical galaxies originate from mergers of spirals.

Numerical simulations performed during the past twenty years verify
that {\it major} mergers of {\it gas-rich} disk galaxies can plausibly
account for these phenomena and elucidate the underlying physics.  In
\citet{hopkins:groups.qso}, we describe the phases of evolution
characterizing these mergers and their relationship to quasars,
starbursts, and elliptical galaxies; we briefly summarize these phases
here.  Tidal torques excited during a merger drive rapid inflows of
gas into the centers of galaxies
\citep{barnes.hernquist.91, barneshernquist96},
triggering starbursts \citep{mihos:starbursts.96,springel:multiphase}
and feeding rapid black hole growth \citep{dimatteo:msigma}.  Gas
consumption by the starburst and dispersal of residual gas by
supernova-driven winds and feedback from black hole growth
\citep{springel:red.galaxies} terminate star formation so that the
remnant quickly evolves from a blue to a red galaxy.  The stellar
component of the progenitors provides the bulk of the material for
producing the remnant spheroid
\citep{barnes:disk.halo.mergers,barnes:disk.disk.mergers,
hernquist:bulgeless.mergers,hernquist:bulge.mergers} through violent
relaxation, but the highest phase-space density material and central
``cusp'' come from dissipation in the final, merger-induced
starburst.

A major merger is generally required in order for the tidal forces to
excite a sufficiently strong response to set up nuclear inflows of gas
and build realistic spheroids.  Although simulations suggest that the
precise definition of a major merger in this context is somewhat
blurred by the degeneracy between the mass ratio of the progenitors
and the orbit of the interaction
\citep{hernquist.89,hernquist.mihos:minor.mergers,bournaud:minor.mergers},
numerical studies \citep{younger:minor.mergers,younger:antitruncated.disks} and
observations
\citep{dasyra:mass.ratio.conditions,woods:tidal.triggering} find that
strong gas inflows and morphological transformation are typically
observed only for mass ratios below $\sim 3:1$, despite the greater
frequency of higher mass-ratio mergers.  In what follows, unless
explicitly noted, we use the term ``mergers'' to refer specifically to
major mergers (mass ratio $\lesssim3:1$). 

Gas dissipation and star formation in the compact, merger-induced 
central starburst are key ingredients in these simulations and observations. It has 
long been recognized that many properties of ellipticals cannot be reproduced 
by purely dissipationless mergers of late-type galaxies
\citep{ostriker80,carlberg:phase.space,naab:no.merger.ell}: 
in particular, their high central 
phase-space densities and specific frequency of global clusters 
\citep[e.g.][]{schweizer98}, degree of rotational support, anisotropy, and 
minor axis rotation/kinematic misalignment \citep{barneshernquist96,
cox:kinematics,naab:gas}, and isophotal shapes.
However, these arguments do not apply 
when the merger constituents contain gas, which can 
radiate energy and hence increase phase-space densities 
\citep{lake:merger.remnant.phase.space} 
and form new stellar populations. 

Observed absorption-line spectra and the red colors of elliptical
galaxies suggest that their stars formed at high redshift
($z\gtrsim1$) and that little star formation has since occurred in
them.  If these galaxies were produced by mergers, the progenitors of
many present-day ellipticals were high-redshift spirals.  While relatively
little is known about disk galaxies at high redshift, it is likely
that they were more concentrated and more gas-rich than their
low-redshift counterparts.  Indeed, observational evidence
\citep{erb:lbg.gasmasses} indicates that galaxies at redshift $z\sim2$
do have large gas fractions $f_{\rm gas}\sim0.5$, with some
approaching $f_{\rm gas}\sim0.8-0.9$. Thus, any attempt to understand
the formation, properties, and scaling relations of the present-day
population of elliptical galaxies within the context of the merger
hypothesis must consider gas-rich mergers.

In pioneering studies employing simulations that incorporated
gas-dynamics and star formation, \citet{mihos:cusps, mihos:starbursts.94}
demonstrated that
dissipation in disk-disk mergers leads to central starbursts similar
to those observed in LIRGs and ULIRGs,
leaving behind stellar populations that easily
match the central phase space and surface densities of observed
ellipticals.  In particular, \citet{mihos:cusps} (hereafter, MH94)
predicted that this process should imprint an
observable signature in the surface brightness profiles of
ellipticals, in the form of a steep departure from the outer
\citet{devaucouleurs} $\sim r^{1/4}$-law
(which arises from violent relaxation of stars in the pre-merger
disks) in the inner regions: i.e.\ the presence of a central cusp
or extra light above the inwards extrapolation of the outer
profile. Subsequent observations \citep{hibbard.yun:excess.light,rj:profiles}
have now confirmed the existence of such excess light in at least a 
significant fraction of known gas-rich merger remnants. 
In principle, this could provide an indicator of how much
mass was involved in the dissipational starburst, and therefore a
critical constraint on the formation history of the galaxy. 
\citet{hibbard.yun:excess.light,rj:profiles} also made first 
attempts to quantify the fraction of stellar mass or light involved; 
however, without a detailed comparison with theoretical models to 
provide the appropriate ``null'' model, it is not clear how exactly 
to quantify the amount of ``excess'' in the light profiles. 

An estimate of how much gas is required to match the central densities of
ellipticals was provided by \citet{hernquist:phasespace}, who used
N-body simulations and analytic arguments to suggest that, without
such a dissipational component, mergers would be deficient by
$\sim15\%$ of their mass relative to the inward extrapolation of an
$r^{1/4}$ law. \citet{naab:profiles} found a similar deficit by 
comparing a sample of surface density profiles from collisionless 
(gas-free) disk merger remnants and observed merger remnants 
(but with reasonable agreement in 
profile shapes in the outer regions). 
Since true $\lesssim \lstar $ ellipticals have a
variety of profiles reflected in their Sersic indices, $I\propto
\exp{\{-(r/r_{0})^{1/n}\}}$, with $n\sim2-4$, this implies a range of
extra light fractions $\sim1-20\%$ needed to account for realistic
populations of $\lesssim \lstar $ ellipticals. 

At the time of the predictions made by MH94, these
central cusps or extra light components, the relics of
dissipational formation in a gas-rich merger, had not been observed.
However, with the advent of {\em Hubble Space Telescope} observations
of the centers of elliptical galaxies, it is now well-established that
typical $\lesssim \lstar $ ellipticals exhibit such central light
excesses, with mass ranges and radii comparable to those originally
predicted by MH94 \citep{kormendy99,jk:profiles,
ferrarese:profiles,cote:smooth.transition}. 
The situation is somewhat different
for the most massive ellipticals, which appear to have central
``cores'' or ``missing light,'' generally accompanied by slow rotation
and boxy isophotal shapes \citep{faber:ell.centers}.
It is believed, however, that this family
of ellipticals has been further modified by subsequent re-mergers or
``dry mergers'' of two or more (initially cuspy gas-rich merger
remnant) ellipticals \citep{quillen:00,rest:01,ravindranath:01,
laine:03,lauer:centers,lauer:bimodal.profiles,ferrarese:profiles,
cote:smooth.transition,jk:profiles}.  While the central profiles
of these objects are certainly interesting (and are considered in
\citet{hopkins:cores}), they do not directly inform the history of
gas-rich mergers and are therefore not the subject of our
investigation here.

In principle, there is great potential for understanding galaxy
formation histories and structural properties by identifying
observed cusps with those predicted as the relics of gas-rich central
starbursts by MH94.  These cusps could enable observers to determine
individually the masses involved in violent relaxation versus
dissipation in the merger, and to separate the stellar populations
arising from both processes, further informing the origin of 
radial gradients
in stellar age, metallicity, and abundance in ellipticals
\citep[e.g.][]{mihos:gradients}.  The
importance and fractional contribution of dissipation as a function of
elliptical properties such as mass and formation time could be
assessed, and this could further be used as a direct test for the role
of differential dissipation being a driver in e.g.\ the tilt of the
fundamental plane, as suggested by theoretical models
\citep{robertson:fp,dekelcox:fp}. Correlations with global kinematic
properties can constrain and test theories, like those outlined above,
that invoke dissipation to resolve long-standing objections to the
merger hypothesis. The proposed links between ULIRGs, quasars, and
elliptical galaxy formation in mergers
\citep[e.g.][]{sanders96:ulirgs.mergers,hopkins:qso.all,
hopkins:red.galaxies, hopkins:groups.ell,hopkins:groups.qso} similarly
rely on dissipation occurring in mergers and, hence, 
anticipate that
it should have a profound impact
on the centers of ellipticals.
 
Unfortunately, despite the major improvements in both observations of
the centers of elliptical galaxies and numerical treatments of star
formation and gas physics, there has been little attempt to use the
extra light content of ellipticals as probes in the manner
above. Rather, studies have mainly been limited to determining whether
or not some extra light component is evident.  This owes in large
part to a lack of theoretical underpinning: the original MH94 work
simply predicted that such cusps should exist; more detailed
interpretation and modeling was limited by spatial and numerical
resolution, and the inability to characterize gas physics with
realistic empirically calibrated models for star formation and
supernova feedback in the interstellar medium.  There have been
dramatic improvements in these areas in recent years
\citep[e.g.][]{springel:multiphase,springel:models}, but the issue has
not been revisited.

Here, and in a companion paper \citep{hopkins:cusps.ell}, we examine
galaxy cusps or extra light components in both simulations and
observed galaxies, in order to identify the origin of different
components of elliptical surface density profiles, their cosmological
scalings and relevance for the formation history of such galaxies, and
their implications for global galaxy properties.  In the present
paper, we focus on the extra light in our simulations and in known
gas-rich merger remnants, in order to isolate the relevant physics and
to calibrate a means to recover the most important physical quantities
involved.

In \S~\ref{sec:sims} we describe our suite of gas-rich merger
simulations, and in \S~\ref{sec:data} we summarize the
observational data set used in our comparisons.
In \S~\ref{sec:origins} we highlight the
physical origin and nature of cusp or extra light components. In
\S~\ref{sec:fits} we compare different means by which to fit the
surface density profile, and attempt to calibrate fitting methods in
order to recover the physically distinct (dissipational versus
dissipationless) components in merger remnants. In
\S~\ref{sec:resolution} we briefly discuss the effects of resolution
and similar numerical issues involved, before proceeding in
\S~\ref{sec:obs} to compare our simulations and fitted galaxy
decomposition to observed systems.  In \S~\ref{sec:dept} we
investigate how the amount of extra light depends on the gas content
of the merging galaxies.  In \S~\ref{sec:structure} we use these
comparisons to study how structural parameters of the outer stellar
light and inner extra light component scale with e.g.\ galaxy mass and
size. In \S~\ref{sec:evolution} we calculate how these results and
comparisons are affected by the choices of observed wavebands and time
evolution after the merger. Finally, in \S~\ref{sec:discuss} we
discuss our results and outline future explorations of these
correlations.

Throughout, we adopt a $\Omega_{\rm M}=0.3$, $\Omega_{\Lambda}=0.7$,
$H_{0}=70\,{\rm km\,s^{-1}\,Mpc^{-1}}$ cosmology, and normalize all
observations and models shown to this cosmology.  We also adopt a
\citet{chabrier:imf} initial mass function (IMF),
and convert all stellar masses and
mass-to-light ratios to this choice. The exact choice of IMF
systematically shifts the normalization of stellar masses herein, but
does not substantially change our comparisons. All magnitudes are in
the Vega system, unless otherwise specified.

\section{The Simulations}
\label{sec:sims}

Our simulations were performed with the parallel TreeSPH code {\small
GADGET-2} \citep{springel:gadget}, employing the fully conservative
formulation \citep{springel:entropy} of smoothed particle
hydrodynamics (SPH), which conserves energy and entropy simultaneously
even when smoothing lengths evolve adaptively \citep[see
e.g.,][]{hernquist:sph.cautions,oshea:sph.tests}.  Our simulations
account for radiative cooling, heating by a UV background \citep[as
in][]{katz:treesph,dave:lyalpha}, and incorporate a sub-resolution
model of a multiphase interstellar medium (ISM) to describe star
formation and supernova feedback \citep{springel:multiphase}.
Feedback from supernovae is captured in this sub-resolution model
through an effective equation of state for star-forming gas, enabling
us to stably evolve disks with arbitrary gas fractions \citep[see,
e.g.][]{springel:models,
springel:spiral.in.merger,robertson:disk.formation,robertson:msigma.evolution}.
This is described by the parameter $\qeos$, which ranges from
$\qeos=0$ for an isothermal gas with effective temperature of $10^4$
K, to $\qeos=1$ for our full multiphase model with an effective
temperature $\sim10^5$ K. We also compare with a subset of simulations
which adopt the star formation feedback prescription from
\citet{mihos:cusps,mihos:gradients,
mihos:starbursts.94,mihos:starbursts.96}, in which
the ISM is treated as a single-phase isothermal medium and feedback
energy is deposited as a kinetic impulse.

Although they make little difference to the extra light component,
supermassive black holes are usually included at the centers of both
progenitor galaxies.  These black holes are represented by ``sink''
particles that accrete gas at a rate $\Mdot$ estimated from the local
gas density and sound speed using an Eddington-limited prescription
based on Bondi-Hoyle-Lyttleton accretion theory.  The bolometric
luminosity of the black hole is taken to be $L_{\rm
bol}=\epsilon_{r}\dot{M}\,c^{2}$, where $\epsilon_r=0.1$ is the
radiative efficiency.  We assume that a small fraction (typically
$\approx 5\%$) of $L_{\rm bol}$ couples dynamically to the surrounding
gas, and that this feedback is injected into the gas as thermal
energy, weighted by the SPH smoothing kernel.  This fraction is a free
parameter, which we determine as in \citet{dimatteo:msigma} by
matching the observed $M_{\rm BH}-\sigma$ relation.  For now, we do
not resolve the small-scale dynamics of the gas in the immediate
vicinity of the black hole, but assume that the time-averaged
accretion rate can be estimated from the gas properties on the scale
of our spatial resolution (roughly $\approx 20$\,pc, in the best
cases).

The progenitor galaxy models are described in
\citet{springel:models}, and we review their properties here.  For each
simulation, we generate two stable, isolated disk galaxies, each with
an extended dark matter halo with a \citet{hernquist:profile} profile,
motivated by cosmological simulations \citep{nfw:profile,busha:halomass}, 
an exponential disk of gas and stars, and (optionally) a
bulge.  The galaxies have total masses $M_{\rm vir}=V_{\rm
vir}^{3}/(10GH_{0})$ for $z=0$, with the baryonic disk having a mass
fraction $m_{\rm d}=0.041$, the bulge (when present) having $m_{\rm
b}=0.0136$, and the rest of the mass in dark matter.  The dark matter
halos are assigned a
concentration parameter scaled as in \citet{robertson:msigma.evolution} appropriately for the 
galaxy mass and redshift following \citet{bullock:concentrations}. We have also 
varied the concentration in a subset of simulations, and find it has little 
effect on our conclusions (because the central regions of the 
galaxy are, in any case, baryon-dominated). 
The disk scale-length is computed
based on an assumed spin parameter $\lambda=0.033$, chosen to be near
the mode in the $\lambda$ distribution measured in simulations \citep{vitvitska:spin},
and the scale-length of the bulge is set to $0.2$ times this.

Typically, each galaxy initially consists of 168000 dark matter halo
particles, 8000 bulge particles (when present), 40000 gas and 40000
stellar disk particles, and one BH particle.  We vary the numerical
resolution, with many simulations using twice, and a subset up to 128
times, as many particles. We choose the initial seed
mass of the black hole either in accord with the observed $M_{\rm
BH}$-$\sigma$ relation or to be sufficiently small that its presence
will not have an immediate dynamical effect, but we have varied the seed
mass to identify any systematic dependencies.  Given the particle
numbers employed, the dark matter, gas, and star particles are all of
roughly equal mass, and central cusps in the dark matter and bulge
are reasonably well resolved.

We consider a series of several hundred simulations of colliding
galaxies, described in detail in 
\citet{robertson:fp,robertson:msigma.evolution} and
\citet{cox:xray.gas,cox:kinematics}.  We vary the numerical resolution, the orbit of the
encounter (disk inclinations, pericenter separation), the masses and
structural properties of the merging galaxies, presence or absence 
of bulges in the progenitor galaxies, initial gas fractions,
halo concentrations, the parameters describing star formation and
feedback from supernovae and black hole growth, and initial black hole
masses. 

The progenitor galaxies have virial velocities $V_{\rm vir}=55, 80, 113, 160,
226, 320,$ and $500\,{\rm km\,s^{-1}}$, and redshifts $z=0, 2, 3, {\rm
and}\ 6$, and our simulations span a range in final spheroid mass
$\mbh\sim10^{8}-10^{13}\,M_{\sun}$, covering essentially the
entire range of the observations we consider at all redshifts, and
allowing us to identify any systematic dependencies in our models.  We
consider initial disk gas fractions (by mass) of $\fgas = 0.05,\ 0.1,\ 0.2,\ 0.4,\ 0.6,\ 
0.8,\ {\rm and}\ 1.0$ for several choices of virial velocities,
redshifts, and ISM equations of state. The results described in this
paper are based primarily on simulations of equal-mass mergers;
however, by examining a small set of simulations of unequal mass
mergers, we find that the behavior does not change dramatically for
mass ratios below about 3:1 or 4:1. This range is appropriate to the
observations of merging galaxies used in this paper, which are
restricted to major merger events. At higher mass ratios, our experiments 
\citep{younger:minor.mergers}
suggest that gas can be channeled by a minor merger into the 
central regions of the galaxy, with an efficiency that declines in 
approximately linear fashion with the mass ratio. However, the resulting 
starburst forms a small bulge in a disk galaxy, so the decomposition 
into a Sersic profile representing the scattered stars and extra light component 
becomes ambiguous (both physically and observationally). 

Each simulation is evolved until the merger is complete and the remnants are 
fully relaxed, typically $\sim1-2$\,Gyr after the final merger 
and coalescence of the BHs. We then analyze the 
remnants following \citet{cox:kinematics}, in a manner designed to mirror 
the methods typically used by observers. For each remnant we project the 
stars onto a plane as if observed from a particular direction (we consider 
100 viewing angles to each remnant, which uniformly sample the unit sphere). 
Given the projected stellar mass distribution, we calculate the iso-density contours 
and fit ellipses 
to each (fitting major and minor 
axis radii and hence ellipticity at each iso-density contour), 
moving concentrically from $r=0$ until the entire stellar mass 
has been enclosed. This is designed to mimic observational isophotal fitting 
algorithms \citep[e.g.][]{bender:87.a4,bender:88.shapes}. The radial deviations 
of the iso-density contours from the fitted ellipses are 
expanded in a Fourier series in the standard fashion to determine 
the boxiness or diskiness of each contour (the $a_{4}$ parameter). 
Throughout, we show profiles and quote our results in 
terms of the major axis radius. For further details, we refer to \citet{cox:kinematics}.

We directly extract the effective radius $\re$ as the projected stellar 
half-mass radius, and the velocity dispersion $\sigma$ as the average 
one-dimensional velocity dispersion within a circular 
aperture of radius $\re$. Note that this differs from what is sometimes adopted 
in the literature, where $\re$ is determined from the best-fit Sersic profile, but because 
we are fitting Sersic profiles to the observed systems we usually quote both the 
true effective radius of the galaxy and effective radii of the fitted Sersic components. 
Throughout, the stellar mass $M_{\ast}$ refers to the total stellar mass of the galaxy, and 
the dynamical mass $\mdyn$ refers to the 
traditional dynamical mass estimator 
\begin{equation}
\mdyn\equiv k\,\frac{\sigma^{2}\,\re}{G},
\end{equation}
where we adopt $k=4.2$ (roughly what is 
expected for a \citet{hernquist:profile} profile, and the choice that most accurately 
matches the true enclosed stellar plus dark matter mass within $\re$ in our 
simulations; 
although this choice is irrelevant as long as we apply it 
uniformly to both observations and simulations). 
When we plot quantities such as $\re$, $\sigma$, and $\mdyn$, we 
typically show just 
the median value for each simulation across all $\sim100$ viewing directions.
The sightline-to-sightline 
variation in these quantities is typically smaller than the 
simulation-to-simulation scatter, but we explicitly note where it is large.

\section{The Observations}
\label{sec:data}

We compare these simulations with the sample of local remnants of recent gas-rich
mergers from \citet{rj:profiles} (henceforth \rj). 
For these objects, \rj\ compile
$K$-band imaging, surface brightness, ellipticity, and $a_{4}/a$
profiles, where the profiles typically range from $\sim100\,$pc
to $\sim10-20$\,kpc. These span a moderate range in luminosity
(including objects from $M_{K}\sim-20$ to $M_{K}\sim-27$, but with
most from $M_{K}\sim-24$ to $M_{K}\sim-26$) and a wide range in merger
stage, from ULIRGs and (a few) unrelaxed systems to shell
ellipticals. As demonstrated therein, these systems will generally
become (or already are, depending on the classification scheme used)
typical $\sim \lstar$ ellipticals, with appropriate phase space
densities, surface brightness profiles, fundamental plane relations,
kinematics, and other properties.

In terms of direct comparison with our
simulations, the data often cover a dynamic range and have resolution
comparable to our simulations, and
(experimenting with different smoothings and imposed dynamic range
limits) we find it is unlikely that resolution or seeing differences
will substantially bias our comparisons.
We have converted the observations to our adopted cosmology, 
and compile other global parameters 
including e.g.\ kinematic properties and luminosities in other bands 
from \citet{rothberg.joseph:kinematics,rothberg.joseph:rotation} and the literature. 
We estimate stellar masses ourselves in a uniform manner for 
all the objects given the 
mean $K$-band $M/L$ as a function of luminosity from \citet{bell:mfs}, corrected 
for our adopted IMF. 
For objects where optical photometry is available, we have 
repeated our analysis using stellar masses derived from 
total $K$-band luminosities and 
$(B-V)$ color-dependent mass-to-light ratios from \citet{bell:mfs}
or from fitting 
integrated $UBVRIJHK$ photometry of each object to a 
single stellar population with the models of \citet{BC03}, and 
find this makes no difference to our conclusions. 
Of course, since these are young objects and we are not performing 
full SED fitting, there is some uncertainty in these estimates. We find a similar result, however, 
using dynamical masses from the kinematics in \citet{rothberg.joseph:kinematics} -- 
although these too are uncertain 
possibly owing to the incomplete relaxation of the systems. 
Within the considerable scatter in the observed properties, however, 
we find these details makes little difference. 

Throughout, we will usually refer interchangeably to the observed surface
brightness profiles in the given bands and the surface stellar mass
density profile. Of course, stellar light is not exactly the same as
stellar mass, but in \S~\ref{sec:evolution} we consider the differences between
the stellar light and the stellar mass density profiles as a function
of time, wavelength, and properties of the merger remnant, and show
that the $K$-band results introduce little bias (i.e.\ are good tracers of
the stellar mass); the Sersic indices and extra light fractions fitted
to the $K$-band profiles of the simulations are good proxies for the
Sersic index of the stellar mass profile and extra mass/starburst mass
fraction, even close in time to the peak episode of star formation.
As has been noted in other works,
many of these systems have weak color gradients, 
empirically supporting the idea that there is generally only weak
variation in $M/L$ with radius.

It is important to note that while we are not concerned about the
absolute normalization of the profile (i.e.\ mean $M/L$), since we
derive total stellar masses separately from the integrated photometry,
we must account for systematics that might be induced by change in
$M/L$ as a function of radius.  The profiles in optical bands such as
$V$ require more care -- when the system is very young ($\lesssim
1-2$\,Gyr after the major merger-induced peak of star formation),
there can be considerable bias or uncertainty owing to stellar
population gradients and dust. However, once the system is relaxed,
the optical bands also become good proxies for the stellar mass
distribution.

\section{The Physical Origin of ``Extra Light''}
\label{sec:origins}

\begin{figure*}
    \centering
    \plotone{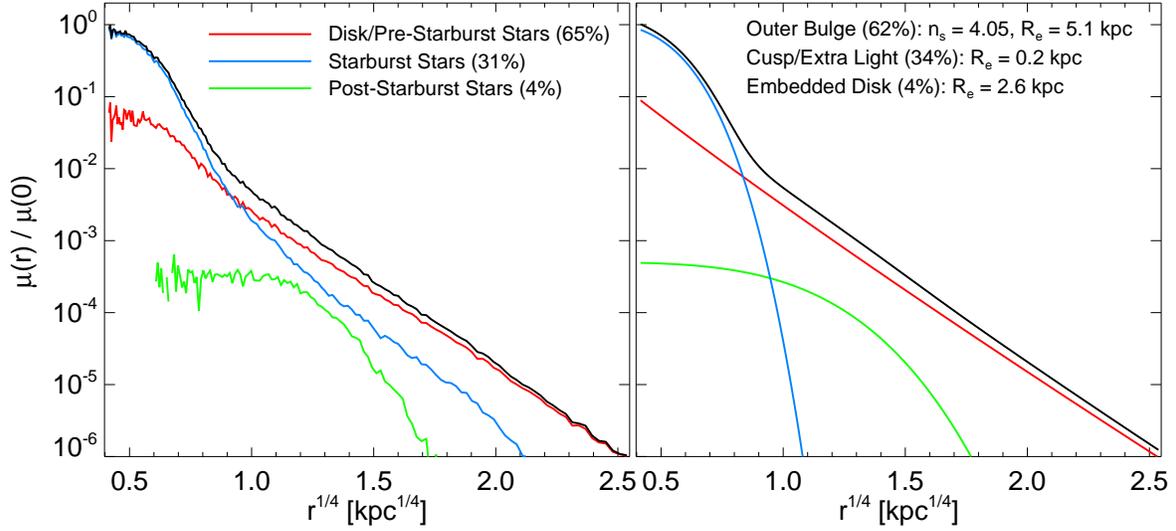}
    \caption{Projected surface mass density of stars 
    in the remnant of a highly gas-rich merger.
    {\em Left:} Total profile (black) is shown with contribution from various physical components: 
    stars in pre-simulation disks and those formed in the simulated disks before the final merger 
    (disk/pre-starburst stars), 
    stars formed in the compact, merger-induced starburst at final coalescence (starburst stars), and 
    stars formed from gas which survives the merger (post-starburst stars). 
    {\em Right:} Sersic profile fits to each of these 
    components, labeled as ``outer bulge,'' ``cusp/extra light,'' and ``embedded disk,'' 
    respectively (sum shown in black). Mass fractions, 
    Sersic indices ($n_{s}$) and effective radii of each component 
    are plotted ($n_{s}=1$ for the embedded disk and cusp/extra light component). 
    \label{fig:origins}}
\end{figure*}

We begin by considering a fiducial case which illustrates many of the 
physical processes relevant to the origin of 
extra light in gas-rich galaxy mergers. 
Figure~\ref{fig:origins} shows the surface mass density of stars (from 
a random sightline) of a typical, but extremely gas-rich, simulation. 
The merging galaxies are similar to $\sim L_{\ast}$ 
disk galaxies (Sc-type galaxies) with a large initial gas 
fraction $\fgas\sim0.8$, and merge on an orbit representative of most 
random encounters. 
Despite the large gas content, the 
total stellar mass profile is quite smooth, at least down to the 
resolution limit of this particular simulation ($\sim50$\,pc). However, 
there is a suggestion of a central extra light concentration or cusp 
within $\sim0.5-1\,$kpc.  (We choose a high gas fraction case merely to 
illustrate the salient features of our findings; as we detail in
what follows, mergers
with lower gas fractions yield qualitatively similar results.)

The simulation allows us to decompose the stellar mass profile according 
to the time when various stellar populations were formed. We 
therefore consider three classes of stars. First, the stars that were present 
in the initial stellar disks at the beginning of the simulation (``disk stars'') 
together with those stars 
formed before the 
final coalescence of the galaxies and starburst phase (``pre-starburst'' stars). 
Although the latter includes 
some stars that originate 
in a small starburst caused by the first passage of the 
two galaxies, most of these stars are formed quiescently in the two 
galactic disks before the disks are destroyed in the final merger. 
We identify these two simulated classes of stars as a single 
population because they are both ``disk'' stars entering the final 
merger, and it is of course arbitrary where (or at what time) we 
initialize our simulations, relative to the final merger, and therefore what fraction of the 
stellar mass forms ``before'' our simulation begins. 
As is clear in the figure, these stars are scattered during the merger 
and undergo violent relaxation, producing an $r^{1/4}$-like surface 
density profile. In detail, these stars relax to a nearly perfect Sersic
profile with a best-fit Sersic index of $n_{s}\approx4.0$, which is shown 
in the right panel of Figure~\ref{fig:origins}. As stars, they are collisionless, 
and therefore remain at large radii and relatively low phase-space densities 
characteristic of the scale lengths and concentrations of the initial 
disks, with an effective radius $\re\approx5\,$kpc. 

Second is the ``starburst'' component, defined somewhat loosely as 
those stars which form within $250$\,Myr of the final merger and 
coalescence of the two nuclei (taken for 
convenience as the merger of the two central black holes, but adopting 
another definition such as the peak in the star formation rate yields a
similar result). This component is much more compact ($\re=0.2$\,kpc), characteristic 
of the central starburst size scales in observed mergers and ULIRGs 
\citep[e.g.][]{downes.solomon:ulirgs,bryant.scoville:ulirgs.co,Genzel01,tacconi:ulirgs.sb.profiles}. 
Because our definition is based on only the time during which the stars 
formed, it 
includes some stars which originate in the merging disks (rather than 
the central, compact starbursts), which can be seen in the tails of this 
component of the surface density profile. However, these contribute
negligibly to 
the surface brightness profile and total mass budget of the starburst component. 
The central component does not follow an $r^{1/4}$ law, but is more 
similar to an exponential ($n_{s}=1$) profile. We will return to the details of the 
exact shape of this central light component in \S~\ref{sec:structure}, but in general expect 
this -- the central starburst is formed {\em in situ} by gas dissipation and star formation, 
and does not undergo violent relaxation to transform it from an exponential-like to 
an $r^{1/4}$ profile. 

Finally, because of the extremely gas-rich nature of this merger, there is 
some amount of cold gas remaining after the final black hole growth and feedback 
phase. This quickly settles into a rotationally supported disk, and forms 
an embedded disk of ``post-starburst'' stars \citep[described in detail in][]{springel:spiral.in.merger}.
Figure~\ref{fig:origins} shows that this 
can be well-fitted by a relatively small exponential disk ($\re\sim3$\,kpc). Although 
this embedded disk is interesting from the perspective of the galaxy 
kinematics \citep[see][]{cox:kinematics}, 
it contains only a tiny fraction $\sim4\%$ of the stellar mass, and, because this is 
concentrated near the total galaxy effective radius, it adds negligibly to the 
surface brightness profile at all radii. 

We therefore confirm the findings of MH94, that gas channeled to the 
central regions of the galaxy forms stars in a compact starburst and leaves a 
central excess in the light profile above the extrapolation of the light of 
the old stellar population (scattered from the disks of the merging galaxies). 
However, even in Figure~\ref{fig:origins}, where the extra light or starburst 
component constitutes a large fraction ($\sim30\%$) of the galaxy stellar mass, 
the overall light profile is quite smooth, and does not show an obvious sharp transition 
to the extra light component. This is unlike
the earlier results of 
MH94, and we discuss some of the reasons for this in \S~\ref{sec:structure}, but they 
primarily relate to our 
improved resolution and treatment of the ISM relative to MH94, as
well as more accurate time integration.

\section{Fitting to Recover the Physically Appropriate ``Extra Light''}
\label{sec:fits}

\begin{figure*}
    \centering
    \plotone{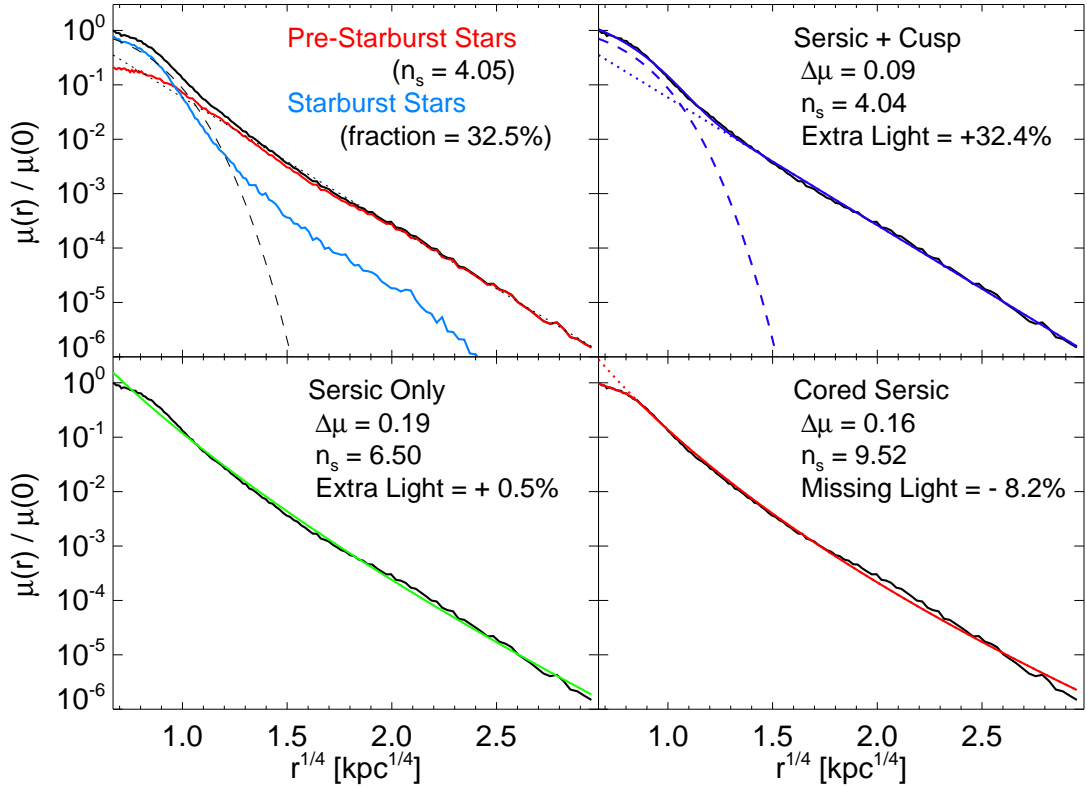}
    \caption{{\em Upper Left:} Surface mass density profile of 
    remnant of a highly gas-rich merger (black),
    decomposed into stars formed prior to the final merger (which 
    are then violently relaxed; red) and stars formed in the merger-driven 
    dissipational starburst (blue). The Sersic fit to each component is shown, with 
    the Sersic index of the pre-starburst component and mass fraction of the 
    starburst component labeled. {\em Upper Right:} Two-component (Sersic plus 
    cusp or extra light; where the cusp is fit with an exponential profile) fit 
    to the total light profile, with the Sersic index of the outer component 
    and mass fraction of the inner ($n_{s}=1$) component, and rms scatter ($\Delta\mu$) about the 
    fit. {\em Lower Left:} Single Sersic function fit to the profile. {\em Lower Right:} 
    Cored Sersic function fit. The two-component fit as we parameterize it accurately 
    recovers the Sersic profile of the violently relaxed component and mass fraction 
    of the starburst component. The other fits give physically misleading results in this 
    case.
    \label{fig:demo.fit.danger}}
\end{figure*}

We would like to be able to use the surface brightness profile of the merger remnant to 
estimate the contribution of the extra light -- ideally as a means of estimating, from 
the amount of extra light, how much of the galaxy mass was involved in a compact 
central starburst. However, as noted in \S~\ref{sec:origins}, the light profiles 
in our simulation remnants are quite smooth, even where the extra light fraction is 
 large. This makes identifying the extra light component a non-trivial procedure, 
which is sensitive to the assumptions made in fitting. 

We begin by reducing a galaxy surface density profile to the two most 
relevant physical components: the ``pre-starburst'' or ``disk'' stars, i.e.\ those formed 
in the rotationally supported disks before the final coalescence of the galaxies, 
and the ``starburst'' stars, formed in the compact, dissipational final starburst. 
We neglect the embedded stellar 
disks which can be formed by gas remaining after the merger; as demonstrated in 
Figure~\ref{fig:origins}, these contribute negligibly to the surface mass density profile 
in even the most gas-rich merger remnants. 

Figure~\ref{fig:demo.fit.danger} shows the surface density of a fiducial gas-rich 
example, decomposed into a pre-starburst and starburst population. Motivated 
by observational methods and our results from \S~\ref{sec:origins}, we 
fit\footnote{Formally, we consider our simulation profiles outside of 
some multiple $\sim3-5$ times the resolution limit, or with a seeing correction 
appropriate for the \rj\ sample (much larger than our resolution limits). We equally sample 
the profile in $\log{r}$ over a dynamic range extending to the largest radii in the \rj\ sample, 
and weight each point equally assuming an intrinsic $\sim0.1\,$mag point-to-point variance 
in the SB profile (the typical magnitude of residuals fitting arbitrary splines to the data). 
We have varied these choices and find that our fits and conclusions are 
not sensitive to them.} 
({\em upper left})
a Sersic profile to the pre-starburst component, and find that it follows a nearly 
perfect $r^{1/4}$ law ($n_{s}=4.05$). 
Loosely motivated by the shape of the starburst stellar profile at small radii and its origin 
in a gas-rich dissipational starburst, we fit the starburst stellar component to an 
exponential Sersic law ($n_{s}=1$). 
We emphasize that neither the simulations nor observations herein 
resolve the innermost regions of the central light where the choice of its Sersic index 
makes a significant difference, so we 
adopt this for simplicity as a standard choice in fitting the 
extra light component (we discuss this choice in greater detail below and \S~\ref{sec:structure}). 
This describes the starburst mass profile quite well where it is important to the overall 
surface density. It does not capture the extended stars at $\gtrsim3\,$kpc 
formed during this time, but as discussed in \S~\ref{sec:origins}, we 
can safely ignore them.
The exponential fit accurately recovers the total 
mass in the starburst component and its effective radius, and the shape of this profile 
where it is relevant to the total surface brightness profile. 

Rather than fit directly to the physical starburst and pre-starburst components, we 
now attempt to fit a two-component model to the observed quantity, the 
total surface brightness profile ({\em upper right}). 
We therefore fit the surface brightness profile to a 
double Sersic model, with an outer component reflecting the pre-starburst stars with a 
free Sersic index, and an inner component reflecting the starburst stars 
with a fixed Sersic index $n_{s}=1$. The total surface brightness profile is then 
\begin{equation} 
I_{\rm tot} = I_{sb}\,\exp{{\Bigl\{}-b_{1}\,{\Bigl(}\frac{r}{r_{sb}}{\Bigr)}{\Bigr\}}}+
I_{b}\,\exp{{\Bigl\{}-b_{n}\,{\Bigl(}\frac{r}{r_{b}}{\Bigr)}^{1/n}{\Bigr\}}},
\end{equation} 
where $r_{sb}$ and $r_{b}$ are the effective radii of the $n_{s}=1$ and 
free $n_{s}$ components (which we identify with the 
``starburst'' and ``old bulge'' or ``pre-starburst'' components, 
respectively), $I_{sb}$ and $I_{b}$ are the corresponding normalizations, 
and $n$ is the Sersic index of the outer bulge or pre-starburst component. The 
constant $b_{n}$ is the appropriate function of $n$ such that 
$r_{sb}$ and $r_{b}$ correspond to the projected half-mass radii. 
Note that changing our choice of $n_{s}=1$ for the inner Sersic component 
will systematically shift the mass balance between the inner and outer components, 
and lower the fitted outer Sersic index (typically increasing the 
inner mass by $\sim30\%$ and lowering the outer Sersic index
$\Delta n_{s}\sim0.25$, respectively, if we adopt $n_{s}=2$). 

We have therefore 
experimented with a range of choices for the inner Sersic index: we find 
that fixing $n_{s}=1$ recovers, {\em on average} (across our 
entire suite of simulations), the correct mass of the 
physical starburst component and Sersic index of the pre-starburst stellar component, 
whereas different choices for the inner $n_{s}$ give systematic offsets in 
these quantities. This is not to say that $n_{s}=1$ is most appropriate for 
every object/sightline or that it yields the most physical results in every case: however, 
if we are interested in recovering the physical extra light component in a 
statistical sense, it is most appropriate. 
Furthermore, even where we arrive at a similar result if 
we fit it with a free Sersic index, we find that this introduces considerably larger uncertainties 
when we fit observed systems (owing to the various parameter degeneracies).

Figure~\ref{fig:demo.fit.danger} shows the outcome of this fit. The resulting model 
of the surface density profile fits the simulation well, with 
a variance of only $\Delta\mu\sim0.09\,{\rm mag\,arcsec^{-2}}$ 
(assuming $\mu\propto-2.5\,\log{I_{\rm tot}}$). 
This is comparable to the point-to-point variance in the profile of 
this simulation if we fit an arbitrary spline to the data, and thus reflects a 
genuinely good fit. We have tested and found that the fit parameters are stable 
with respect to the dynamic range and error weighting, and that the residuals 
(typically of this magnitude, $\Delta\mu\sim0.1\,{\rm mag\,arcsec^{-2}}$) 
are independent of radius (i.e.\ there is no systematic bias in the functional form). 
More important, this fit, despite having no direct information about 
the physical components into which we decompose the brightness profile, 
recovers almost exactly the appropriate parameters for both profiles. The 
best-fit Sersic index ($n_{s}=4.04$ compared to $4.05$) and effective radius of the 
``outer'' or ``bulge'' component are almost perfect matches to those fitted directly 
to the pre-starburst stellar population. Likewise, the inner or extra light component 
is a close match to the physical starburst component, and the fit recovers 
the extra light fraction accurately ($32.4\%$ compared to $32.5\%$; a much smaller 
difference than reasonable uncertainties in our physical definition of the starburst 
component). We have repeated this fitting procedure for several hundred simulations, and 
find in almost all cases similar good agreement between the best-fit components 
and the physical decomposition of the galaxy stellar mass profile. 

However, there are a number of local minima in the fitting, and care must be taken to 
fit the most appropriate physical model. We have, in some sense, tuned our two-component 
parameterization (choosing the functional form and fixing the inner $n_{s}=1$) 
so that it, on average, accurately recovers the distinct physical components in our 
simulated merger remnants. Other parameter choices may systematically 
fail (or fail in certain regimes of either high or low true starburst mass fraction) to 
recover a physically meaningful decomposition.

For example, Figure~\ref{fig:demo.fit.danger} 
also shows the results of fitting a pure Sersic function to the entire surface density profile
({\em lower left}). 
There is a reasonable fit to the entire profile with a single Sersic index $n_{s}=6.50$, 
very different from the Sersic index which describes either the pre-starburst or 
starburst light components. Likewise, if we consider the excess light to be that 
light in the real profile above the prediction of the best-fit Sersic model, we would 
infer only a tiny extra light fraction $\sim0.5\%$, essentially consistent with no 
starburst component (given the fit uncertainties). 
Although the fit is technically worse, with variance $\Delta\mu=0.19$, the difference is not 
dramatic, and by many observational standards would be considered a good fit. 
Clearly, however, the fitted results have no physical meaning in this case. 

The problem becomes even worse if we add a degree of freedom and fit a 
``cored Sersic'' profile ({\em lower right}), of the form 
\begin{equation}
I = I'\,{\Bigl[}1+{\Bigl(}\frac{r_{b}}{r}{\Bigr)}^{\alpha}{\Bigr]}^{\gamma/\alpha}\,\exp{{\Bigl[}
-b_{n}\,{\Bigl(}\frac{r^{\alpha}+r_{b}^{\alpha}}{r_{e}^{\alpha}}{\Bigr)}^{1/(\alpha\,n)}{\Bigr]}}
\end{equation}
\citep[e.g.][]{graham:core.sersic}, where $r_{b}$ is the core break radius within which the profile 
breaks to a power law of slope $\gamma$, $r_{e}$ is the effective radius and 
$n$ the Sersic index of the outer light profile, and $\alpha$ is a parameter describing 
how rapidly the break occurs. Figure~\ref{fig:demo.fit.danger} shows the results 
of this fit, which is again good in a purely statistical sense, albeit worse than our 
best fit Sersic+extra light fit ($\Delta\mu=0.16$\,mag). However, here the fit parameters 
become even more unphysical -- the best fit Sersic index is a very steep $n_{s}=9.5$ 
and one actually infers that the system is a {\em cored} galaxy, with {\em missing} light 
relative to the best-fit Sersic profile. 

The reason for these catastrophic failures is that the extra light component blends smoothly 
with the outer pre-starburst light profile. By increasing the central surface brightness, 
the extra light component makes the overall profile appear steeper (concave up 
in the $\mu-r^{1/4}$ projection), owing to the rise at small $r$. However, the cusp itself does 
not continue to rise steeply inwards (in most cases), so after steepening the best-fit 
Sersic index to fit the outer part of the extra light component, one is often forced to 
infer the existence of a core in the central regions. Again, these fits have no direct 
physical meaning, but they are not terrible matches to the light profile. This emphasizes 
that whenever fitting a parameterized profile to the data, one must take care to 
adopt a physically motivated procedure.
Fortunately, there are some indications, from the purely 
observational point of view, that the Sersic only and cored Sersic fits are inappropriate 
(when applied in this manner and into the central regions of gas-rich merger remnants).  

First, 
they begin to fail at large radii -- however, this is where the 
true nature of the Sersic profile of the outer light component is most prominent, so 
any failure at large $r$ should be especially worrisome. Furthermore, when we examine
the kinematics (e.g.\ ellipticity, boxy/diskiness, rotation properties) along the 
major axis, one can often see a transition occur in these properties where 
the extra light begins to dominate the profile (see \S~\ref{sec:obs}), whereas one would expect 
no such change if a continuous Sersic profile was the 
meaningful choice. Finally, 
when fitting a cored Sersic profile in an inappropriate case, one often infers a large 
missing light fraction, relative to what is observed in genuine cored 
galaxies \citep[e.g.\ massive, boxy, slow-rotating ellipticals; see][]{jk:profiles}.
One typically expects 
missing light masses in real cores of $\lesssim1\%$ the galaxy mass, 
comparable to the central black hole masses \citep[e.g.][]{milosavljevic:core.mass}. 

Note that we are not saying that a pure Sersic or cored Sersic profile is always an 
unphysical parameterization of the galaxy light distribution -- however, for 
gas-rich merger remnants, we know in our simulations and have good reason to 
believe observationally that there is some excess light component. In these cases, 
the results of these fits are demonstrably 
unphysical and can be misleading. 

We caution that our adopted 
fitting functions and their physical interpretations are appropriate 
only for the galaxy spheroid, and should not be applied at radii where 
a galactic disk dominates the light profile. Implicitly, our 
choice of functional form also has the advantage of providing a 
check that stellar disks and 
distinct kinematic components do not contribute 
significantly to the surface brightness profile. 
Because it includes a Sersic and an exponential component, if 
a traditional disk plus bulge decomposition of the light profile were required (if 
such a decomposition were the best fit to the observed profile), we would see 
this in the fit -- our intended ``extra light'' component would in fact dominate the light profile 
at {\em large} radii, with a larger effective radius and substantially lower effective surface 
brightness than the fitted bulge. 
We discuss these cases in more detail in 
\citet{hopkins:cusps.ell}, but note here that 
only a few of the simulated or observed systems show this behavior -- in 
other words, they are primarily true ellipticals, not S0 galaxies (which should not 
be surprising, since they are primarily remnants of major $1:1$ mergers). 
In our simulations, we can confirm from the stellar 
kinematics that such fits are only preferred when indeed a rotationally supported disk 
is a significant fraction of the light at large radii -- i.e.\ so long as we restrict ourselves to 
considering galaxy spheroids, there is no ambiguity that our fits are physically distinct from and 
statistically superior to simple disk plus bulge decompositions. 
Ideally, in the cases where a traditional disk/bulge decomposition is required (i.e.\ there 
is a prominent large-scale disk), we would 
fit a three-component model: an outer exponential disk, and a two-component (dissipational 
and dissipationless) bulge. Unfortunately, even with the ideal data quality in our simulations, 
the degeneracies involved in fitting for three independent components are large, and in 
cases where a disk dominates the profile at large radii, there is no robust ``lever arm'' 
on the outer bulge profile shape. We therefore generally exclude these cases from 
our comparisons (although we still can and do 
compare these profiles directly to observed systems), but emphasize that 
including or excluding them 
from our simulation sample or the observed \rj\ sample makes no difference to 
our conclusions. 

We also emphasize that although there are some superficial similarities 
between our adopted parametric profile decomposition and 
that in e.g.\ \citet{cote:virgo} and \citet{ferrarese:profiles}, 
the two are in detail significantly different and address 
very different spatial scales and physical properties of the galaxies. 
Typically, the ``outer profile'' we refer to extends to and beyond (in 
our simulations) the limits of our ground-based photometry, corresponding 
to physical radii of $\sim20-100\,$kpc, and our ``inner profile'' refers to 
the residual from a central starburst at scales where a significant 
fraction of the galaxy mass becomes self-gravitating (see \S~\ref{sec:structure}), 
at $\sim0.5-1\,$kpc. We stress again that we are not resolving 
inwards of the central $\sim30-50\,$pc, and our modeling should 
not be extrapolated to within these radii without considerable care. 
In contrast, the ``outer profile'' in \citet{ferrarese:profiles} is 
based on the HST ACS profiles, which extent to outer 
radii $\sim1$\,kpc, and their ``inner profiles'' typically dominate 
the light profile at very small radii $\sim0.01-0.02\,R_{e}$ ($\sim10-40\,$pc 
for most of their sample). 
This is more akin to separating our 
``inner'' component itself into multiple sub-components -- i.e.\ a starburst 
stellar component that blends (as we have shown) relatively smoothly 
onto the outer, violently relaxed stars and an innermost nuclear 
component. The authors themselves address this, and denote these 
nuclear excesses as central stellar clusters. 
Such systems may indeed be present (and could be formed 
in the same dissipational starburst which we model): but if so they are distinct subsystems 
sitting on top of the starburst light component, 
which we do not have the ability to model or resolve in our simulations or the 
observations (the typical nuclear star cluster scales are well below 
the resolution limits of the observations from \rj). 
Therefore, while the two approaches may yield complementary constraints 
and some similar conclusions, we caution that our 
results are not directly comparable and are specifically designed 
to trace distinct physical structures.

\section{The Effects of Resolution}
\label{sec:resolution}

\begin{figure}
    \centering
    \plotone{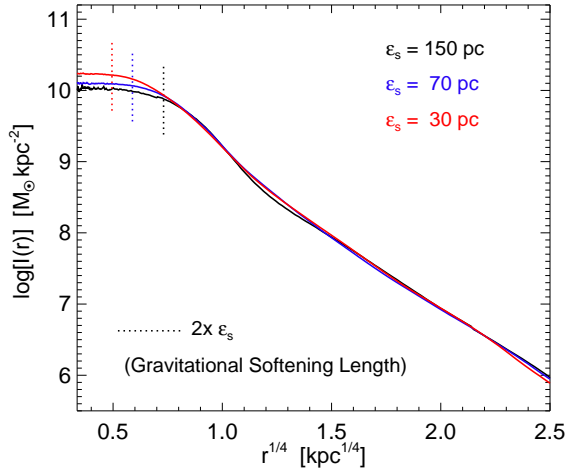}
    \caption{Surface brightness profile as a function of spatial resolution in 
    a resolution study. The profiles are converged outside of roughly a 
    softening length.
    \label{fig:res1}}
\end{figure}

Given the small physical sizes of the starburst or extra light components in the 
simulations, it is important to verify that finite 
spatial resolution is not biasing our conclusions. In general, we have experimented 
with a wide range of simulation spatial and mass resolution scales and particle 
numbers, and find that all of the results in this paper (except where otherwise 
explicitly noted) are robust to the effects of resolution.
However, we briefly present a detailed resolution study to examine these 
issues. 

Figure~\ref{fig:res1} plots the mean (averaged over $\sim100$ sightlines) 
surface mass density profile in a series of otherwise identical simulations 
with increasing spatial resolution, where the 
gravitational softening length decreases from $\approx150\,$pc to $30\,$pc. 
Note that the SPH gas smoothing length can be (and is) much smaller than 
this, especially in the dense, central regions of the galaxy during the 
starburst phase. 

Unsurprisingly, the profiles flatten within several softening lengths. 
However, they are well converged at larger radii (all the differences between 
the mean profiles plotted at $>5$ softening lengths are within
typical sightline-to-sightline variations). The primary difference seen in higher 
resolution simulations is that the profile continues to rise to smaller and smaller radii. 
Note that there is still some well-resolved curvature to the profile -- the central 
extra light or starburst component is not rising as steeply as a 
pure power-law, for example. However, we expect from this trend that 
the profiles, with further improved resolution, should continue rising to 
small radii $\lesssim30-50$\,pc. Below these scales, we caution that, regardless of the 
numerical simulation resolution, our description of star formation and the ISM 
is no longer appropriate. Recall, we describe the ISM and star formation in a 
statistically volume-averaged sense, representative of a balance between 
the multiphase components of the gas. By these smallest scales, we approach the physical 
sizes of molecular clouds and individual stellar clusters, and therefore 
cannot make physically meaningful claims without incorporating these 
gas phases and physics in a complete manner in the simulations.

\begin{figure}
    \centering
    \plotone{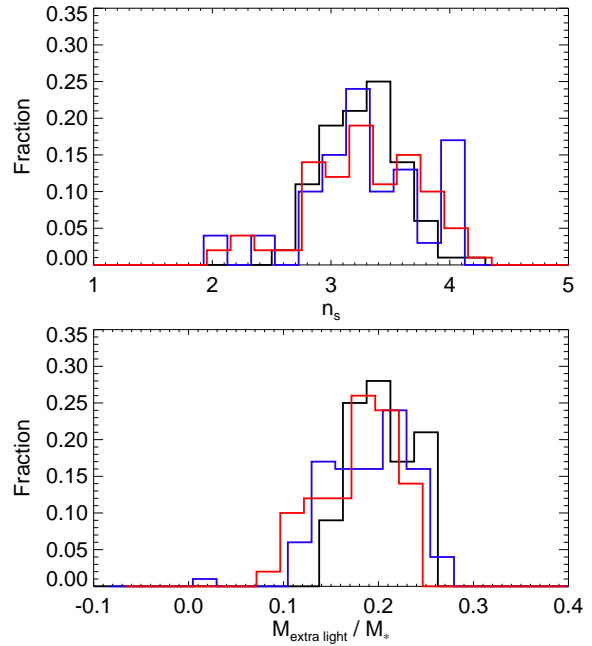}
    \caption{Distribution (across $\sim100$ sightlines) of outer Sersic index and 
    inner/extra light mass fraction fitted to the resolution study profiles in 
    Figure~\ref{fig:res1} (black, blue, and red for $\epsilon_{s}=150,\ 70,\ 30$\,pc, 
    respectively). The fits are reasonably well-converged even at 
    relatively poor ($150\,$pc) spatial resolution.
    \label{fig:res2}}
\end{figure}

Given these issues, it is clear that simulation resolution does affect the 
shape of the central regions of the extra light components, and hence
we cannot make strong conclusions regarding their innermost shapes.
However, we also see that the 
outer profile (the pre-starburst component) is well-converged, and that the 
simulations exhibit a transition to an extra light component at similar radii 
for all resolutions. The resolution limits {\em do not} affect the radii where the 
excess light begins to depart from the outer profile ($\sim0.5-1\,$kpc), 
or most of the range of the observed \rj\ profiles with which we compare 
($\gtrsim100\,$pc).
Figure~\ref{fig:res2} shows the distribution of 
Sersic indices fitted to the outer component of these simulations (across 
all sightlines), and the distribution of the fitted extra light component 
fraction. Within the errors, these distributions are more or less fully 
converged even for our lowest-resolution simulation. 
Similarly, if we directly extract the ``starburst'' and ``pre-starburst'' components 
from the simulations, we find that their total mass fractions, 
effective radii, and surface brightness 
profiles (outside a few softening lengths) are converged. 
We also show in \S~\ref{sec:dept} that the size-mass relation of the 
extra light components in the simulation is robust over the entire 
range of resolutions we have studied. We therefore conclude that 
resolution limitations, while constraining our ability to follow the profile of the 
extra light to the smallest radii, do not bias our overall estimates of the 
galactic structure or 
contribution of the extra light to the total mass.

\section{Comparison with Observations: The Extra Light in Local Merger Remnants}
\label{sec:obs}

Bearing these cautions in mind, we apply our fitting procedure (calibrated from 
our simulations) to the sample of $\sim50$ local merger remnants from \rj\
which spans a representative range in their post-merger stages (from 
LIRG and ULIRG systems still undergoing some star formation to 
shell ellipticals which are completely relaxed). 

\begin{figure}
    \centering
    \plotter{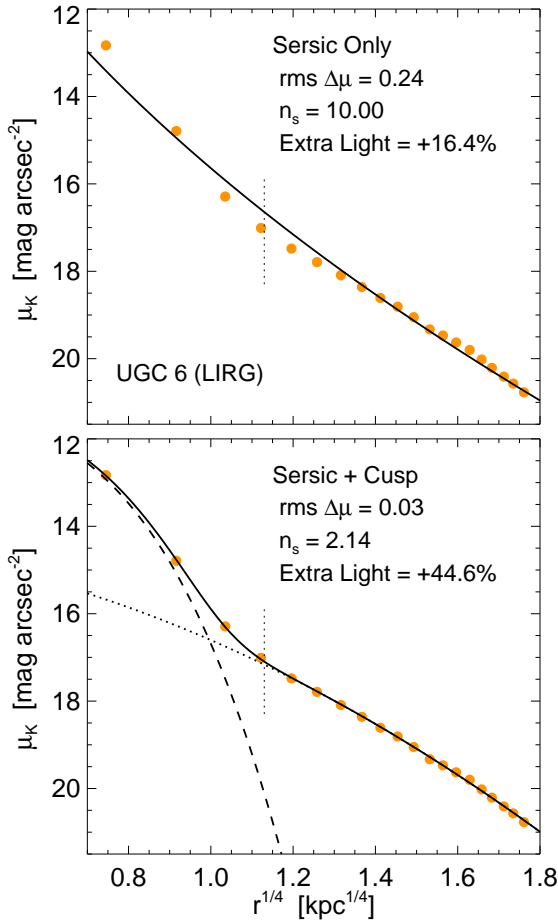}
    \caption{Observed $K$-band surface brightness profile (points) of the LIRG merger remnant 
    UGC 6. {\em Top:} Single Sersic function fit (solid line) from \rj, with the 
    variance about the fit $\Delta\mu$, Sersic index, and extra light fraction. 
    Dotted vertical line marks the half-light radius. 
    {\em Bottom:} Our two component fit (total, inner, and outer as solid, dashed, and dotted). 
    The fit quality is dramatically improved, and the outer Sersic index better describes the 
    shape of the light profile at large radii. 
    \label{fig:ugc6}}
\end{figure}

We first illustrate the caveats from \S~\ref{sec:fits} with a representative 
example. Figure~\ref{fig:ugc6} shows the $K$-band luminosity profile of 
UGC6, a merger remnant and LIRG in the sample of \rj. \rj\ 
fit their profiles to single Sersic laws, and their best fit is shown -- they find a 
very large Sersic index $n_{s}=10$. 
\rj\ note that such a high Sersic value is probably unphysical, and parameters 
derived from it should be regarded with caution. Supporting this, 
\citet{naab:profiles} note that collisionless merger remnants, which 
can be reasonably approximated by a single Sersic law, do not 
result in such large Sersic indices. 
\rj\ also make a first attempt to quantify the extra light in their objects by 
comparing the total luminosity to the inward extrapolation of a 
$r^{1/4}$-law (different from the 
best-fit free Sersic index fit), similar in spirit to our approach but with 
less allowance for the detailed structure of the inner and outer 
portions of the galaxy.  
It is clear that the observed 
profile indeed shows considerable structure not captured in this fit (overall 
it is rather poor, with $\Delta\mu=0.24$\,mag), and that the
profile at large $r$ is in fact concave down in this projection (indicating 
$n_{s}<4$), not concave up ($n_{s}>4$). 
Furthermore, the status of this object 
as a LIRG indicates that the merger must have been highly gas-rich; such 
gas-rich events are expected to produce more disk-like outer light profiles, 
not large Sersic indices. 

Fitting instead to a Sersic+extra light component yields a much better 
fit in a statistical sense, even accounting for the added degrees of 
freedom ($\Delta\mu=0.03$), as well as much more physical 
parameters -- the outer Sersic index $n_{s}=2.1$ matches the concavity of 
the light profile and is expected in a gas-rich event, the profile is matched 
near the effective radius, and the shape of the extra light component is 
well-fitted. The resulting extra-light component is much larger with this fit, 
$\sim45\%$ (compared to $\sim16\%$) -- this may seem high relative to 
what is typically quoted for such systems, but this has to do with our 
choice of fitting method. Indeed, such large fractional extra light components 
are reasonably common in simulations of very gas-rich mergers expected to produce 
LIRGs (see the example in \S~\ref{sec:origins}), and it is also observationally 
quite reasonable to see $\sim$ half the $K$-band luminosity of a LIRG originate 
in a relatively compact $\sim$ kpc nucleus (note though, as demonstrated 
in \S~\ref{sec:evolution}, that the brightest LIRG and ULIRGs are the objects 
for which, even in $K$-band, there may be more extra {\em light} 
than extra {\em mass}, owing to stellar population 
effects).

\begin{figure*}
    \centering
    \plotone{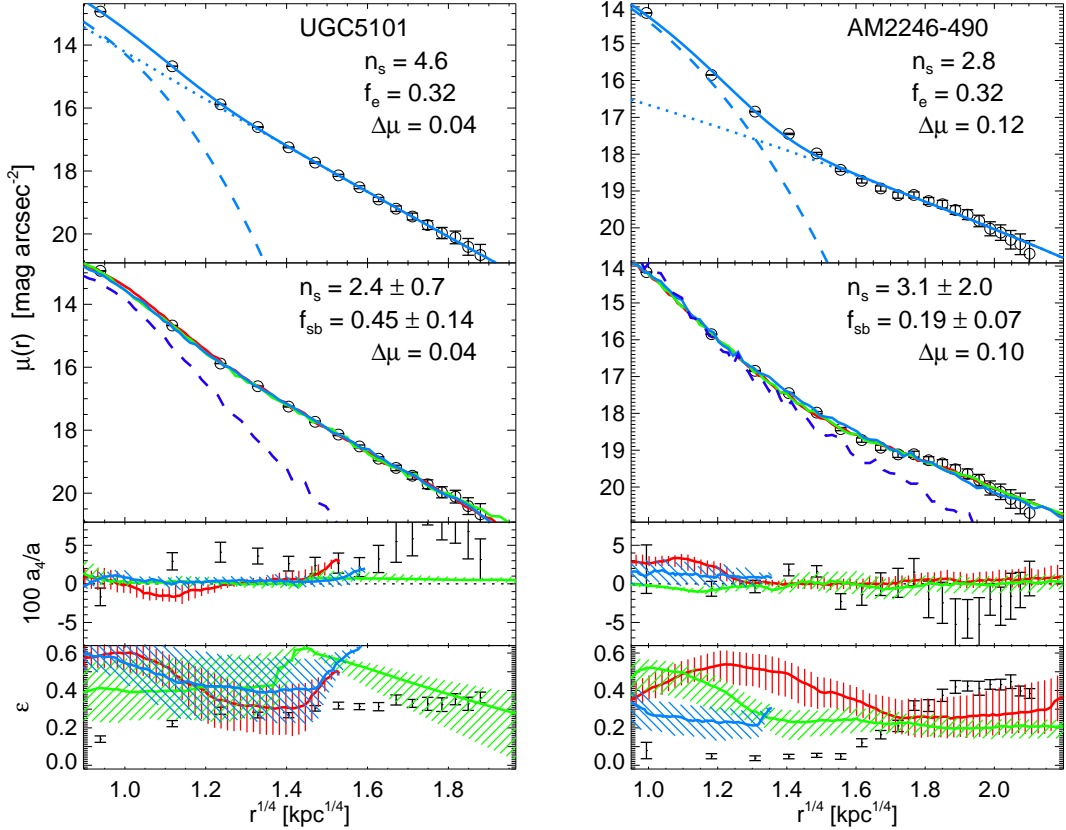}
    \caption{Observed, deconvolved major-axis $K$-band light profiles of 
    gas-rich merger remnants from \rj\ (points with error bars). 
    {\em Top:} Fitted two-component model (inner, outer, and total as dashed, dotted, 
    and solid lines), with outer Sersic index ($n_{s}$), extra light fraction 
    ($f_{e}$), and variance ($\Delta\mu$) shown. 
    {\em Middle:} Direct comparison with simulated 
    surface brightness profiles. The three simulations 
    which match most closely are shown (red, green, and blue lines from best to 
    third-best fit). Dashed line shows the physical starburst component of the 
    best-fitting simulation. Range in outer $n_{s}$ and physical starburst mass 
    fraction ($f_{\rm sb}$) for the best-fitting simulations are shown, with the variance of the observations 
    with respect to the simulated profile. 
    {\em Bottom:} Observed disky/boxy-ness ($a_{4}$) and ellipticity profiles, 
    with the median (solid) and $25-75\%$ range (shaded) corresponding profile 
    from the best-matching simulations above. Note that these are not fitted for in any sense. 
    This panel shows the results for the ULIRGs, which require large gas 
    contents and large extra light masses. 
    \label{fig:rj1}}
\end{figure*}
\begin{figure*}
    \centering
    \plotone{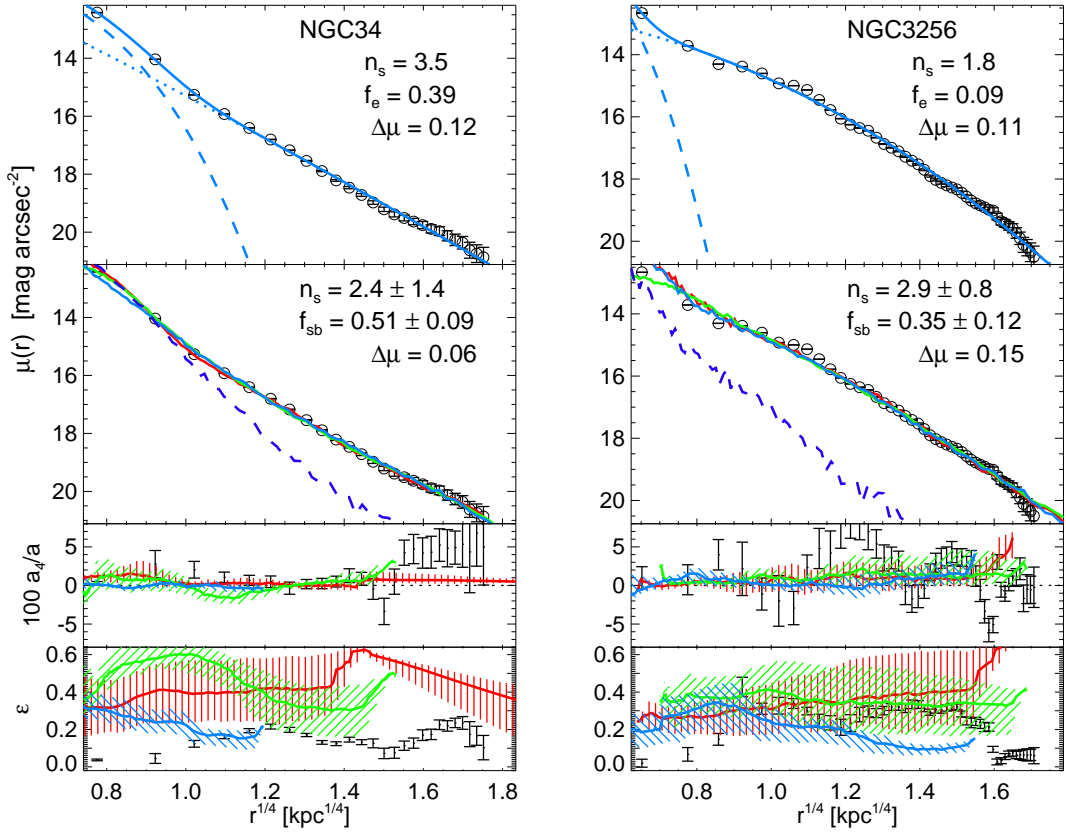}
    \caption{As Figure~\ref{fig:rj1}. Two LIRGs, which also require large 
    gas content.
    \label{fig:rj2}}
\end{figure*}
\begin{figure*}
    \centering
    \plotone{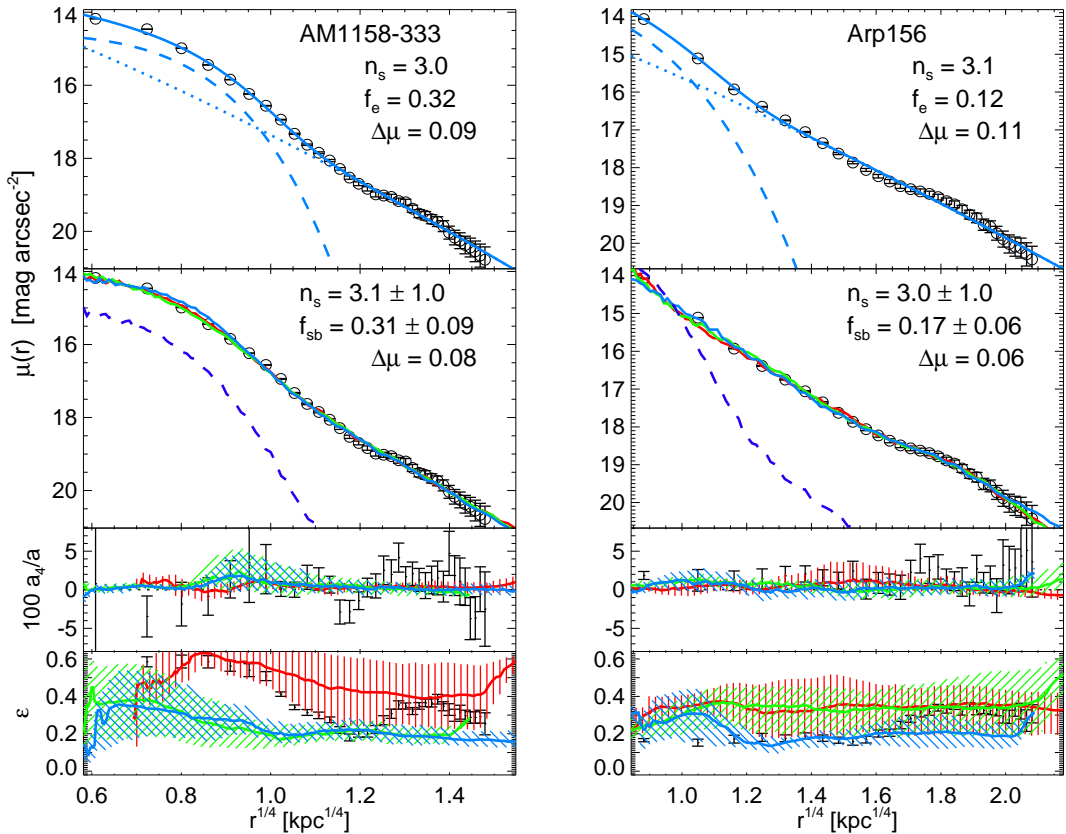}
    \caption{As Figure~\ref{fig:rj1}. Two merger remnants with very prominent 
    (deviant from  the outer Sersic law) central excess light components.
    \label{fig:rj3}}
\end{figure*}
\begin{figure*}
    \centering
    \plotone{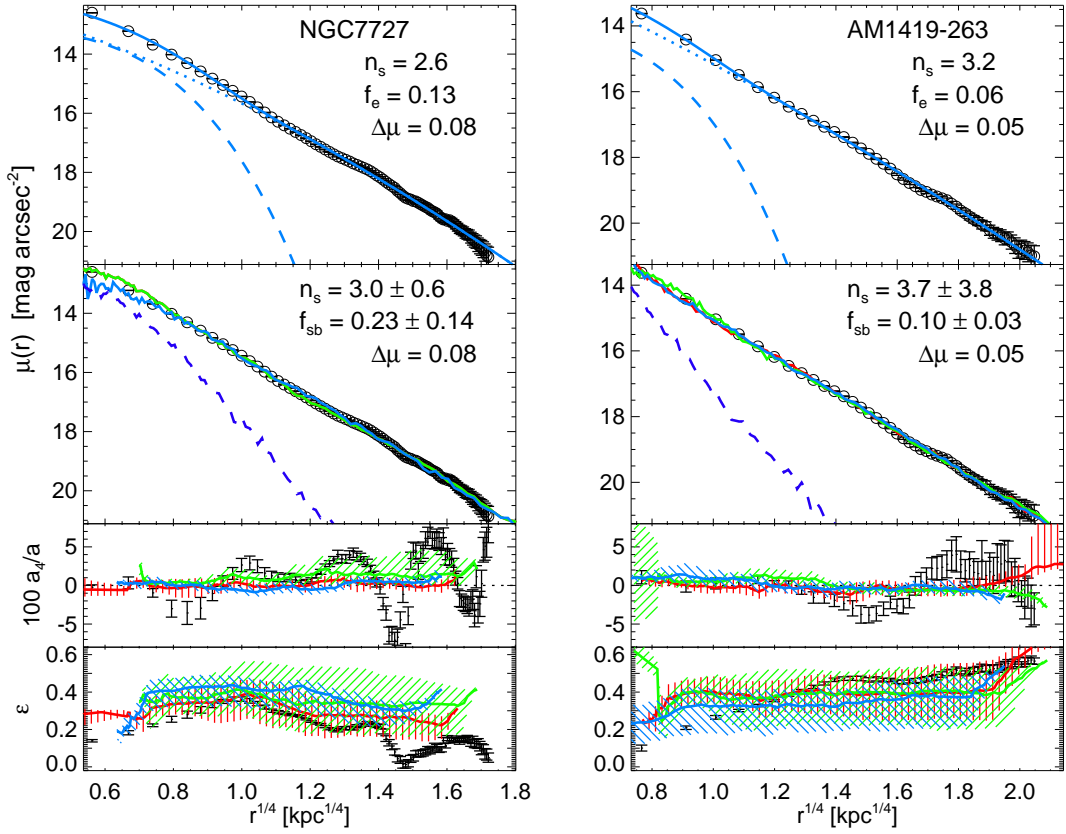}
    \caption{As Figure~\ref{fig:rj1}. Two merger remnants with 
    less prominent  
    central excess light components.
    \label{fig:rj4}}
\end{figure*}
\begin{figure*}
    \centering
    \plotone{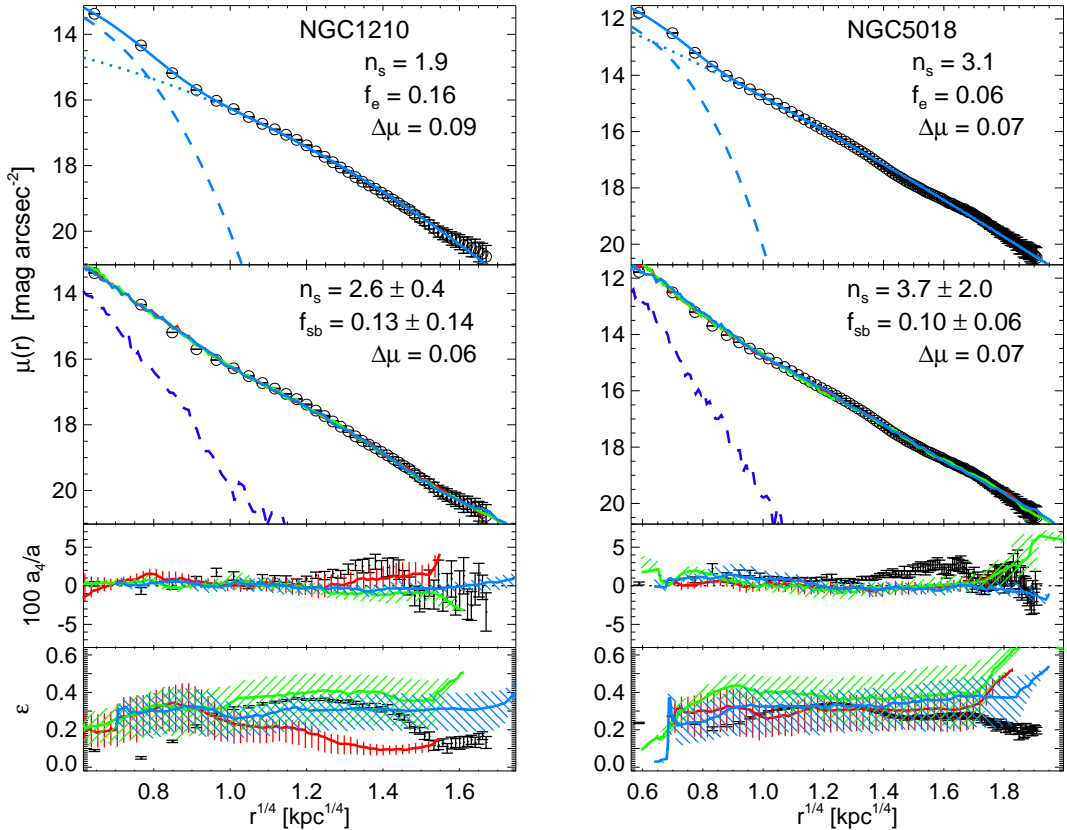}
    \caption{As Figure~\ref{fig:rj1}. Two shell ellipticals, with 
    relatively low excess light fractions. 
    \label{fig:rj6}}
\end{figure*}
\begin{figure*}
    \centering
    \plotone{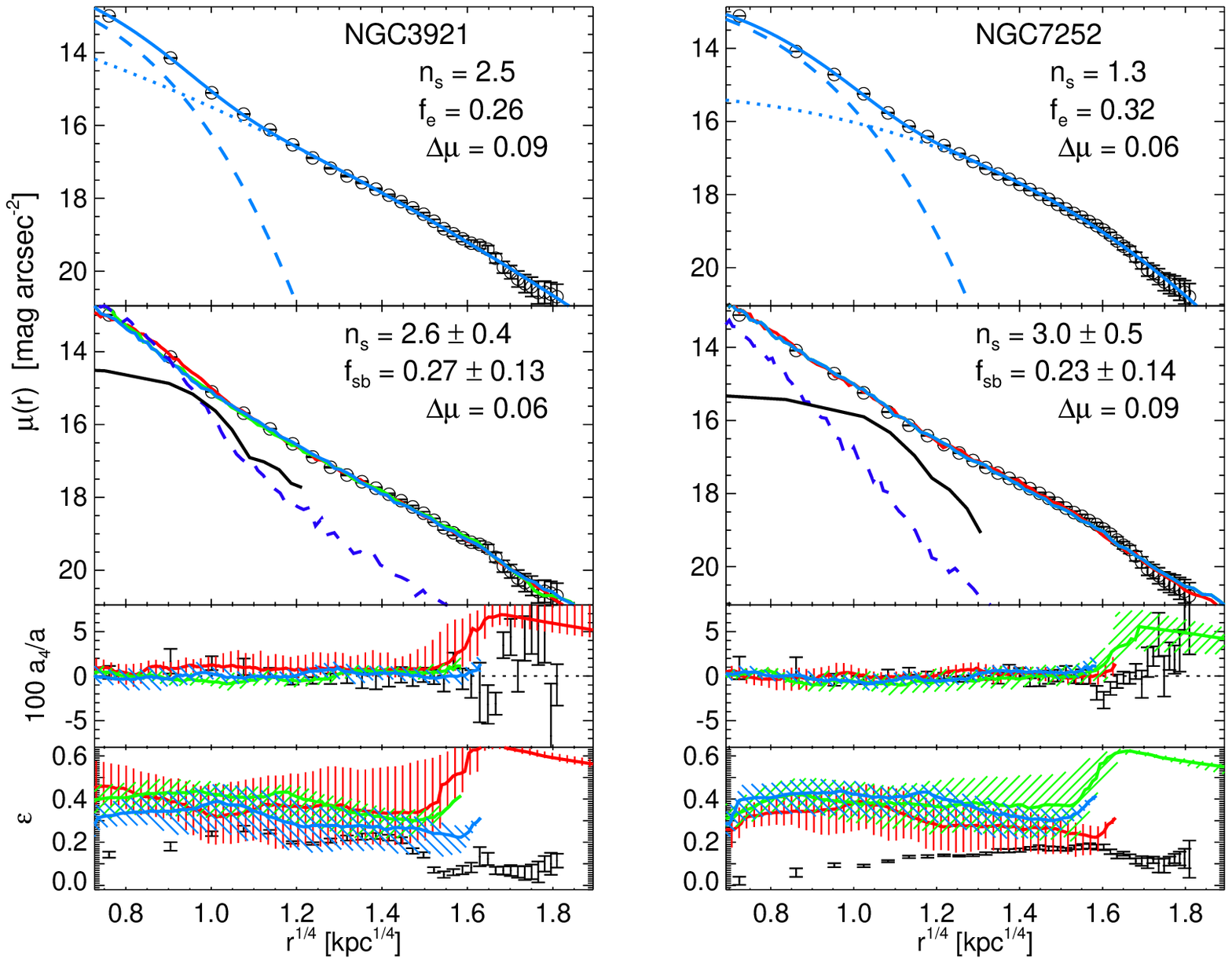}
    \caption{As Figure~\ref{fig:rj1}. Two merger remnants from the study of 
    \citet{hibbard.yun:excess.light}. The central gas identified in that work is 
    shown (black solid line in middle). 
    These are remnants, so the central excess/starburst component
    as we identify it (and in the corresponding 
    simulations) has {\em already formed}, and blends 
    smoothly with the outer profile. The gas remaining corresponds to 
    gas surviving the starburst (not part of the excess light), forming 
    embedded disks (see Figure~\ref{fig:origins}). 
    \label{fig:rj5}}
\end{figure*}

Figures~\ref{fig:rj1}-\ref{fig:rj6} expand on these results
by showing a comparison 
with a representative subset of the \rj\ sample.
(For comparison with the entire \rj\ sample, see 
\S~\ref{sec:appendix} and Table~\ref{tbl:rj.fits}.) In each, we 
show the observed $K$-band luminosity profile, with our best-fit 
two-component model, and note the fitted outer Sersic index, 
extra light fraction, and variance about the best-fit. 
We also directly fit our simulations to the observed profiles - i.e.\ considering 
the surface brightness profiles of our entire library of simulations, and find 
those which match the observed systems most closely. Because our simulations 
sample a finite, discrete number of total galaxy masses, we allow the 
precise normalization of the simulation profiles to  vary freely to match the 
observed data points (although we insist that the best-fit simulated galaxy mass 
be within a factor of $\sim$ a few of the estimated true galaxy mass, to 
make our comparisons as physically robust as possible). 
We show the three best-fit simulations to each profile (with the 
variance of the data about the profile), along with the 
Sersic index directly fitted to the pre-starburst stellar component of 
each and the mass fraction in the starburst component. We plot 
these components for the best-fit case. Errors refer to the range of a given 
parameter (e.g.\ $n_{s}$ of the pre-starburst component) in all simulations 
which are a comparably good match to the observations. 

We also 
show $a_{4}$ and ellipticity profiles of observations and simulations, 
although we emphasize that these are not fitted, and exhibit much greater 
sightline-to-sightline variation than the surface brightness profiles (thus should 
be considered somewhat less robust measures of agreement). We 
caution that, as demonstrated in \citet{cox:kinematics} and \citet{naab:gas}, 
subtle features such as the isophotal shape $a_{4}$ are much more 
sensitive (compared to the surface brightness profile) 
to e.g.\ the orbital parameters and merger mass ratio. 
Furthermore, it is likely that seeing effects bias the observed 
ellipticity and $a_{4}$ near the resolution limit, and this may be 
significant to radii as large as $\sim1\,$kpc; 
we also note that if the central regions have higher light-to-mass ratios 
owing to a nuclear starburst, they will appear more spherical than they truly are
(although discrepancies 
in these shapes amount to 
small $\sim0.1\,{\rm mag\,arcsec^{-2}}$ effects in the brightness profile). 
For example, the observed ellipticity profiles here all go to zero at small radii 
$\lesssim$kpc, as do ground-based profiles of many nearby 
young or shell ellipticals -- but in sufficiently high-resolution {\em HST} 
observations, it has been demonstrated that 
high ellipticities at $\sim$\,kpc scales continue (on average) 
at least down to $\lesssim100\,$pc \citep[e.g.][]{lauer:centers}, similar to 
what is predicted in our simulations. 
Nevertheless, 
as is also seen in \citet{naab:gas}, the agreement between 
the observations of \rj\ and our simulations is quite good 
at most radii, even in the isophotal shape and ellipticity profiles. 

Both methods we have considered 
(fitting a parameterized profile and directly comparing to the 
simulations) have advantages and disadvantages -- a parameterized fit 
is of course independent of the simulations, and conclusions can be drawn 
from it even if the simulation physics are incomplete. On the other hand, there 
is always bias introduced in some parameterized fit, and certain features 
in the light profiles (e.g.\ variation at large radii from shells or ripples, and 
embedded disks at intermediate radii) can throw off our fitting routines, but 
are accurately reflected in the direct simulation mass profiles. Reassuringly, 
in almost all cases the parameters describing the best-fit simulations are 
the same (within the errors) as those from our parameterized fits. The 
variance about the best-fit simulations is also comparable to that from the 
fitted profiles -- in a number of cases, the direct simulation profiles 
match the observed systems even {\em better} than a parameterized fit. 
This suggests that our fitting procedure is physically robust
and that our simulations are reliably modeling real galaxy mergers 
and their extra light distributions.

\section{Dependence of Extra Light on Merger Gas Content}
\label{sec:dept}

\begin{figure*}
    \centering
    \plotone{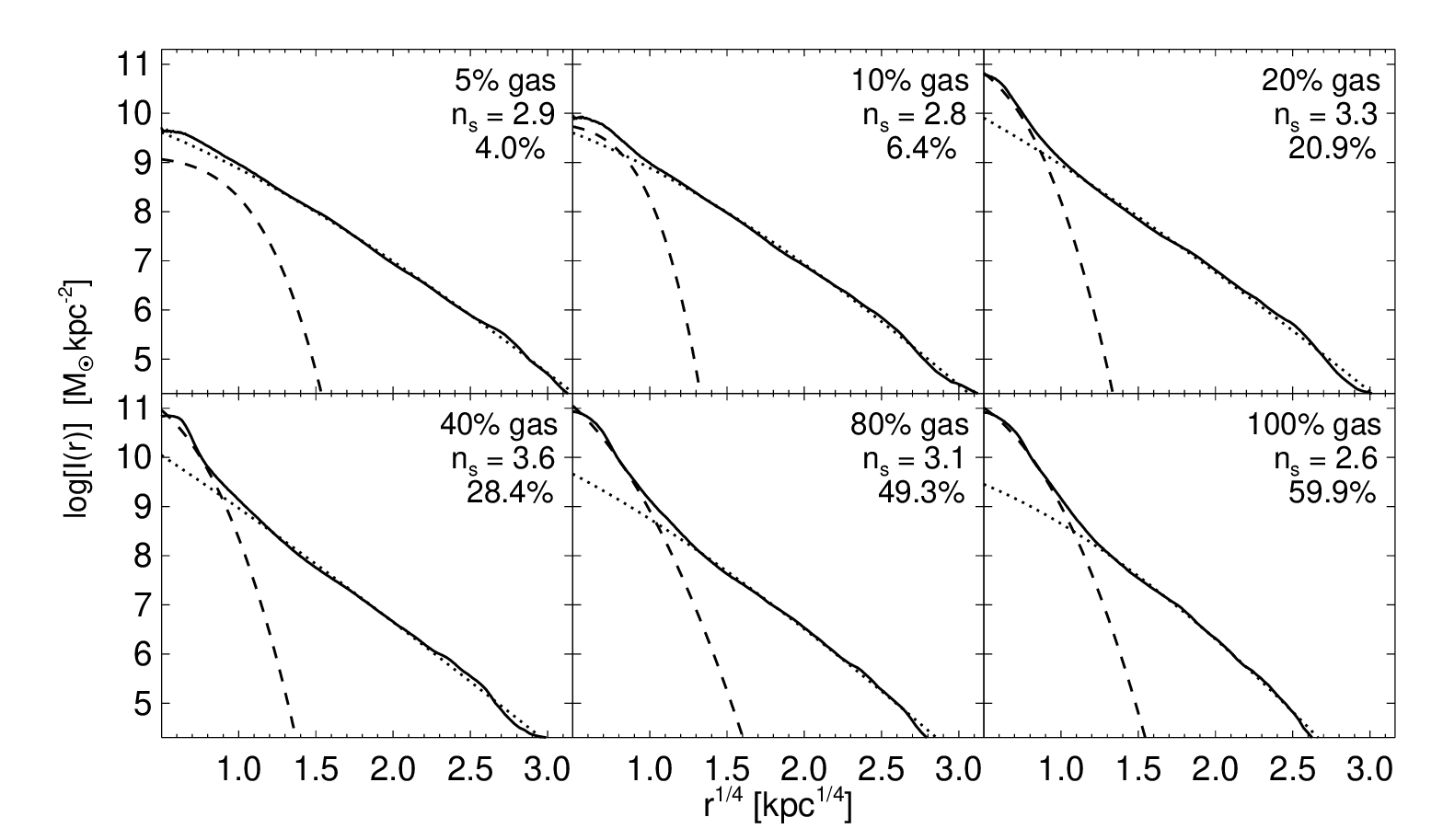}
    \caption{Median surface brightness profile (solid), 
    with our inner (dashed) plus outer (dotted) component decomposition, 
    of a typical simulation 
    as a function of initial gas fraction. Outer Sersic index and 
    extra light fraction for each are shown. All else being equal, the strength 
    of the extra light component increases systematically with 
    progenitor gas content. This does not necessarily mean that the excess light 
    departs more sharply from the outer profile shape. 
    \label{fig:fgas}}
\end{figure*}

Having analyzed a large number of simulations and observed 
merger remnants, we can now study the properties of 
extra light or starburst components in a global sense. 
We expect based on our above analysis that 
the extra light should be a reasonable proxy for the 
central, dissipational starburst component, and therefore should 
scale with the gas supply of the merging galaxies.
Figure~\ref{fig:fgas} shows this explicitly for a systematic 
survey of gas fractions.
The mergers are otherwise identical, and 
similar to our $\sim L_{\ast}$ fiducial case in \S~\ref{sec:origins}, but 
vary in initial gas fraction from $\fgas=0.05-1$. We plot 
the sightline-averaged total surface brightness profile and 
median best-fit extra light and outer/pre-starburst Sersic components 
in each case. It is clear that the extra light component systematically 
increases with increasing gas fraction, as expected, typically 
containing $\sim1/2$ the initial gas mass (although this exact value 
depends on a number of other conditions, as shown below). 
The profile of the scattered, violently relaxed stars, on the other hand, 
is basically independent of gas fraction, as expected if these are simply 
being scattered from a rotationally supported disk to a 
partially violently relaxed (Sersic-law $n_{s}\sim2.5-3.5$) object. 

\begin{figure*}
    \centering
    \plotone{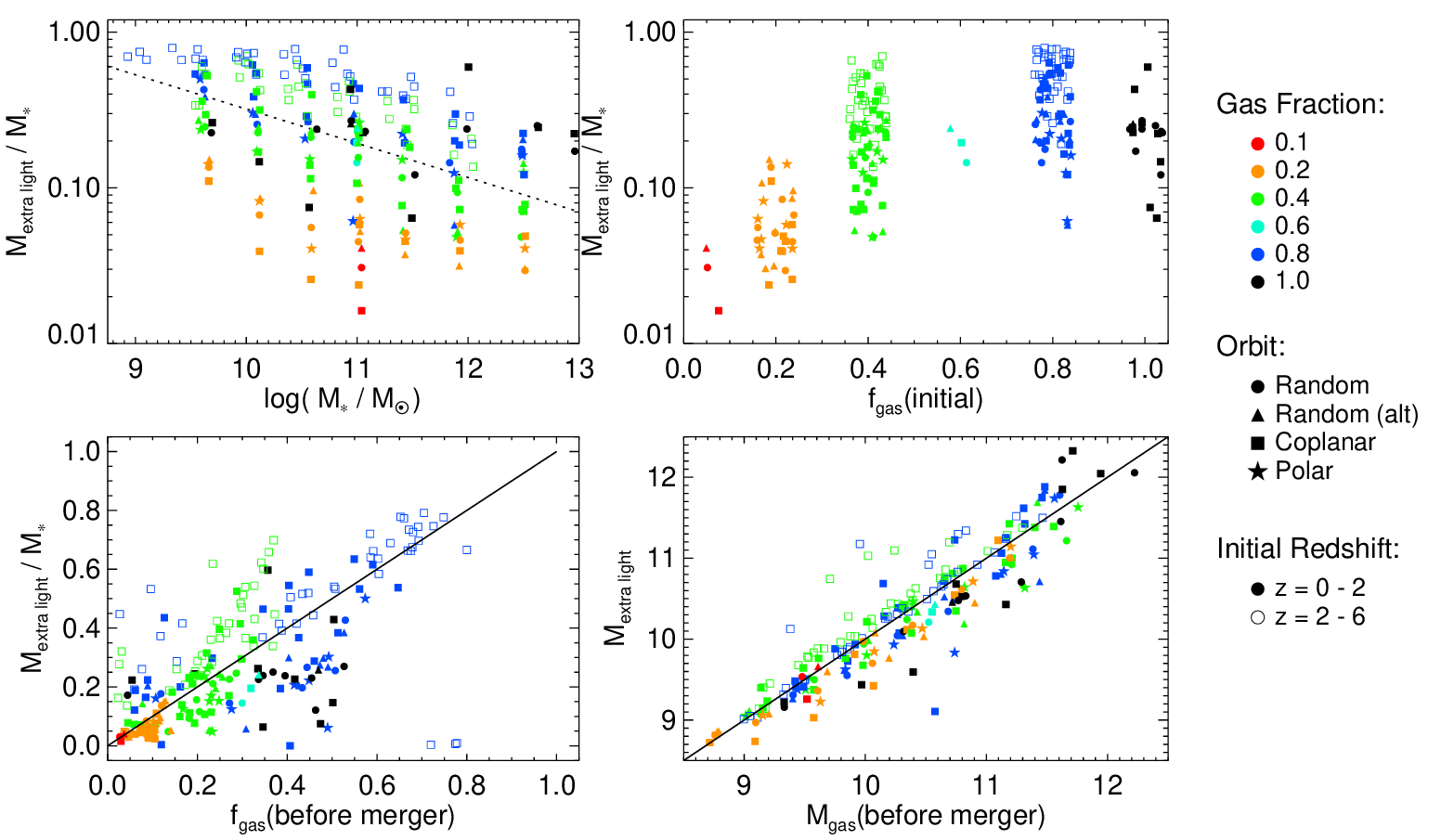}
    \caption{Fitted extra light fraction/mass as a function of stellar mass 
    ({\em top left}), 
    initial (simulation) gas fraction ({\em top right}), gas fraction involved in the 
    final coalescence/merger ({\em bottom left}), and total gas mass in the 
    final merger ({\em bottom right}). 
    Shown are the results for simulations: color 
    encodes gas fraction (red, orange, green, cyan, blue, and black 
    for initial $f_{\rm gas}\sim0.1,\ 0.2,\ 0.4,\ 0.6,\ 0.8,\ 1.0$), 
    symbol shape encodes orbital parameters (circles and triangles 
    are typical random orbits, squares coplanar, stars polar), and symbol 
    fill encodes initial redshift (filled for $z<1$ typical progenitor disks/halos; 
    open for $z>3$ compact progenitor disks/halos). The plotted 
    simulation values are the median across $\sim100$ sightlines;
    sightline-to-sightline differences are illustrated in Figure~\ref{fig:res2}. 
    \label{fig:mass.vs.fgas}}
\end{figure*}

\begin{figure*}
    \centering
    \plotone{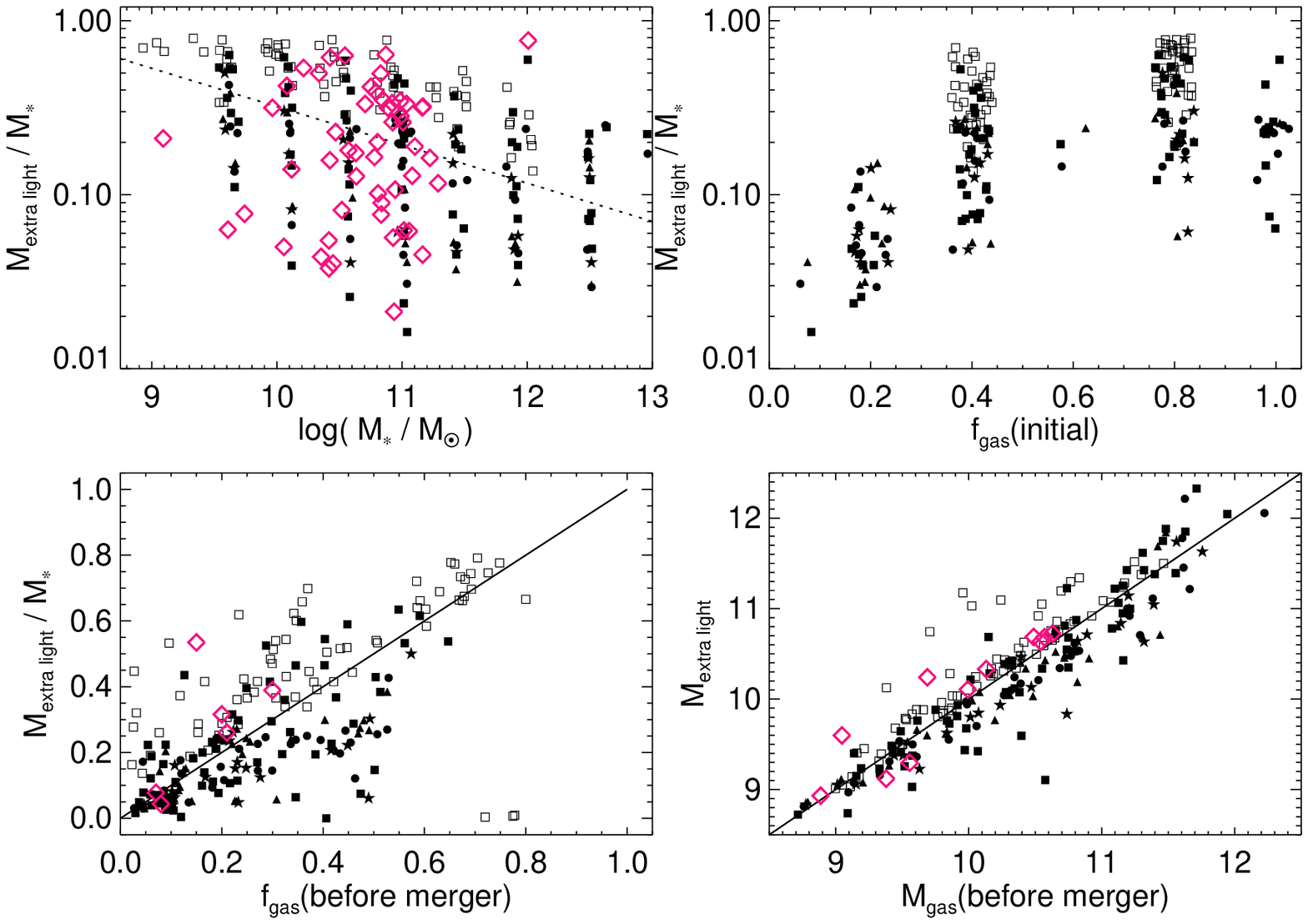}
    \caption{As Figure~\ref{fig:mass.vs.fgas}, but with the simulations 
    shown in black and the observed systems from the sample of 
    \rj\ (magenta diamonds). 
    Where multiple-component stellar population 
    analysis allows an estimate of the mass fraction formed in the merger-induced 
    starburst, we have plotted the observed systems in the lower panels. 
    \label{fig:mass.vs.fgas.obs}}
\end{figure*}

Figures~\ref{fig:mass.vs.fgas} \&\ \ref{fig:mass.vs.fgas.obs} 
extend this comparison to our entire suite of 
simulations and the fitted observations. We plot the extra light fraction 
as a function of the stellar mass ({\em upper left})
or initial gas fraction of the simulation ({\em upper right}). As 
expected, the mean extra light fraction increases with initial gas content, 
but there is a significant residual dependence on the galaxy stellar mass 
and orbital parameters. 
At fixed initial gas fraction, the extra light fraction declines with stellar mass, 
in a roughly power-law fashion as $f_{\rm extra}\propto M_{\ast}^{-0.15}$. As 
demonstrated by \citet{robertson:fp}, this is a 
consequence of the scaling of star formation efficiency with galaxy 
stellar mass. Higher mass galaxies have higher surface densities, and 
therefore exhaust their gas more rapidly -- i.e.\ by the time the final 
merger and starburst occur, a typical $\mstar\sim10^{12}\,\msun$ galaxy 
will have a smaller remaining gas fraction available to participate in the 
compact starburst than a typical $\mstar\sim10^{10}\,\msun$ galaxy, even 
when both are initialized with the same gas fraction. Similarly, the 
orbital parameters determine how long it will take the galaxies to merge, and 
therefore how much of the gas will be consumed in the disks before the final 
merger event. 

The physically relevant quantity is therefore not the gas fraction at the 
beginning of our simulations (which, as noted before, is somewhat 
arbitrary), but the gas fraction of the system just before the final 
merger/coalescence of the galaxies. If we consider the gas 
fraction $\sim0.2$\,Gyr before the peak of the final starburst or coalescence of the 
black holes (roughly the ``beginning'' of the 
starburst, in most systems), Figure~\ref{fig:mass.vs.fgas} ({\em lower panels})
shows that there is a good, linear correlation between 
this pre-merger gas fraction and the extra light fraction in the remnant 
($f_{\rm extra}\approx f_{\rm gas,\ pre-merger}$). Accounting for this, there is 
no significant dependence on galaxy mass or orbital parameters. The only systems 
which deviate significantly tend to be pathological, either extremely gas 
rich $\fgas=1$, very high redshift $z=6$, or e.g.\ perfectly co-planar mergers. In 
all of these cases, the systems can retain a large amount of gas even after the 
merger and form relatively large new gaseous disks \citep{robertson:disk.formation}, which are not 
accounted for in our fitting. Even in these situations, the deviations are not large -- 
plotting the total mass in the extra light component against the gas mass present 
just before the final merger shows a tight correlation, as expected. 

Similarly, we find that changing properties of the simulations such as the 
presence or absence of an initial bulge, the concentration of the progenitor 
halos and disks, the presence or absence of a supermassive 
black hole, and the treatment of star formation and the ISM equation of state 
can all appear to change (albeit by a smaller amount than directly 
changing the gas fraction of the merger) the final mass in the starburst or extra 
light component. In all these cases, however, the effect is indirect -- these 
choices influence how efficiently gas is consumed and/or expelled before the 
final merger, and therefore how much is available to participate in the
starburst. For a fixed gas mass at the time of the final starburst, the extra 
light component mass is independent of these effects.

Figure~\ref{fig:mass.vs.fgas.obs} summarizes the results of our fitting to the 
entire sample of \rj. We note that one should regard the stellar mass 
estimates (based just on the $K$-band luminosities) 
as somewhat uncertain -- however, using dynamical 
masses from \citet{rothberg.joseph:kinematics} yields a similar result 
and using a more sophisticated 
stellar mass estimator makes little difference within the considerable scatter. 
The observed systems occupy a similar locus in the 
$f_{\rm extra}-\mstar$ plane ({\em upper left}) 
to our simulations. There are a couple with very low $f_{\rm extra}$ -- however, 
without exception, these also have rather low best fit outer Sersic indices $n_{s}\approx1.0$, 
and are probably better classified as S0 or spheroidal galaxies, for which our 
fitting routines are not appropriate and the best-fit $f_{\rm extra}$ is not necessarily physically 
representative. 

We cannot, of course, assign a meaningful initial gas fraction to the observed systems. However, 
it is in principle possible, by studying the stellar populations in sufficient detail, to estimate the 
mass fraction which formed in a recent, central starburst event (as opposed to the more 
extended quiescent star formation history). Unfortunately, there are still a number of 
degeneracies, and this requires detailed observations, but it has been attempted 
by \citet{titus:ssp.decomp,schweizer:7252,schweizer96,schweizer:ngc34.disk,
reichardt:ssp.decomp,michard:ssp.decomp} 
for several of the observed systems. Adopting their 
estimates for the mass fraction in the newly formed stellar populations, we compare
our fitted extra light fractions to the gas mass which participated in the starbursts 
in these objects. We find a similar tight correlation -- although there are only a few objects 
for which sufficiently accurate stellar populations are available to allow this 
comparison, they all suggest that our fitted extra light component is indeed a good proxy for 
the mass fraction which was involved in the central, merger-driven starburst. 

In \citet{hopkins:cusps.ell}, we compare the properties of progenitor disks required 
to form realistic merger remnants and ellipticals to the observed properties of disks, 
and find good agreement. Here, we briefly note that the implied gas fractions 
in Figure~\ref{fig:mass.vs.fgas.obs} 
agree reasonably well with those observed in local disks of the same stellar mass 
\citep[for stellar masses $\sim0.1-1\,\mstar$, where most of our sample lies, 
observations imply typical disk gas fractions $\sim0.2$ at $z=0$ to $\sim0.3$ at $z=1$, 
with a factor $\sim2$ intrinsic scatter at both redshifts;][]{belldejong:disk.sfh,shapley:z1.abundances}. 
Likewise, we obtain the fits to the observed surface brightness profiles in 
Figures~\ref{fig:rj1}-\ref{fig:rj5} given progenitor disks that obey 
the baryonic Tully-Fisher and size-stellar mass relations 
\citep[observed to evolve only weakly from 
$z=0-1$; e.g.][]{belldejong:tf,flores:tf.evolution,trujillo:size.evolution}. We have 
performed some limited studies of initial conditions which do not resemble 
observed disks (e.g.\ initial disks that are very compact for their mass), 
and find that these fare poorly at simultaneously 
matching the observed light profiles and stellar masses, or require 
unlikely gas fractions \citep[see also][]{hopkins:cores}. In short, the 
observed profiles are consistent with the expected remnants of 
typical local gas-rich mergers. 

\section{Structure and Size of the Extra Light Component}
\label{sec:structure}

\begin{figure}
    \centering
    \figexpand
    \plotone{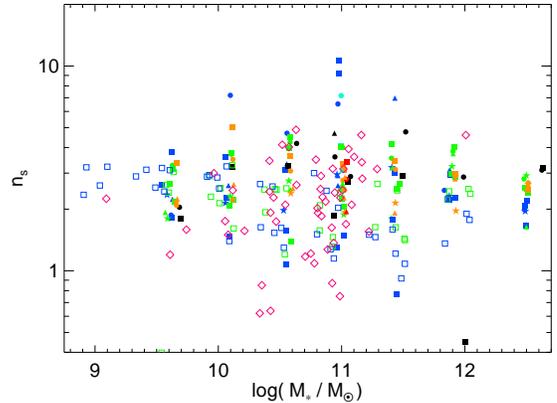}
    \caption{Outer Sersic indices in observed and simulated gas-rich merger 
    remnants, using our 
    two-component decomposition. 
    Points are as in Figure~\ref{fig:mass.vs.fgas}. 
    Gas-rich merger remnants have characteristic outer profile Sersic index
    $n\sim2-3$, without a strong systematic dependence on mass or other properties. 
     \label{fig:sersic.vs.mass}}
\end{figure}

In Figure~\ref{fig:sersic.vs.mass}, we 
examine how the Sersic indices of the outer or pre-starburst light 
component vary with galaxy properties. Interestingly, there is no 
significant trend with galaxy mass. This appears contrary to the conclusions of 
\citet{prugniel:fp.non-homology,graham:bulges,ferrarese:profiles}; 
however, we emphasize that we are fitting 
only gas-rich major merger remnants. Their samples include a number of 
spheroidal galaxies and ``pseudobulges'' at low masses (at the lowest 
$n_{s}$ values) and cored, boxy, slowly rotating ellipticals at 
the highest masses (and highest $n_{s}$ values). To the extent that 
gas-rich merger remnants dominate the population at somewhat intermediate 
masses, the distribution of Sersic indices we recover is consistent with their 
estimated trends. Supporting this, \citet{jk:profiles} find a similar, relatively 
constant distribution of Sersic indices if they consider just the elliptical 
galaxies in their sample which have some central excess light. 
Furthermore, our fitting procedure should not be directly compared to that 
in \citet{prugniel:fp.non-homology} 
and others, because we fit an outer bulge and inner extra light component 
simultaneously, whereas they fit profiles to a single Sersic index.

Physically, this independence of $n_{s}$ on 
other parameters is expected -- the gravitational physics which scatter the 
stellar disks in violent relaxation are self-similar. It is only the gas physics of 
dissipation and star formation, responsible primarily for the extra light component, 
that break this self-similarity. 
This is explicitly clear if we compare our simulations to those 
in \citet{naab:profiles}, who find that in simulated 
collisionless (gas-free) disk merger 
remnants -- i.e.\ systems for which the entire profile is by definition part of 
the ``outer,'' violently relaxed component -- there is no significant 
dependence of the Sersic index on mass, effective radius, or merger mass 
ratio. At the lowest masses, some of our simulated remnants 
have very low Sersic indices $n_{s}\sim1$ -- these are generally low mass, 
extremely gas rich systems which form large disks after the merger, and should 
not therefore be considered typical spheroids. 

We compare with the Sersic indices of the \rj\ sample. They 
occupy a similar locus and have a similar distribution to the 
simulations. There are a few objects with extremely low Sersic indices; 
as mentioned in \S~\ref{sec:dept} and \citet{naab:profiles}, 
these are probably systems that are 
actually disk-dominated, and have small extra light components. 
For the systems that are genuinely bulge-dominated, the distribution of 
Sersic indices observed is consistent with the simulations. We also compare
with the Sersic indices of relaxed ellipticals that have central cusps or 
extra light components, from \citet{jk:profiles}. These trace an 
almost identical distribution to the simulations and the \rj\ sample. 
Here, we do not see the anomalous objects with very low $n_{s}$ because 
the systems are specifically selected to be bulge-dominated. 
In both the observed merger remnants and ellipticals, there is no trend of 
$n_{s}$ with galaxy mass, as predicted by the simulations, a point 
discussed further in \citet{hopkins:cusps.ell}. 

\begin{figure}
    \centering
    \plotter{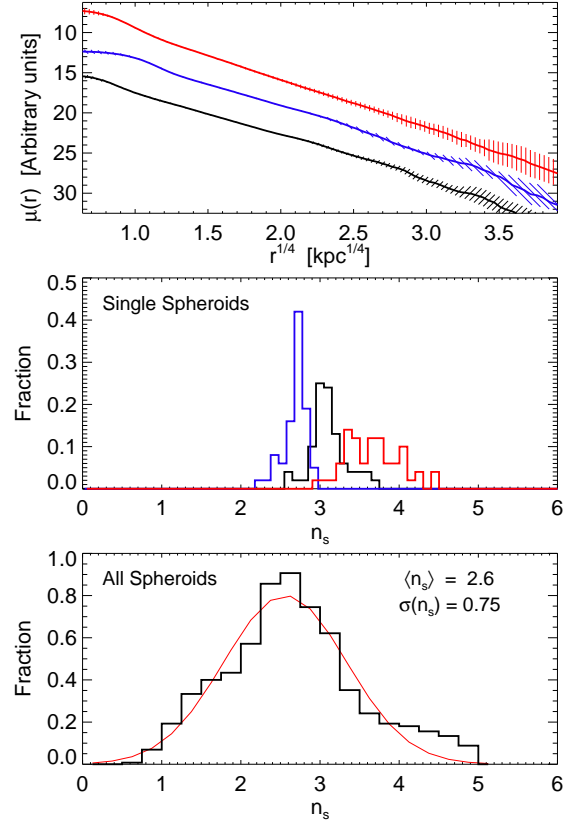}
    \caption{{\em Top:} Surface brightness profiles of three random 
    ($\sim\lstar$, $f_{\rm gas}\sim0.2-0.4$) simulations
    with slightly different $n_{s}$ values. Solid lines show the median 
    profile over $\sim100$ sightlines, shaded range shows the $25-75\%$ range 
    across sightlines. 
    {\em Middle:} Distribution of fitted outer $n_{s}$ values for each of the simulations. 
    {\em Bottom:} Distribution across all sightlines and simulations of outer 
    Sersic index in our simulated gas-rich merger remnants. A Gaussian fit 
    to this distribution is also shown.
    \label{fig:sersic.distrib}}
\end{figure}

Figure~\ref{fig:sersic.distrib} shows the full distribution of Sersic indices in 
the simulation outer bulge or pre-starburst components. 
The distribution is smooth and can be approximated by a simple Gaussian with 
median Sersic index $n_{s}\sim2.6$ and $1\sigma$ dispersion of $\Delta n_{s}=0.75$. 
The distribution of Sersic indices owes primarily to galaxy-to-galaxy 
variations: the figure shows the range of surface brightness profiles 
and fitted $n_{s}$ values 
for three random simulations, over $\sim100$ sightlines. There 
is relatively little sightline-to-sightline variation in the surface brightness 
profile, and in general a small dispersion in Sersic index (dispersion of 
$\Delta n_{s}\sim0.1-0.3$ across viewing angles). 
Again, this is similar to the simulated distribution in collisionless disk-disk 
merger remnants in \citet{naab:profiles}, which include a wider range in 
e.g.\ merger mass ratios, but do not include gas and therefore should 
be strictly self-similar.

\begin{figure}
    \centering
    \plotter{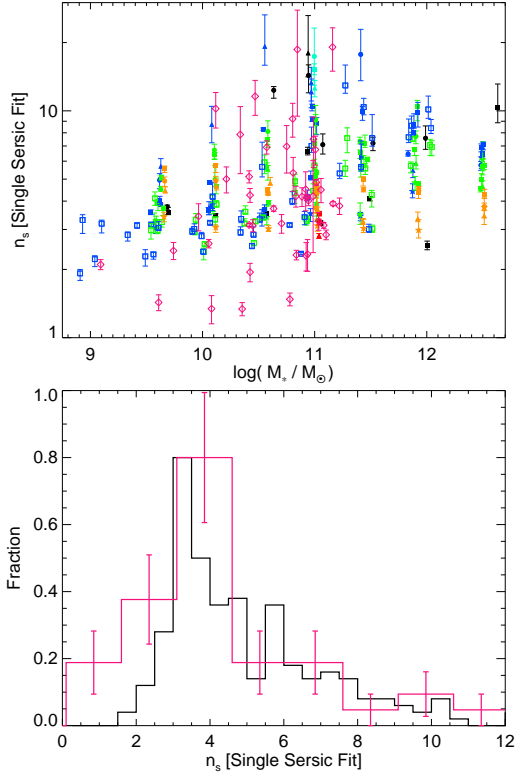}
    \caption{{\em Top:} Best-fit Sersic indices obtained from simulated and observed \rj\ merger 
    remnants (style as Figure~\ref{fig:sersic.vs.mass}) when fit to a single Sersic profile 
    (as opposed to our two-component fit). {\em Bottom:} Distribution of these single Sersic 
    profile indices (histograms with and without error bars are for the \rj\ observed sample 
    and simulations, respectively). Fitting to a single Sersic law yields higher Sersic indices 
    and shows a dependence of $n_{s}$ on galaxy stellar mass, radius, and surface brightness 
    similar to what is ``typically'' observed. As shown in Figure~\ref{fig:demo.fit.danger}, however, 
    this does not reflect the true shape at large radii (and should not be used to estimate e.g.\ the 
    true extra light fraction), but 
    rather indirectly reflects the contributions of extra light changing the profile shape at small radii. 
    \label{fig:single.sersic}}
\end{figure}

For comparison with what is typically measured, Figure~\ref{fig:single.sersic} 
shows the distribution of 
Sersic indices measured if we simply fit a single Sersic profile to the 
entire surface brightness profile of each of the
simulated and observed merger remnants. Fitting a single Sersic profile, we see 
that the distribution of Sersic indices is systematically pushed to higher 
values, with a median $n_{s}\sim4.0-4.5$ for both the simulations and observations, 
and a much larger ``tail'' towards high Sersic indices $n_{s}>5$. Fitted in this 
way, Sersic index does appear to depend (albeit weakly) on galaxy mass, 
as well as effective radius and surface brightness, similar to what has been 
seen in samples of ellipticals (when fit in this manner) as discussed above. 
We caution, however, that as demonstrated in Figure~\ref{fig:demo.fit.danger}, these 
fits are not physically representative of the outer light profile or central extra light. 
The presence of an extra light component makes the total central light profile 
steeper, leading to the higher $n_{s}\sim4$ typical of ellipticals that are fit 
to single Sersic indices -- as compared to the $n_{s}\sim2-3$ that genuinely 
characterizes the outer, violently relaxed stellar populations in our simulations. Because, 
as we have shown, this outer profile is roughly self-similar, the 
dependence of the single Sersic index on mass and other galaxy properties 
seen here owes to differences in the typical masses, shapes, and spatial extent 
of the extra light components, {\em not} to differences in the outer profiles. 
Even allowing for a fit to the full profile, \rj\ found (similar to what we see here) 
that the correlation between luminosity and Sersic index had a large scatter, and 
the correlations between e.g.\ size and Sersic index were even less significant. 
In short, the behavior of a single Sersic law fit in the simulations and observations is similar to what 
has generally been observed in elliptical galaxies, but when fit in this manner this 
``typical'' behavior reflects a combination of the inner extra light structure (which 
depends on the degree of dissipation and therefore other galaxy properties) 
and outer, violently relaxed stars (which are approximately self-similar), rather 
than any single, robust physical component of the galaxy. 


We can, in principle, also attempt to measure the Sersic indices 
specifically of the cusp or 
extra light components. Unsurprisingly, 
they typically show disky $n_{s}\sim1-2$ profiles. However, we caution 
that the inner regions of the cusp are strongly affected by our 
resolution limits. It is guaranteed that near the resolution limit the surface 
density profile will become flat, which results in a lower Sersic index. 
Therefore, although the general trend that the extra light component tends to 
have a diskier profile than the outer light appears to be true in 
the observations as well, 
the structure in the inner regions of the simulated extra light components 
should not be considered robust. As discussed in \S~\ref{sec:fits}, our 
choice to fit the inner extra light component to an $n_{s}=1$ law 
is not motivated by the actual shape at small radii (well inside the 
effective radius of the extra light), which neither simulations nor 
observations resolve \citep[and high-resolution observations of ellipticals 
suggest can have irregular structure owing to e.g.\ star clusters and 
the effects of black hole dynamics; see e.g.][]{lauer:centers}; instead, we chose this 
because it fits reasonably well to where the extra light maps on to the 
outer profile and provides the most reliable recovery of the true physical 
starburst component.

\begin{figure*}
    \centering
    \plotone{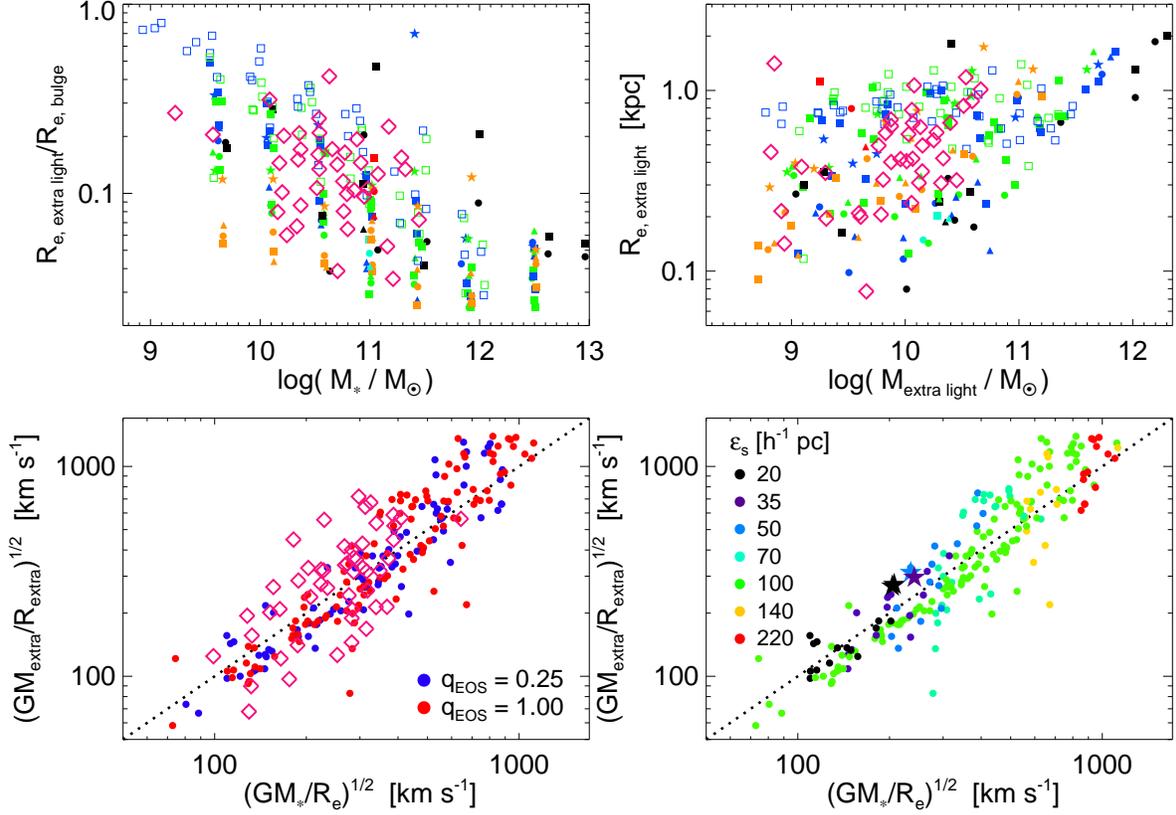}
    \caption{{\em Top Left:} Ratio of the half-light radius of the 
    fitted extra light component to that of the fitted outer Sersic component, as a 
    function of galaxy stellar mass. Points show simulations and 
    observations, as in Figure~\ref{fig:mass.vs.fgas}. {\em Top Right:} Size-mass
    relation of the extra light. 
    {\em Bottom Left:} Potential of the extra light component 
    (here defined as $\sqrt{G\,M/R}$) as a function of that of the entire galaxy. 
    Dotted line denotes a linear relation (indicating that the extra light is self-gravitating). 
    In this space, there is no systematic dependence of extra light component 
    size on e.g.\ galaxy mass, orbital parameters, or merger redshift. Colors here 
    denote the effective ISM pressurization and equation of state: red for the 
    full \citet{springel:multiphase} model, blue for a nearly isothermal model. 
    {\em Bottom Right:} Same, but colors denote simulation spatial resolution, 
    from $\sim300\,$pc to $\sim30$\,pc. Stars denote the resolution study simulations 
    from Figure~\ref{fig:res1}. 
    \label{fig:sizes}}
\end{figure*}

More robustly, we can study the effective radii of the extra light or 
starburst components in the simulations and observations. Figure~\ref{fig:sizes} 
compares the spatial extent of the extra light component to the properties of the 
galaxy. The extra light component appears to be fractionally smaller in more 
massive galaxies (relative to the effective radius of the galaxy), in both 
simulations and observations. Looking at this in detail, it is clear that this 
is a consequence of the trend in Figure~\ref{fig:mass.vs.fgas} -- more massive 
galaxies tend to have less mass in their extra light components. 
The extra light itself appears to follow a rough size-mass relation, shown 
in Figure~\ref{fig:sizes} of the form $R_{e,\ \rm extra\ light}\propto M_{\rm extra\ light}^{0.33}$. 
Note that by size of the extra light component, we refer explicitly to its half-mass 
projected effective radius: this is not necessarily the same size scale at which the 
extra light component begins to dominate the total surface density profile of the galaxy. 
The appropriate comparison is with the effective radii of the extra light components fitted 
to the observations -- where we see good agreement. By mass of the extra light 
component, we simply refer to our fitted extra light fraction times the total estimated 
stellar mass of the galaxies. 

Considering the evolution of the starburst component in the simulations, it 
appears to follow a general formation mechanism, as outlined in \citet{mihos:starbursts.96}. 
In the final stages of the merger, 
the gravitational torques acting on the gas remove its angular momentum, and it 
essentially begins a free-fall towards the central regions of the galaxy. 
The cooling time of the gas is almost always shorter than the dynamical time 
at this stage, so the process is not pressure supported and is primarily a gravitational 
collapse. This continues until the collapsing gas becomes self-gravitating: at this point, 
the gas shocks and establishes a quasi-hydrostatic central mass concentration, and 
becomes stable against further collapse. The stability criterion is simply that the 
effective equation of state of the gas be $\gamma>4/3$, which is true at this time 
because the gravitational compression of the gas is effectively adiabatic. Furthermore, 
in the classical \citet{mckee.ostriker:ism} picture of the multiphase 
ISM, the densities are sufficiently high that 
the system is subject to
thermal instability.  In this state, the hot and warm 
diffuse phases of the ISM can radiate and cool efficiently into cold, dense 
clouds, which then form stars, whose supernovae and stellar wind feedback 
reheat the diffuse ISM. The important point is that, in this regime, cooling does not allow the 
system to lose effective pressure support, because the cooling of the diffuse ISM is balanced 
by energy injection from star formation and supernovae in the cold phase of the ISM. 
Regardless of the cooling time, then, one expects that the system cannot collapse further 
once it becomes self-gravitating, and will stall at this radius and continue rapidly forming stars 
until the gas supply is exhausted. 

Studies of the central regions of starburst systems support this picture -- typically, 
observations find a dense concentration of molecular clouds forming stars at a high rate, 
with very high effective temperatures and pressures of the cumulative (diffuse) ISM gas 
\citep[e.g.][]{sanders96:ulirgs.mergers,
solomon.downes:ulirg.ism,bryant.scoville:ulirgs.co}. 
Even if the system can contract, it does so on the local dynamical time, which, 
for a self-gravitating system, is comparable to the star formation timescale if the 
system obeys a Kennicutt-Schmidt type star formation law \citep{kennicutt98}, and so 
the system will exhaust gas in star formation in a self-similar manner as it contracts. 

The expectation of these physical models is then that the effective radii of the 
extra light components will be determined by the condition that they become 
self-gravitating: 
\begin{equation}
\frac{G\,M_{\rm extra}}{R_{\rm extra}} =\alpha \frac{G\,M_{\rm tot}}{R_{e}}
\end{equation}
where $\alpha\sim1$ is a constant which depends on the exact shape of the 
galaxy mass profile and stellar-to-dark matter mass ratio as a function of radius. 
Figure~\ref{fig:sizes} plots this relationship, for both the simulations and the observed 
sample of \rj. We find that indeed, the objects follow a tight, linear 
correlation as predicted, with a best-fit normalization $\alpha\approx1$, independent of 
the absolute value of galaxy mass, orbital parameters, or the initial redshift of the 
merging galaxies.

If the spatial 
extent of the extra light component were not set by the self-gravitation condition, but 
by, for example, the pressure support of the ISM or some competition between 
the star formation and cooling timescale, one would expect 
the size to be systematically sensitive to the treatment of star formation and 
supernova feedback and the corresponding ISM equation of state. Figure~\ref{fig:sizes} 
compares the size-mass relation for two values of our equation of state parameter 
$\qeos$, described in \S~\ref{sec:sims}. Our fully multiphase model for the 
ISM, with a high effective pressure, corresponds to $\qeos=1$, and a 
model much closer to a completely isothermal ISM to $\qeos=0.25$. Both cases 
follow an identical size-mass relation, suggesting that it is indeed simply the self-gravitation
condition that sets the size of the starburst component. Note that the {\em absolute} 
sizes of the extra light or starburst components tend to be smaller in the 
$\qeos=0.25$ case. This is because the more isothermal equation of state allows 
star formation to proceed more efficiently in the pre-merger disks, exhausting more 
of the gas supply before the final merger and leaving an extra light component 
of lower total mass. But, for fixed gas mass just before the final merger stages or 
(equivalently) fixed starburst mass fraction, the two equations of state yield 
extra light components of the same physical size. 

It is also natural to wonder whether or not the sizes of the cusps are sensitive to 
our simulation resolution. Figure~\ref{fig:sizes} compares the points on this 
effective size-mass relation as a function of the simulation resolution. From 
our lowest resolution simulations with gravitational softening length 
$\epsilon_{s} \approx 300$\,pc, to our highest resolution simulations with 
$\epsilon_{s} \approx 30$\,pc, the systems obey the same correlation. 
This is reassuring 
in that our resolution limits, while such that we do not 
resolve the inner regions of the extra light, do not cause us to systematically 
overestimate the size of the extra light distribution. We have also tested 
a series of simulations of varying integration accuracy, and find similar results.

\section{Stellar Populations and Evolution of Extra Light Components with Wavelength and Time}
\label{sec:evolution}

\begin{figure}
    \centering
    \plotone{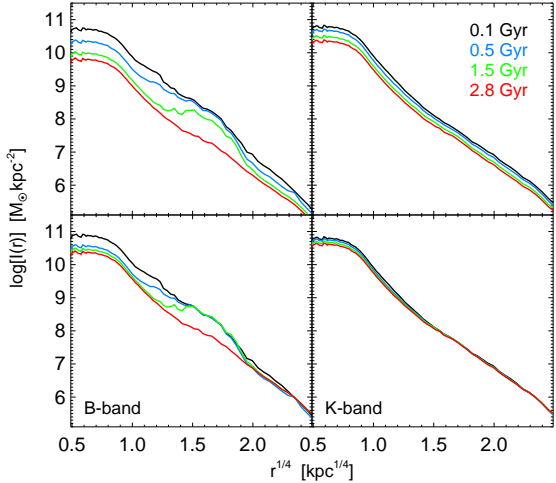}
    \caption{Surface brightness profile of an extremely gas-rich ($f_{\rm gas}=0.8$)
    merger remnant 
    in $B$-band ({\em left}) and $K$-band ({\em right}), as a function of time 
    after the coalescence of the two merging nuclei (colors, as labeled). {\em Top:} Absolute 
    values, which show fading owing to passive evolution. {\em Bottom:} Profiles 
    rescaled to the same value at $\sim30\,$kpc, to show the change in profile shape. 
    There is little evolution in $K$-band, but substantial time dependence in $B$-band. 
    In this case the system forms an embedded disk 
    (around $t\sim0.5$\,Gyr) from gas which survives the merger, 
    which is prominent in $B$-band (the ``bump'' at $\sim4-15$\,kpc)
    owing to its young age but negligible in 
    $K$-band (it contains only $\sim4\%$ of the galaxy stellar mass). The young 
    stellar populations in the embedded disk have faded in $B$-band by $\sim3$\,Gyr. 
    \label{fig:mu.evol}}
\end{figure}

Figure~\ref{fig:mu.evol} plots the evolution of the surface brightness profile of a 
highly gas-rich merger (initial $f_{\rm gas}=0.8$) remnant with time, in the 
observed $B$-band and $K$-band, ignoring (for now) the effects of dust obscuration. 
We calculate the observed 
surface brightness profile in these bands using the stellar population 
synthesis models of \citet{BC03}, given the ages and metallicities 
determined self-consistently for the stellar particles formed in our simulations, 
assuming a \citet{chabrier:imf} IMF (although this choice 
primarily affects only the total luminosity, not the shape of the 
light profile). The initial stars in the simulation disks (in this case 
a relatively small fraction $1-f_{\rm gas}=0.2$) are given pre-merger 
age, metallicity, and enrichment distributions appropriate for 
the disk stellar mass according to the 
median best-fit $\tau$ model star formation histories fit to local 
observed disks in \citet{belldejong:disk.sfh} (thereby are 
guaranteed to lie on the observed star-forming galaxy 
mass-metallicity relation). We discuss the issues of stellar population 
gradients in much greater detai in \citet{hopkins:cusps.ell}, but 
note that we have also experimented with other stellar population 
models or fitted star formation histories, 
and with including initial gradients in the disk stellar populations, 
and find these choices make little difference 
(the central age, metallicity, and gradients of interest are 
largely set by the dissipational starburst).

The figure demonstrates that, although there can be significant 
overall fading in the near IR with time after the merger (in this 
case $\sim2$\,magnitudes over $\sim3$\,Gyr), especially for such a gas 
rich system where a large fraction of stars are formed in the final starburst, 
this has little {\em differential} effect with radius. The shape 
of the surface brightness profile in 
$K$-band is nearly time-independent, after the final coalescence of the 
two galaxies. 
In optical bands such as $B$, however, there can be a much more 
dramatic time dependence. In this case, not only does the nucleus 
fade considerably relative to the outer profile, but one can see 
a prominent embedded disk appear at $\sim0.5$\,Gyr and 
fade out by $\sim3$\,Gyr after the merger. 

Because the system is so gas-rich, 
a significant quantity of gas survives the merger and cools to re-form an 
embedded disk. This process takes $\sim$ a couple $\times 10^{8}\,$yr after the 
coalescence, hence the disk is not as prominent immediately 
following the merger. 
By mass, the disk represents only $\sim$ a few percent of the 
galaxy stellar mass, similar to the case in Figure~\ref{fig:origins}. However, because 
its stellar populations are very young, it appears prominently in $B$-band, 
constituting $\sim30-40\%$ of the total light and making the system appear to be a clear 
S0 rather than a true elliptical. By $\sim3$\,Gyr after the merger, the embedded disk 
(while still forming stars at a low rate) has aged sufficiently so that its 
mass-to-light ratio is not much different from the rest of the galaxy, and 
(given its small mass fraction, which is concentrated near $R_{e}$, where most 
of the spheroid mass also lies) it disappears again. 
Note that, to illustrate this, we have ignored dust, which will preferentially 
lie in the gas-rich star forming region of the disk, and make the difference 
between $B$ and $K$ bands less obvious. Still, the presence of 
such disks in optical (and not NIR light) is not uncommon in 
gas-rich merger remnants: the profile and evolution above is 
similar to that seen (and inferred) for NGC 34 \citep{schweizer:ngc34.disk}. 

The example above and observed systems emphasize that, 
at times close to the merger, the $K$-band is a much more 
robust tracer of the stellar mass distribution.
Our experiments suggest that we can have reasonable 
confidence that the $K$-band profiles of the observed systems will 
not change substantially with time -- therefore our results would 
likely be unchanged if we analyzed the merger remnants at a time 
when they were more relaxed. Of course, then, it would not be clear that 
the systems are indeed gas rich merger remnants. The most substantial 
change would probably be a reduction in the scatter about the mean trends 
defined in Figures~\ref{fig:mass.vs.fgas} and \ref{fig:sizes}; we typically see little evolution in 
the median surface brightness profiles after the merger, but unrelaxed features can cause 
considerably larger sightline-to-sightline variation.

\begin{figure*}
    \centering
    \plotone{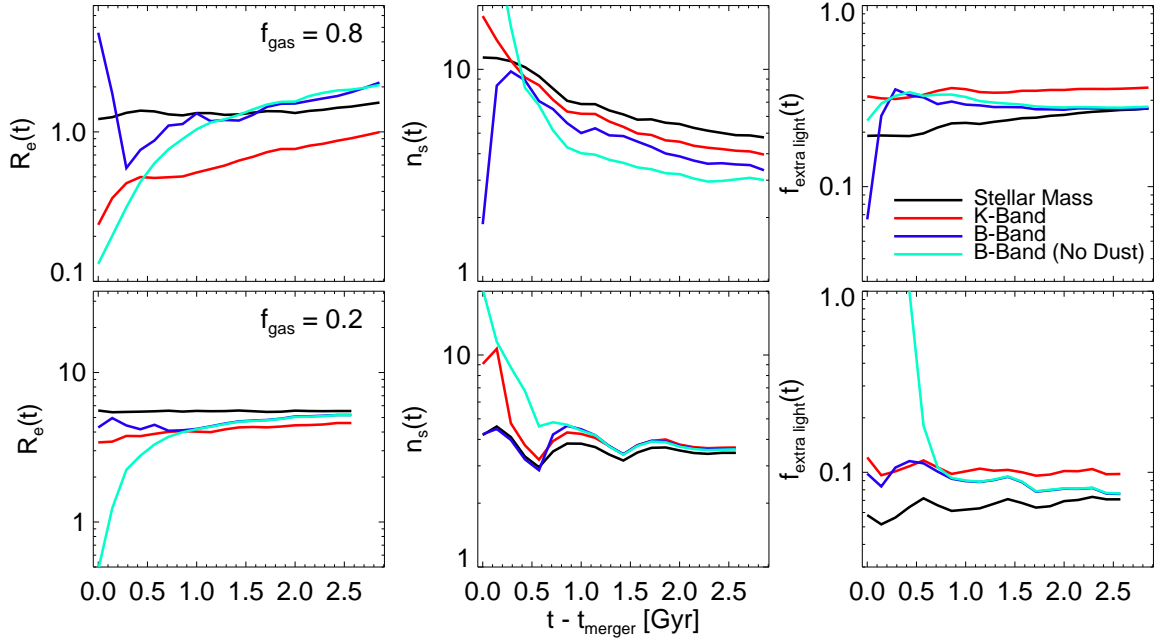}
    \caption{Evolution of surface brightness profile parameters as a function of time 
    after the coalescence of merging nuclei, for a highly gas-rich major merger ({\em top}) 
    and a less gas-rich major merger ({\em bottom}), in terms of the stellar mass 
    profile, observed $K$-band (including stellar population age and metallicity 
    effects and dust obscuration), observed $B$-band, and $B$-band ignoring dust 
    obscuration (as labeled). Values shown at each time are medians across $\sim100$ sightlines, 
    the sightline-to-sightline variance is generally slightly smaller than the offsets in the plotted 
    curves. {\em Left:} Half-light radius (directly from the light profiles). {\em Middle:} Outer 
    profile Sersic index $n_{s}$. 
    {\em Right:} Extra light fraction. 
    The evolution in $n_{s}$ is largely real (and does not generally follow systematic 
    patterns), and owes to relaxation of shells and other features; other evolution owes to 
    stellar populations and dust. 
    \label{fig:param.evol}}
\end{figure*}

Figure~\ref{fig:param.evol} shows the evolution with time of the best-fit 
surface brightness profile parameters for two more representative cases, one with a 
large initial gas fraction and one with a small initial gas fraction. Unlike 
Figure~\ref{fig:mu.evol} which was just a qualitative illustration, we 
include dust obscuration by calculating the column densities through the 
gas in our simulations in the manner of \citet{hopkins:lifetimes.methods,hopkins:lifetimes.letter}. 
We show the evolution of the absolute effective radius (i.e.\ independent 
of any fitting), the best-fit outer Sersic 
index, and the best-fit extra light fraction with time after the final coalescence, in terms of 
stellar mass, $K$-band light, $B$-band light, and $B$-band light in the 
absence of dust obscuration. 

The effective radius and extra light fraction in terms of stellar mass show
little evolution with time. In other words, the profile is relaxed out to $\sim R_{e}$ 
in a short time ($\sim$ the dynamical time). The behavior in $n_{s}$, however, is 
more sensitive to the outer regions where the dynamical time is long, and therefore 
more varied. We see in Figure~\ref{fig:param.evol} that, in one case, $n_{s}$ declines 
slowly with time -- this is caused by a large sloshing from the merger, which moves 
stars to large radii and forces a fit with a more extended envelope (higher $n_{s}$), until 
the sloshing relaxes. In the other case, there is a clear ringing or oscillation in 
$n_{s}$. This is caused by material on nearly radial orbits creating shells, as described in 
detail in \citet{quinn.84,hernquist.quinn.87} and \citet{hernquist.spergel.92}. 
The oscillations occur as the material in shells moves in and out to 
large radii. As more orbits are completed, the shell stars phase-wrap, smoothing their 
projected density distribution, and so the ringing in $n_{s}$ damps out. These 
behaviors are not unusual, but there is considerable variation in the effects on $n_{s}$ 
with time -- it is difficult to define a typical behavior for the relaxation of $n_{s}$. 
It is also clear that the degree to which $n_{s}$ might be different after the system is 
relaxed to that measured in the recent merger remnant stage is highly variable -- 
there is no well-defined systematic bias (or even a typical direction of such a bias, to 
higher or lower $n_{s}$) in $n_{s}$ at early times versus late times. More likely, we 
expect that there will be less scatter in $n_{s}$ in the 
merger remnant sample of \rj\ once the objects have all relaxed 
\citep[see also][]{naab:profiles}.

The $K$-band profile reflects these trends -- there is 
some (generally weak, even in the extreme high $f_{\rm gas}$ case) 
time evolution owing to the weak dependence of $K$-band $M/L$ on stellar age. 
The sense of this is what is generally expected -- since stars in the 
center of the remnant are formed in the final starburst and younger, they have 
larger $L/M$ at early times, and more of the light appears to come from inner radii, 
biasing $R_{e}$ to slightly smaller values. The effect is generally applicable for
only $\sim$ a couple $\times 10^{8}$\,yr after the merger, however, considerably 
shorter than the likely ages of most of the objects in the observed sample (excepting, 
perhaps, those still in the ULIRG/LIRG class). After this time, the age difference 
is small relative to the absolute ages of the populations, and makes little difference to 
$M/L$ as a function of radius.

That is not to say, however, that the $K$-band is free of bias. The measured quantities 
reflect the stellar mass profile with little additional time dependence, but 
are systematically offset by a small amount. This is because both of the examples 
shown have strong metallicity gradients, with higher metallicities in their central 
regions. In $K$-band, $L/M$ rises with metallicity, so this bias goes in the 
same direction as the age bias, but is more permanent. The metallicity gradient does 
not fade in the same manner as an initial age gradient. There is therefore a 
systematic bias towards underestimating $R_{e}$ and overestimating 
$f_{\rm extra light}$ by roughly $\sim20\%$. The magnitude of this bias, however, is 
directly sensitive to the metallicity gradient, and observed ellipticals tend to have 
gradients ranging from nil to comparable strength to those above (discussed in 
detail in \citet{hopkins:cusps.ell}), so the bias will range from object to object in this range. 
It is therefore difficult to systematically correct our analysis of the observed systems -- 
but even if we uniformly apply the most extreme correction for this fact, it makes no 
difference to our conclusions, and the offsets implied are considerably smaller 
than the object-to-object scatter in these parameters. In fact, considering an ensemble 
of simulations, one is far more likely to introduce bias in $R_{e}$ and $f_{\rm extra\ light}$ 
by not including sufficient dynamic range in the surface brightness profile used for fitting. 

In $B$-band, the comparison is roughly as expected. If one ignores dust, there is an 
extremely strong time dependence of all the fitted parameters until $\sim1\,$Gyr after 
the merger. Dust, however, negates much of this, because the youngest stellar populations 
introducing the greatest bias also tend to be the most obscured. To the extent that this 
is typical, the biggest effects in $B$-band are also in the first $\sim$ couple
$\times 10^{8}$\,yr, where 
stellar populations are young and substantial dust exists which can make the 
best-fit parameters highly variable and sensitive to viewing angle. Interestingly, at 
late times after the merger, the optical light profile reflects the stellar mass profile with 
{\em less} bias than the NIR profile. This is because, in $B$ and other optical bands, the 
trend of $L/M$ with metallicity is opposite that with age: $L/M$ decreases with metallicity. 
Therefore the effects of age gradients and metallicity gradients actually tend to 
cancel one another out at late times. This suggests that, at early times after a merger 
(typical of the systems in the \rj\ sample studied here) the NIR is a better tracer of 
the surface brightness profile, considerably less sensitive to 
age effects and dust-induced sightline-to-sightline 
variation. Once the systems are fully relaxed, however (provided there has been no new 
infall of gas and dust or new star formation event), optical light becomes similarly robust, and 
can actually trace the mass distribution with slightly less systematic bias. In practice, however, 
even evolved ellipticals often have dust lanes or central dust concentrations, 
so care must be taken regarding their effects.

\section{Discussion}
\label{sec:discuss}

We have studied the origin and properties of extra light or cusps
in mergers of gas-rich galaxies using a large suite of both numerical
simulations and local observed merger remnants. We confirm the
original prediction of \citet{mihos:cusps}
with our improved numerical models: namely, that
tidal torques in major mergers of gas-rich disks channel gas into the
central regions of the galaxy, where it forms a dense central
starburst. The starburst leaves a central light excess with a high
phase space density, making the remnant more compact, reconciling
the physical and phase space densities of disks and elliptical
galaxies, and imprinting radial gradients into the remnant
\citep{mihos:gradients}.

Stars in the remnant can be separated into three distinct populations. First, 
those stars which form in the progenitor disks before the final merger and coalescence 
of the two galaxies. The final merger scatters the orbits of these 
stars as 
they experience violent relaxation. They dominate the light, even in extremely gas-rich 
merger remnants, outside of $\sim0.5-1\,$kpc, and develop a Sersic-law 
$n_{s} \sim2.6\pm0.7$ profile 
owing to their partial violent relaxation. These stars are effectively the 
dissipationless component of the merger, and behave to lowest order much 
as they would entirely in the absence of gas. For example, the effective 
radius of just these stars reflects that of the progenitor disks (it is substantially larger than 
that of a typical elliptical of the same mass). 

Second is the starburst population, 
formed in the central gas concentration during the final 
stages of the merger. Since dissipation 
allows the gas to lose energy rapidly, this component is very compact, and 
dominates the light inside a small radius $\lesssim0.5-1\,$kpc, comparable to the 
observed scales of central starbursts in merging LIRGs and ULIRGs. These 
stars {\em do not} undergo violent relaxation -- because the dynamical times in the 
center of the galaxy are short, the final coalescence has largely completed in these 
central regions by the time the starburst nears maximum. In other words, these 
stars form in a compact, dissipational starburst in a nearly fixed background potential 
set by the dissipationless component of the merger. We show that the size of this 
dissipational component is set primarily by the radius at which it becomes 
self-gravitating. The gas is then generally stable against further collapse (even with 
cooling) and rapidly forms stars. Since the star formation time scale and subsequent 
collapse are both regulated by the dynamical time (and the system is self-gravitating), 
the central component systematically processes its gas as it contracts. 
It's worth noting that merger-induced starbursts may not be the only source of 
dissipation \citep[for example, stellar mass loss may replenish the gas supply and 
lead to new dissipational bursts, see e.g.][]{ciottiostriker:recycling}, but for 
our purposes, all dissipational star formation will appear similar when observed 
and have the same effects (we are essentially measuring the integrated amount of 
dissipation). 

Third, some gas survives the merger. This is primarily material which
has been moved to large radii temporarily, either blown out by a
combination of supernova and AGN feedback or thrown out in tidal
tails. This gas will then slowly settle back in, against the
background of a largely relaxed remnant.  Since there are no strong
tidal torques remaining, the material quickly settles into a small,
rotationally supported disk, and typically forms embedded kinematic
components \citep[embedded disks, kinematically decoupled cores, etc.; see][]{hernquist:kinematic.subsystems,
hoffman:prep}. While potentially important for the kinematics of the
remnant (rotation and isophotal shapes, in particular), this component
rarely contributes
substantially to the surface density profile of
most objects, and therefore is not of immediate interest. 
Furthermore, this component has little effect on the key
predictions here, regarding e.g.\ the degree of central dissipation
and central mass concentration.  Some caution should be taken, though,
since in optical bands a younger embedded disk can appear much more
prominent (for example, up to $\sim20\%$ of the $B$-band light, despite being
only $\sim2\%$ of the stellar mass).

In principle, observed merger remnant light profiles can be decomposed
into a central concentration tracing the dissipational starburst and
an outer Sersic profile reflecting the dissipationless, violently
relaxed stars. However, we show that any attempt to infer the extra
light content of a galaxy requires care. Fitting the
light profile of a typical, albeit very gas-rich merger remnant to a
pure Sersic law or a cored Sersic law, for example, and comparing the
central regions with the extrapolation of the outer profile yields
physically meaningless values of both the outer Sersic index and the
extra light fraction (in the cored Sersic case, the fit actually is
biased towards inferring that these gas-rich merger remnants have {\em
missing} light in their centers, i.e.\ some kind of central mass
deficit). The values do not at all reflect the {\em physical} values:
i.e.\ the amount of mass involved in the dissipational
starburst or the profile of the violently relaxed (non-starburst)
stars.  That is not to say, however, that these are uniformly poor
fits to the profiles -- if one wishes to use such profiles to
recover e.g.\ the effective radius or total light content of the
galaxy, the bias is not severe \citep[but
see][]{boylankolchin:mergers.fp}.  However, it highlights the fact
that any parameterized fit will have degeneracies
and can recover systematically different values even for a nominally
similar parameter (such as the outer Sersic index).

By using our simulations as a testing ground, we can calibrate such a fit in order to 
design a parametric decomposition 
of the profile that recovers the physical values of 
interest. We find that a simple means of doing so is to fit the surface brightness profiles 
to the sum of an inner exponential (Sersic $n=1$) and outer Sersic (free $n$) 
profile. The choice of $n=1$ for the inner regions is {\em not} meant to say that this 
reflects the true shape of the central extra light extrapolated inwards 
to very small radii, which can be complex 
\citep[with e.g.\ stellar clusters and features at 
$\ll 50\,$pc][]{lauer:centers}. In fact, this fitting procedure should not be used or 
extrapolated to within $\sim30-50\,$pc, which neither our simulations nor the observations 
to which we compare typically resolve (indeed, in experiments with some well-resolved 
profiles, we find that including these small radii when fitting
can lead to misleading results and a 
much higher rate of catastrophic failures of the fit owing to the
presence of extremely small-scale features that are unimportant for
the overall profile). 
The reason for the choice of $n=1$ for 
the inner component is that it, on average, yields the correct decomposition in both total 
mass and radius between the starburst and non-starburst stars, and because it 
minimizes the degeneracy with the outer Sersic index (fitting an inner $n=4$, for example, 
would introduce a large degeneracy with an outer profile that had a similar 
$n\sim4$ profile). 

We apply this decomposition to our simulations and to a large sample
of gas-rich merger remnants observed by \rj, ranging
from ULIRGs to shell ellipticals. We also directly fit each of the
observed profiles to a suite of simulations -- i.e.\ determine the
simulation mass profile which most closely matches that observed. We
find that, in all cases, we have simulations which provide good
matches to the observed systems, to better than the typical
point-to-point variance inherent in the simulation surface brightness
profiles ($\Delta\mu\lesssim0.1$).  We also find that the physical
starburst components in these best-fitting simulations are closely
related to those that we fit directly to the observed profiles,
lending further support to our procedure for decomposing the profiles.

Given our analysis, we can then study how the properties of the
two components scale. For
the outer profiles, we find that their Sersic indices are
remarkably constant as a function of stellar mass or any other
properties. Indeed, there appears to be a typical Sersic index
$\sim2.5-3$ for gas-rich merger remnants, with a scatter of
$\sim0.7$ about this median. We emphasize that this outer Sersic
index is only meaningful in the sense of reflecting those violently
relaxed stellar populations recovered in our two component
decomposition. Fitting the entire profile to a single Sersic index can
yield a very different result, and can introduce systematic trends (if
e.g.\ the typical extra light fraction or size changes with mass).
Given our attempt to carefully separate these components, then, this
should not be surprising: the dissipationless component is simply
acting under the influence of gravity, and is therefore completely
self-similar across scales \citep[see also][]{naab:profiles}. 
Nevertheless, we make a prediction that
{\em gas-rich merger remnants} should have this approximately constant
Sersic index distribution. Subsequent gas-poor remergers may have
different Sersic indices, as might pseudobulges or other low-mass
bulges (and together these may drive a systematic dependence of Sersic
index on mass, owing to the cosmological dependence of formation
history on mass); but this particular class of ellipticals should have
roughly a fixed Sersic index distribution.

The properties of the extra light component also scale in a regular
fashion. For a given initial simulation gas fraction, the extra light
content of the remnant systematically decreases with mass. This is
expected, since the star formation efficiency is higher in larger
mass systems, so they consume more of whatever gas they have at
earlier stages (e.g.\ on first passage), before the final merger. What
the extra light mass reflects is the gas content available and
channeled to the center during the final coalescence of the two nuclei.
Comparing the estimated extra light masses from our two-component
decomposition to this physical value, we find a good
correlation. There is substantial scatter -- for a given object, the
inferred extra light fraction based on fitting the profile can be
misleading (in the sense of not reflecting the true starburst mass
fraction) by a factor $\sim$ a couple, but, on average, the appropriate
value is recovered.

In a few observed systems, detailed stellar population studies have
enabled estimates of how much mass was recently formed in a
merger-induced starburst (as opposed to more extended star formation
prior to the merger). For these cases, then, we can directly compare
our estimated extra light component masses to the young stellar
populations, and find that indeed they trace one another to within a
factor $\sim2-3$, comparable to the expected scatter in both
estimators. Again, this suggests that we can in fact infer physical
decompositions from observed merger remnant profiles.

There are a number of interesting applications of this decomposition,
beyond noting the importance and physical nature of extra light
in recent gas-rich merger remnants. In \citet{hopkins:cusps.ell} we
apply this analysis to large samples of old, relaxed
elliptical galaxies with
central cusps, and consider the role of the extra light in shaping the
global kinematic properties of the galaxies. But we show here that one
can in fact use this light component to infer something physical about
the formation of the galaxy. Furthermore, we demonstrate in detail how
the different components relevant to the surface brightness profiles
of gas-rich merger remnants form and evolve. We also note that, with
this physical decomposition in place, we often infer extra light
fractions/masses and radii which are much larger than previous
estimates
(i.e.\ that extra light is typically $\sim1-5\%$ of the galaxy
light and becomes important only within $\sim0.05\,R_{e}$). This is
because the extra light component, as it becomes larger, typically
blends in more smoothly with the outer profile, and does not
necessarily appear as a sharp departure from the outer light
profile. We see this in both simulations and observations -- but in
almost every such case there is additional evidence that the
transition to extra light is real, including changes in the stellar
populations, ellipticity, and boxy or diskiness of the remnants at
these radii.

We find that gas-rich merger remnants do have excess light properties
similar to those in gas-rich merger simulations. Given a careful
two-component decomposition of the surface brightness profile (as
opposed to fitting the entire profile to a single Sersic index,
for example), we identify a statistically significant extra light
component in {\em every} gas rich merger remnant observed (and
simulated), with mass fractions spanning a wide range
$\sim3-30\%$. There are good simulation analogues to each observed
merger remnant, and they similarly contain this range of masses
involved in their final merger-induced starbursts. These extra light
masses correspond to reasonable, expected initial gas fractions for
the merging disks (although there is no one-to-one
correspondence). They are also comparable to the estimated fractions
in \citet{hernquist:phasespace}, namely the dissipational mass
fraction needed to explain the discrepancies between the maximum phase
space densities and surface brightness of ellipticals and their
progenitor spiral galaxies. We therefore confirm in observations of
gas-rich merger remnants the long-standing theoretical prediction that
sufficient dissipation is required to explain this discrepancy
in the central profiles of ellipticals.

We have studied these properties and identified robust trends across a
large suite of simulations, in which we vary e.g.\ the galaxy masses,
initial gas fractions, concentrations, halo masses, presence or
absence of bulges, presence or absence of black holes, feedback
parameters from supernovae and stellar winds, orbital parameters and
disk inclinations, and mass ratios of the merging galaxies.  This
range of parameters allows us to identify the most important
physics. Most of these choices, for example, affect the surface
brightness profile, extra light mass and radius of the extra light,
concentration and effective radius of the remnant, and even its
ellipticity and isophotal shape only indirectly. Ultimately, what
determines the structure of the remnant (insofar as the properties we
have considered) is, to first order, how much mass is in the
dissipationless (violently relaxed) component versus the
dissipational/starburst component at the time of the final coalescence
of the merging galaxies.  Therefore, varying e.g. the orbital
parameters or initial galaxy structure can alter the remnant
substantially, but predominantly only insofar as it changes the amount
of gas which will be available at the time of the final coalescence of
the galaxy nuclei (i.e.\ how much mass ends up in the starburst
component, as opposed to being violently relaxed in the
merger).

In the simplest possible scenario outlined above, the central light and dissipationless 
component arise simultaneously in a single gas-rich merger. In practice, 
the situation need not be so simple. It is well-established that spheroids 
undergoing subsequent mergers will conserve rank order in particle (stellar) 
binding energy and (correspondingly) mass profile shape \citep{barnes:disk.disk.mergers}. 
As a consequence, we expect that re-mergers or ``dry'' mergers will conserve 
the central light excesses (starburst component) 
originally established in a gas-rich merger, even if the entire galaxy profile expands 
by a typical factor $\sim2$ as it doubles in mass. In \citet{hopkins:cores}, we explicitly 
confirm in numerical simulations that the distinction between the remnant 
excess light and dissipationless outer profile is conserved in 
successive re-mergers, and that our methodology continues to reliably separate 
the original dissipational (compact, merger-driven starburst) and dissipationless 
(disk) components. This applies just as well to bulges in early-type galaxies: 
regardless of their formation mechanism (whether formed ``in situ'' by disk instabilities 
or minor mergers, or formed by earlier major mergers with a re-accreted disk), 
they will fundamentally be composed of a dissipationless (scattered) component 
and a dissipational (starburst driven by energy 
and angular momentum loss in gas) component, and these components 
will be separately conserved in subsequent mergers. For any merger remnant or 
elliptical galaxy, therefore, the observed central or excess light component 
should most generally be thought of as the sum of all compact starburst components 
formed in the history of the galaxy, and the outer light likewise as the sum of 
dissipationlessly scattered disk stars. In other words, if a spiral is transformed to a 
progressively earlier type (and, eventually, a true elliptical) by, say, a rapid series of 
many $5:1$ mergers, each of which disrupts some disk stars (scattering them into 
a spheroid) and causes some gas to lose angular momentum and produce a 
small nuclear starburst (building up the central light component), then the final 
inner and outer light components will reflect the sum of the dissipational and 
dissipationless events. 

Our physical interpretation of the extra light, and 
our conclusions and comparisons in this work, are therefore 
not changed by the merger histories of observed galaxies, nor by 
e.g.\ whether or not the progenitor disks have bulges -- it is simply 
possible that some of the central mass concentration was built up in multiple previous 
events along with the most recent merger (but the total gas/dissipational content 
involved is conserved). Of course, in {\em modeling} 
those progenitor bulges, this raises the important question of how their structure 
should be initialized: some choice must be adopted (just as if one were to construct 
an ``initial'' elliptical to model a ``dry'' merger) for the initial mass fraction in a very compact 
dissipational component (which will, then, remain compact or ``extra'' light in the final 
merger remnant) versus the more extended dissipationless component (which 
will become part of the dissipationless component of the final remnant). 
Fortunately, to the extent that (at least ``classical'') bulges obey the same parameter correlations 
as ellipticals, this suggests that we can estimate such properties and what they 
imply for the original progenitors and their gas content by restricting our 
study to bulge-dominated systems or pure ellipticals.

We note that there is considerable room for progress in modeling the 
extra light component itself -- including its kinematics, the shape of the 
surface brightness profile at small radii ($\ll100\,$pc), and the structure of 
galactic nuclei near a central black hole. Observations are rapidly making 
progress in this area, and revealing new insights into the formation histories 
of elliptical galaxies. Unfortunately, modeling these radii in a meaningful 
sense will require fundamental improvements in numerical simulations. At present, 
our spatial resolution reaches $\sim30\,$pc. In principle, it would not be hard to 
improve this to $\lesssim10\,$pc. However, at these scales, we approach the 
sizes of structure in the ISM -- i.e.\ individual giant molecular 
clouds, star forming regions, and massive star clusters or galactic stellar nuclei. 
Without simulations which can self-consistently form these structures (i.e.\ include 
the multiple gas phases of the ISM and their exchange), as well as resolve e.g.\ 
individual supernova blastwaves and remnants, improved spatial resolution 
has no physical meaning. It is therefore an important and ambitious goal that the next generation 
of studies of galactic nuclei move beyond the sub-resolution prescriptions 
necessary when modeling large scales and attempt to include star formation, 
supernova feedback, and realistic, resolved ISM structure in simulations 
of galaxy mergers.  

In earlier work \citep[e.g.][]{hopkins:qso.all,hopkins:red.galaxies,
hopkins:groups.ell,hopkins:groups.qso} we developed a model linking
starbursts, quasar activity, the growth of supermassive black holes,
and the origin of ellipticals through evolutionary phases of the same
events, driven by mergers of gas-rich galaxies.  There is much
observational support for the various links in this chain.  ULIRGs are
invariably associated with gas-rich mergers
\citep[e.g.][]{sanders96:ulirgs.mergers} and have bolometric
properties similar to bright quasars
\citep[e.g.][]{sanders:review,sanders88:warm.ulirgs},
suggesting that ULIRGs evolve into quasars.  By the Soltan argument
\citep{soltan82,hopkins:bol.qlf}, the bulk of the cosmic mass density in
supermassive black holes was accumulated during periods of bright
quasar activity.  Observed correlations between supermassive black
holes and properties of their host ellipticals \citep{magorrian,
FM00,Gebhardt00} demonstrate
that they formed together, not independently.  To the extent that
gas-rich mergers were responsible for growing most of the mass in
supermassive black holes, these correlations strongly endorse the view
that ellipticals formed originally in gas-rich mergers of spirals.
(With the possibility that they could have been modified subsequently
in gas-free mergers with other ellipticals that did not trigger quasar
activity or lead to black hole growth, and that modest black hole growth 
can occur with non-merger induced fueling \citep{hopkins:seyferts}.)

In addition to these observational lines of evidence, there are simple
{\it physical} arguments that support these connections.  The bulk of
the stellar mass in ellipticals was likely assembled dissipationlessly
through violent relaxation \citep{lynden-bell67}, but the growth of
supermassive black holes probably involved accretion of gas
\citep{lynden-bell69}.  As we argue in e.g.
\citet{hopkins:groups.qso}, the condition that supermassive black
holes and ellipticals originate together requires their assembly in
{\it gas-rich} mergers, which contain a galaxy's worth supply of both
gas and stars.  Our analysis here lends additional credence to this
hypothesis: the outer components of the light profiles of ellipticals
were indeed put into place through self-similar gravitational physics,
but the compact, inner starburst populations resulted from gas
dissipation.  The same gas dissipation that yielded this inner
component also provided the material to grow supermassive black holes
in ellipticals in a self-regulated manner, accounting for the
similarities between the fundamental plane of ellipticals
\citep{dressler87:fp,dd87:fp} and the black hole fundamental plane
\citep{hopkins:bhfp.theory, hopkins:bhfp.obs}.  Observationally, then, we
see that this blending of dissipationless stellar dynamics and gas
dissipation is reflected not only in correlations between 
supermassive black holes and their hosts, but in the detailed
structure of elliptical galaxies as well.

\acknowledgments We thank our referee, Thorsten Naab, for comments 
contributing to the content and presentation of this paper. This work
was supported in part by NSF grants ACI 96-19019, AST 00-71019, AST
02-06299, and AST 03-07690, and NASA ATP grants NAG5-12140,
NAG5-13292, and NAG5-13381. Support for 
TJC was provided by the W.~M.\ Keck 
Foundation.

\bibliography{/Users/phopkins/Documents/lars_galaxies/papers/ms}

\begin{appendix}
\section{Fits to the Sample of \rj}
\label{sec:appendix}

\begin{figure*}
    \centering
    \ascaleup
    \plotone{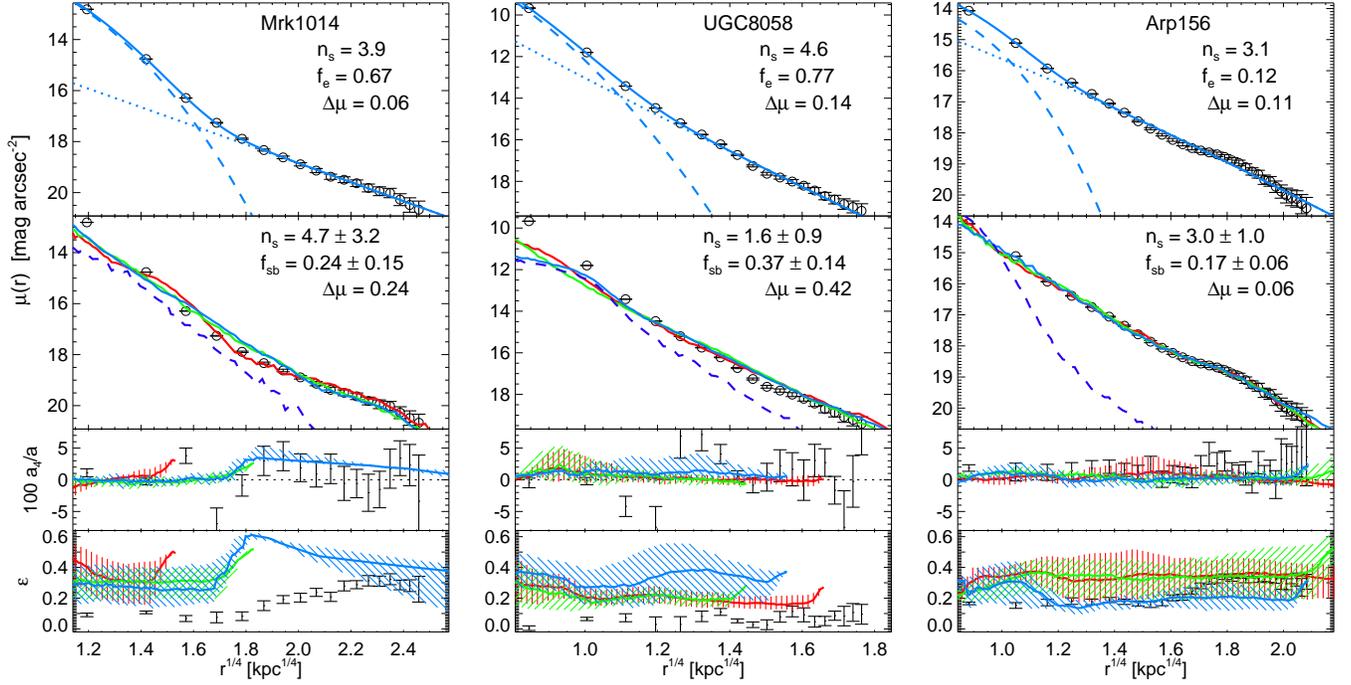}
    \caption{As Figure~\ref{fig:rj1}. Figures~\ref{fig:rj.all.1.ps}-\ref{fig:rj.all.18.ps} show 
    the results for all objects in the \rj\ sample. Objects are sorted from most to least 
    luminous in $K$-band. Note that Mrk1014 (previously unpublished) 
    and UGC8058 (Mrk231) shown here are contaminated by 
    central AGN, giving rise to the large discrepancies and poor fits seen.
    \label{fig:rj.all.1.ps}}
\end{figure*}

\begin{figure*}
    \centering
    \ascaleup
    \plotone{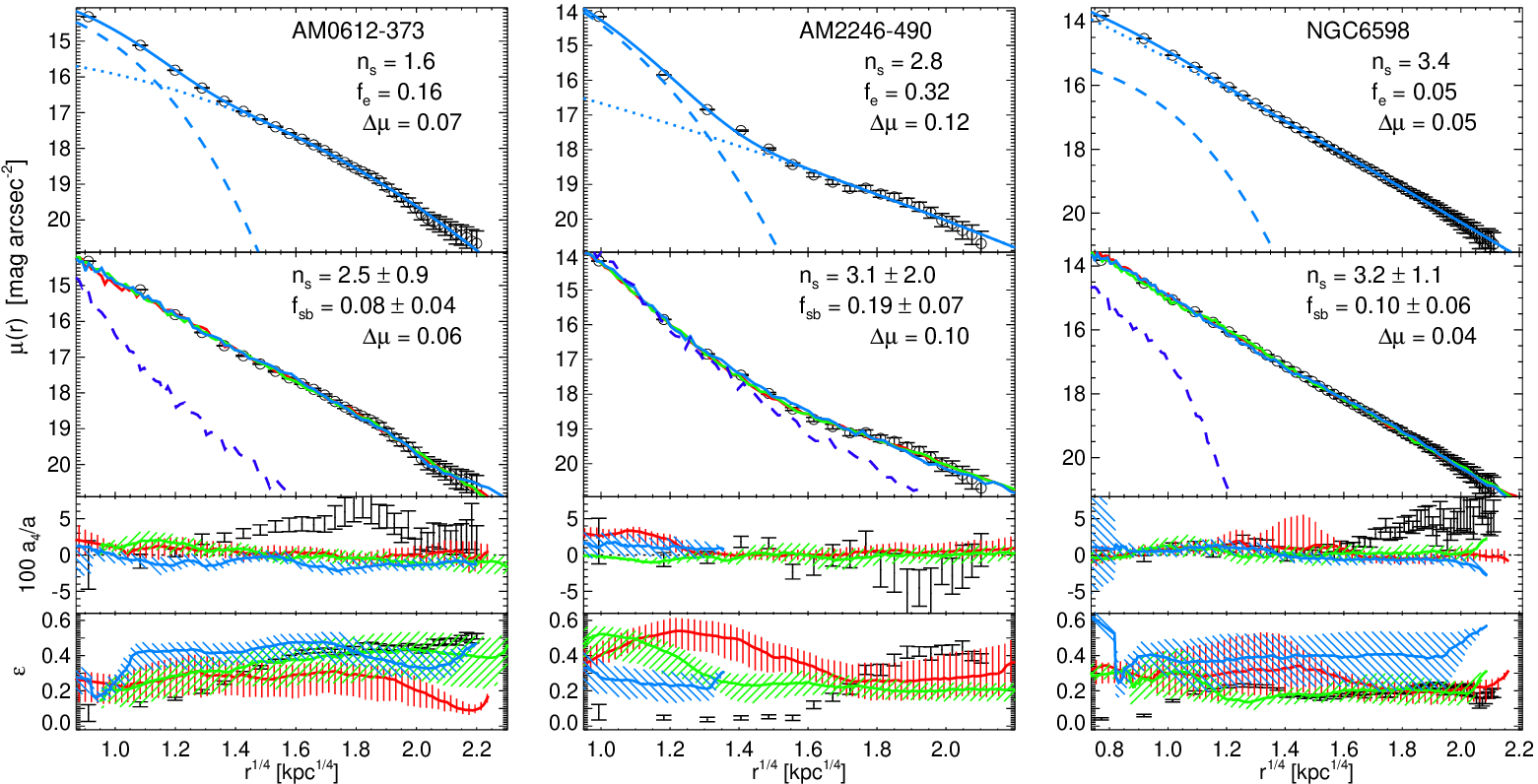}
    \caption{Figure~\ref{fig:rj.all.1.ps}, continued. 
    \label{fig:rj.all.2.ps}}
\end{figure*}

\begin{figure*}
    \centering
    \ascaleup
    \plotone{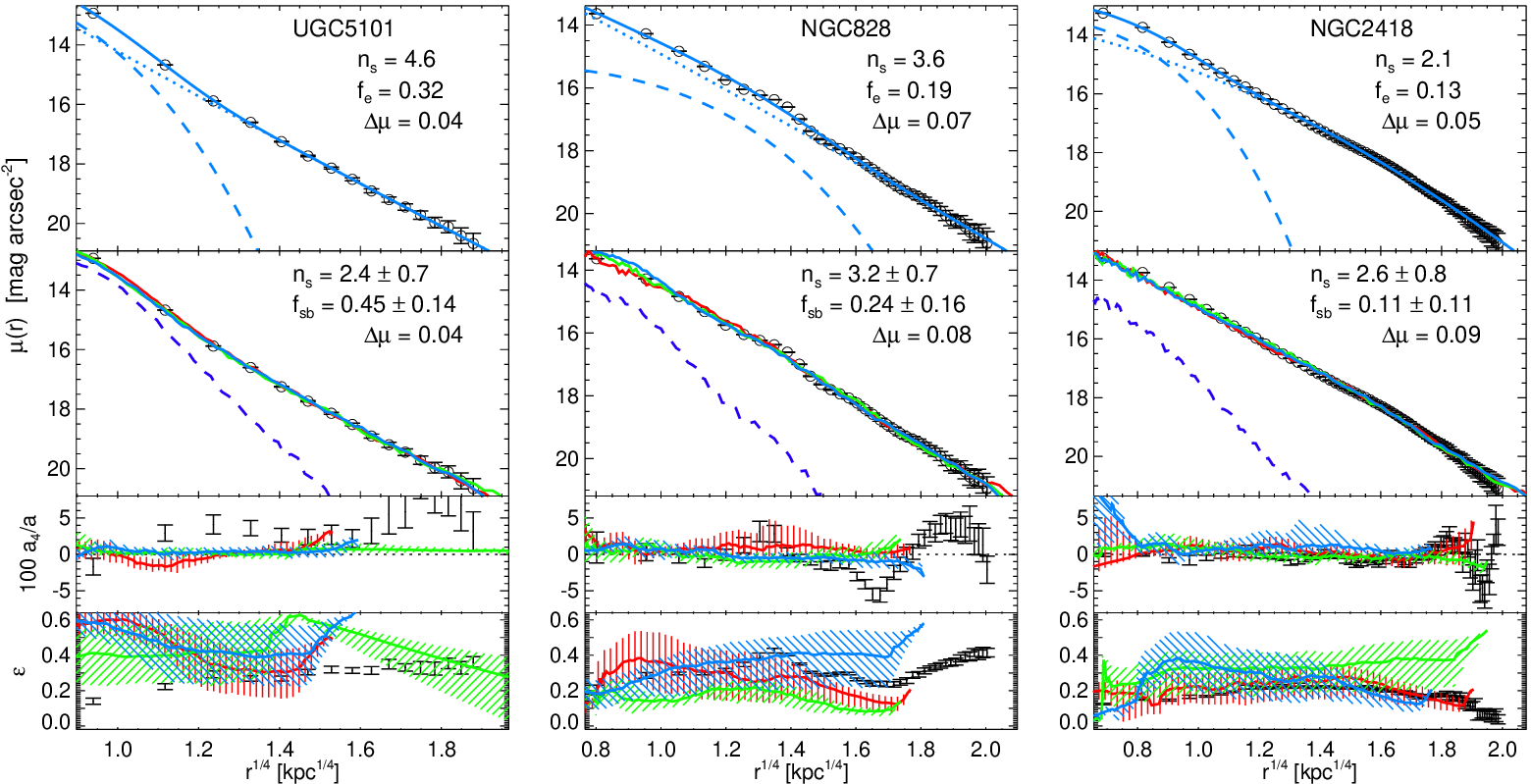}
    \caption{Figure~\ref{fig:rj.all.1.ps}, continued. 
    \label{fig:rj.all.3.ps}}
\end{figure*}

\begin{figure*}
    \centering
    \ascaleup
    \plotone{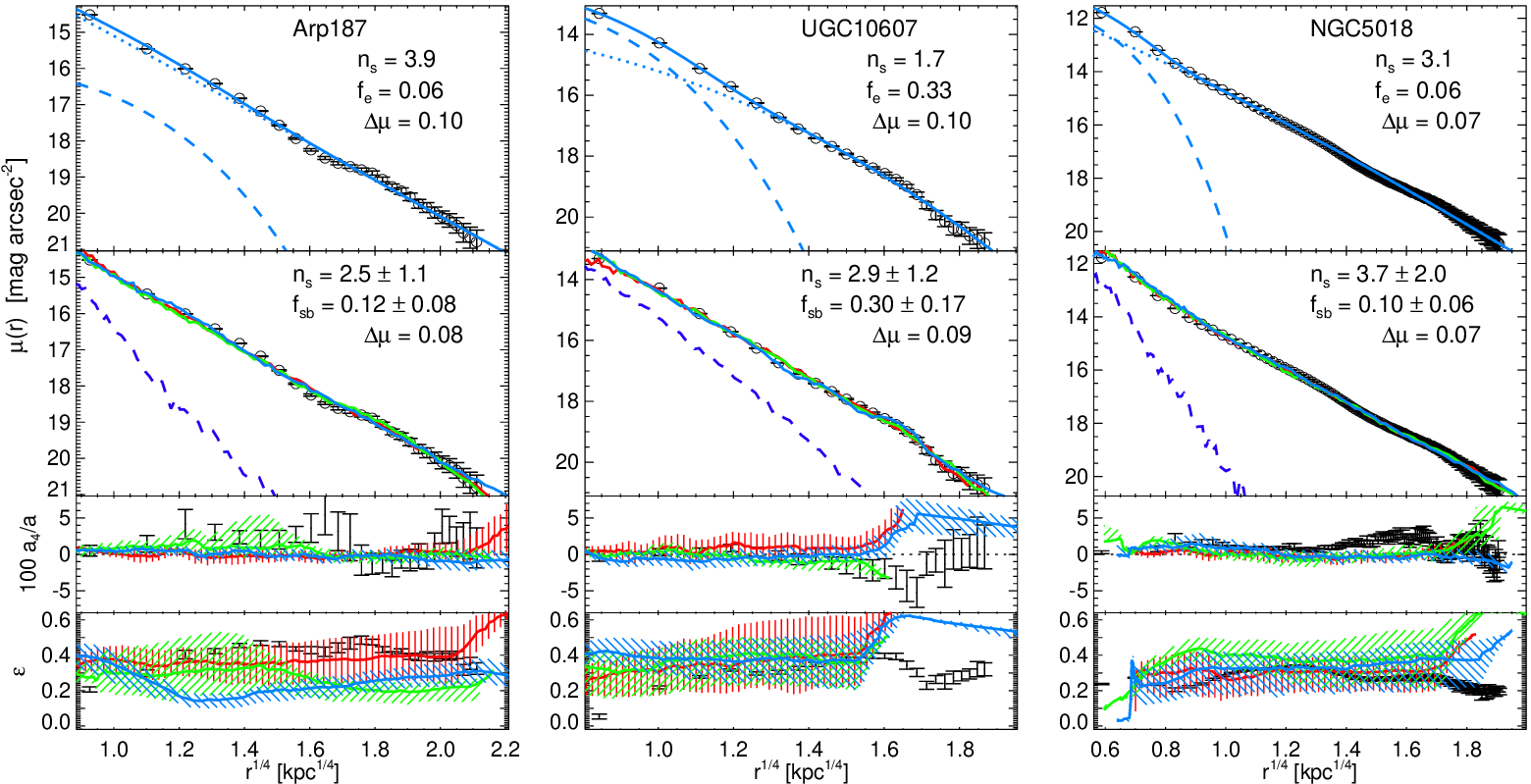}
    \caption{Figure~\ref{fig:rj.all.1.ps}, continued. 
    \label{fig:rj.all.4.ps}}
\end{figure*}

\begin{figure*}
    \centering
    \ascaleup
    \plotone{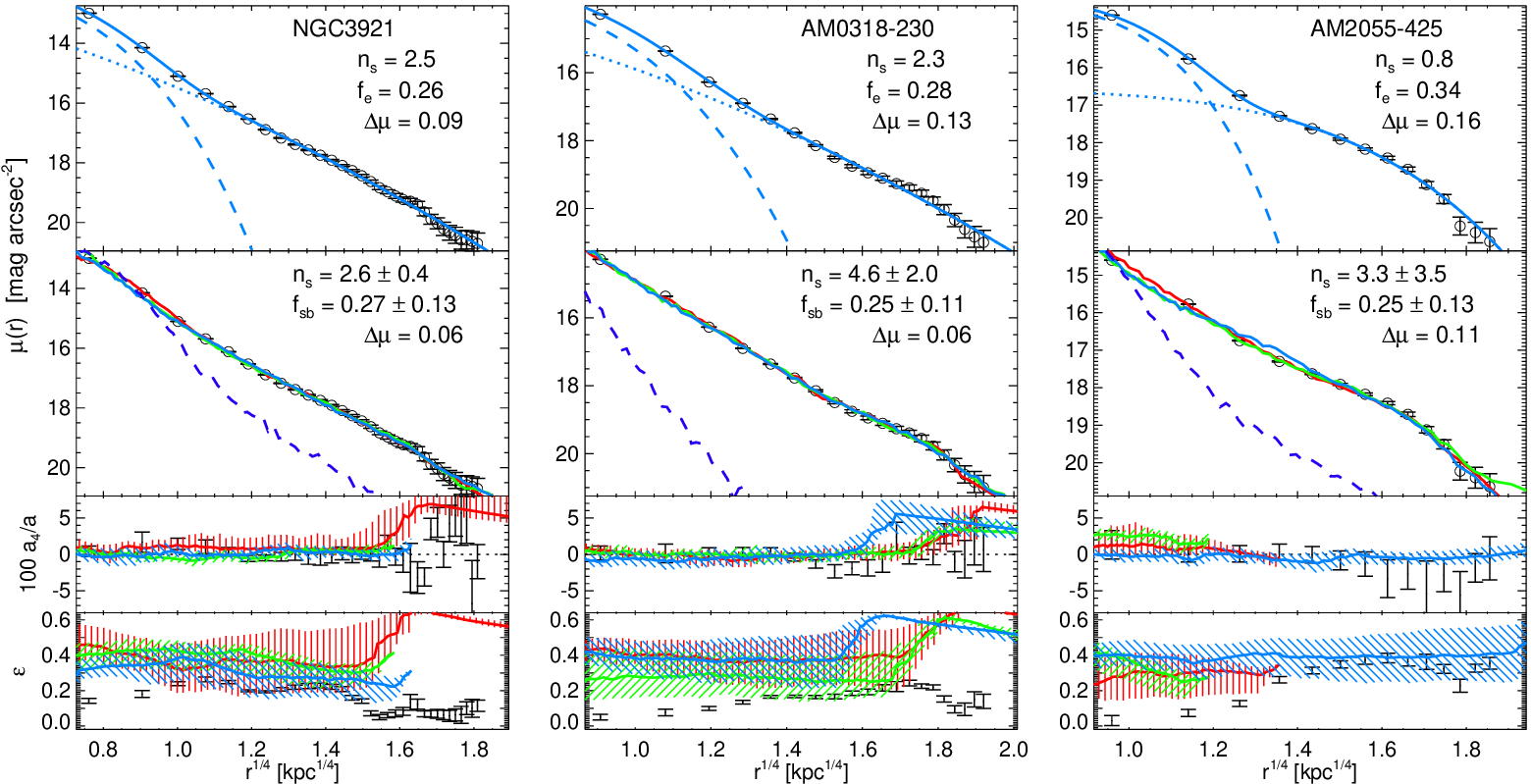}
    \caption{Figure~\ref{fig:rj.all.1.ps}, continued. 
    \label{fig:rj.all.5.ps}}
\end{figure*}

\begin{figure*}
    \centering
    \ascaleup
    \plotone{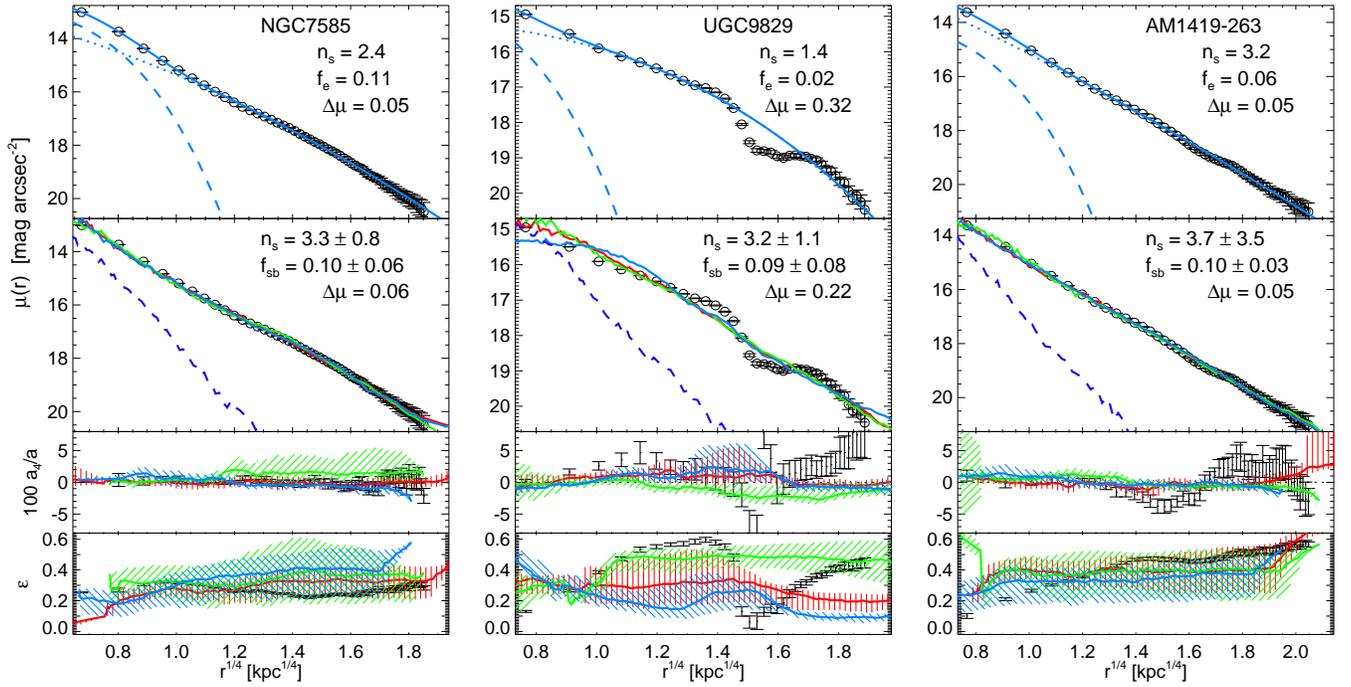}
    \caption{Figure~\ref{fig:rj.all.1.ps}, continued. Unrelaxed features 
    (including remaining spiral and tidal structure) in UGC9829 prevent a reliable fit. 
    \label{fig:rj.all.6.ps}}
\end{figure*}

\begin{figure*}
    \centering
    \ascaleup
    \plotone{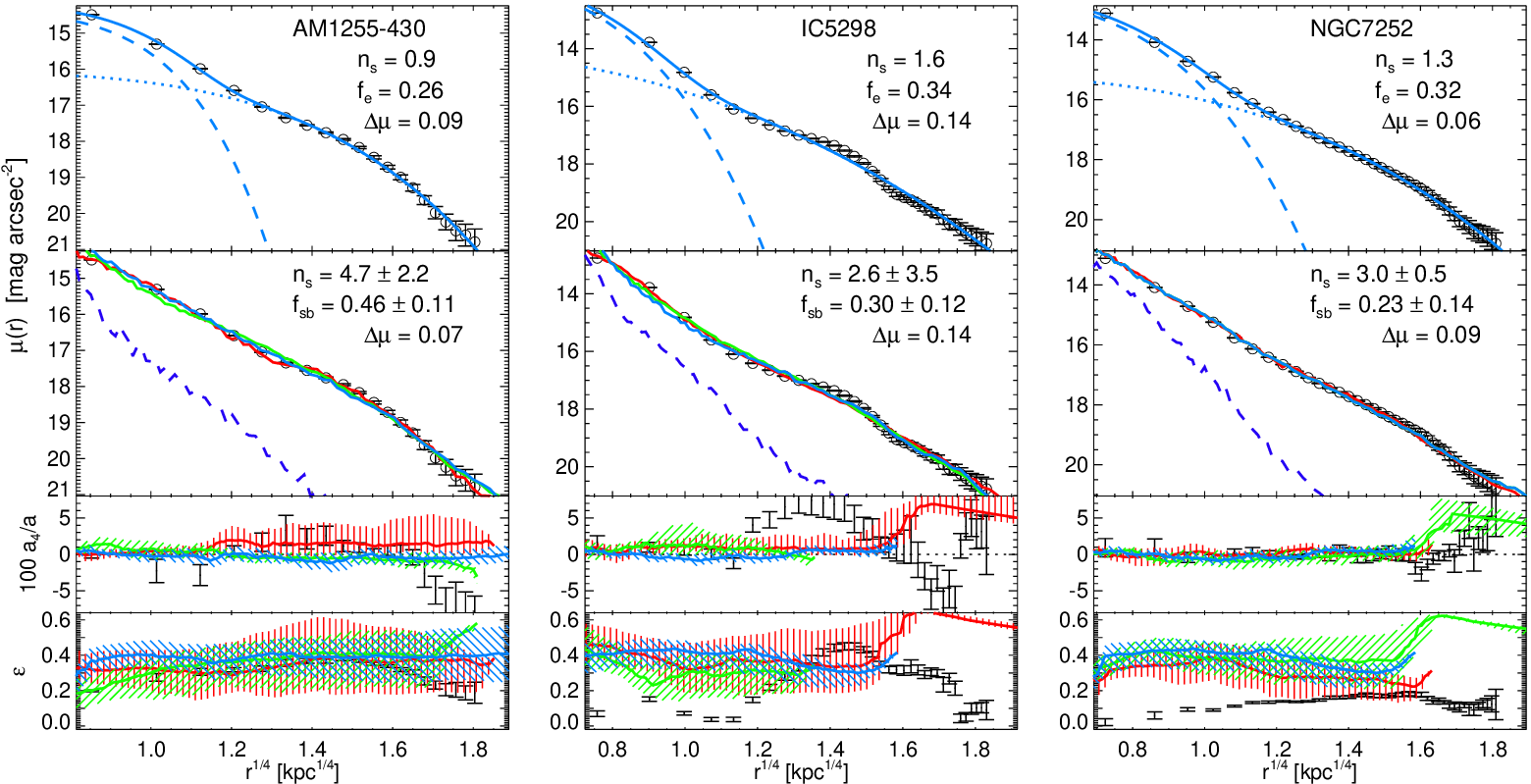}
    \caption{Figure~\ref{fig:rj.all.1.ps}, continued. 
    \label{fig:rj.all.7.ps}}
\end{figure*}

\begin{figure*}
    \centering
    \ascaleup
    \plotone{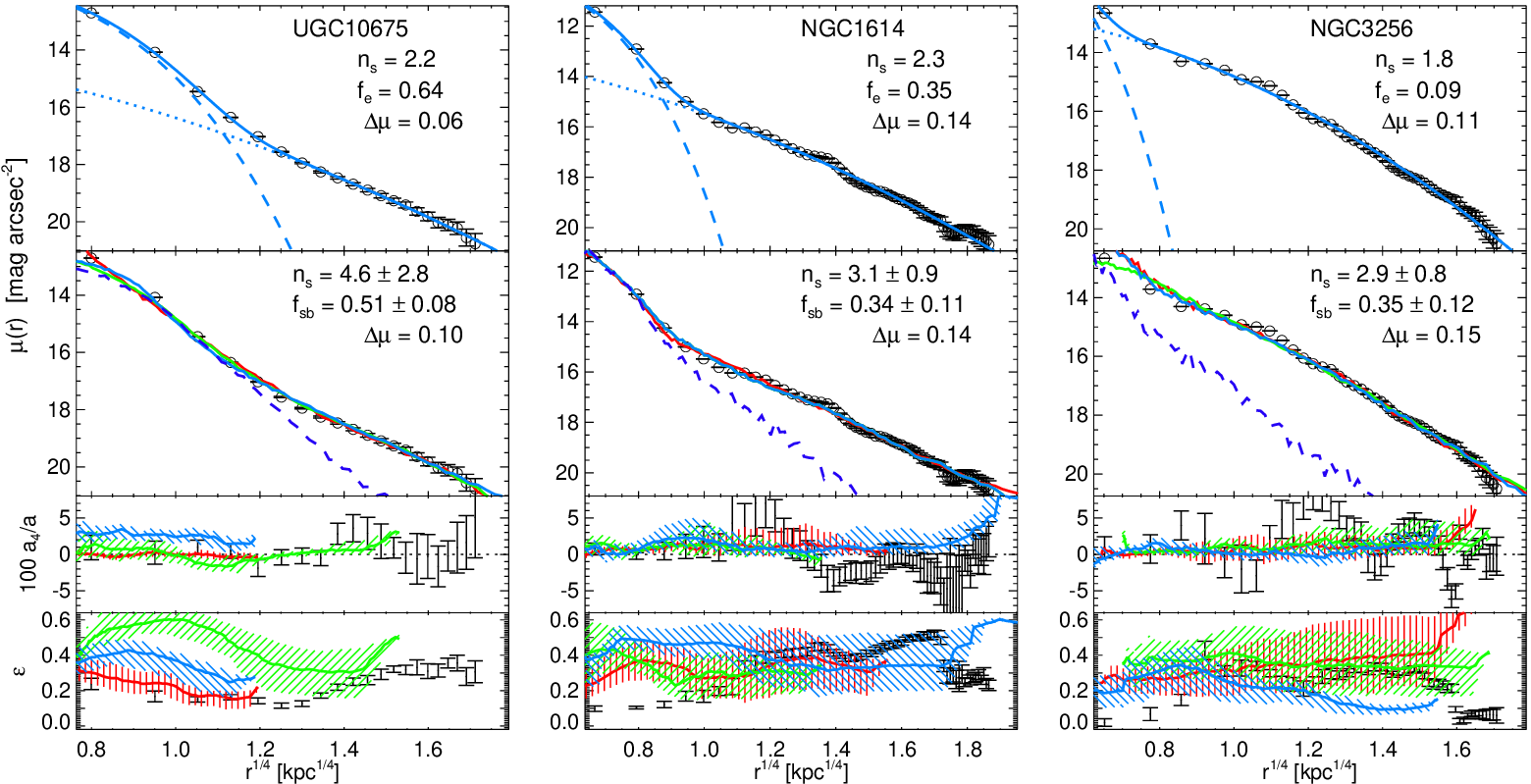}
    \caption{Figure~\ref{fig:rj.all.1.ps}, continued. 
    \label{fig:rj.all.8.ps}}
\end{figure*}

\begin{figure*}
    \centering
    \ascaleup
    \plotone{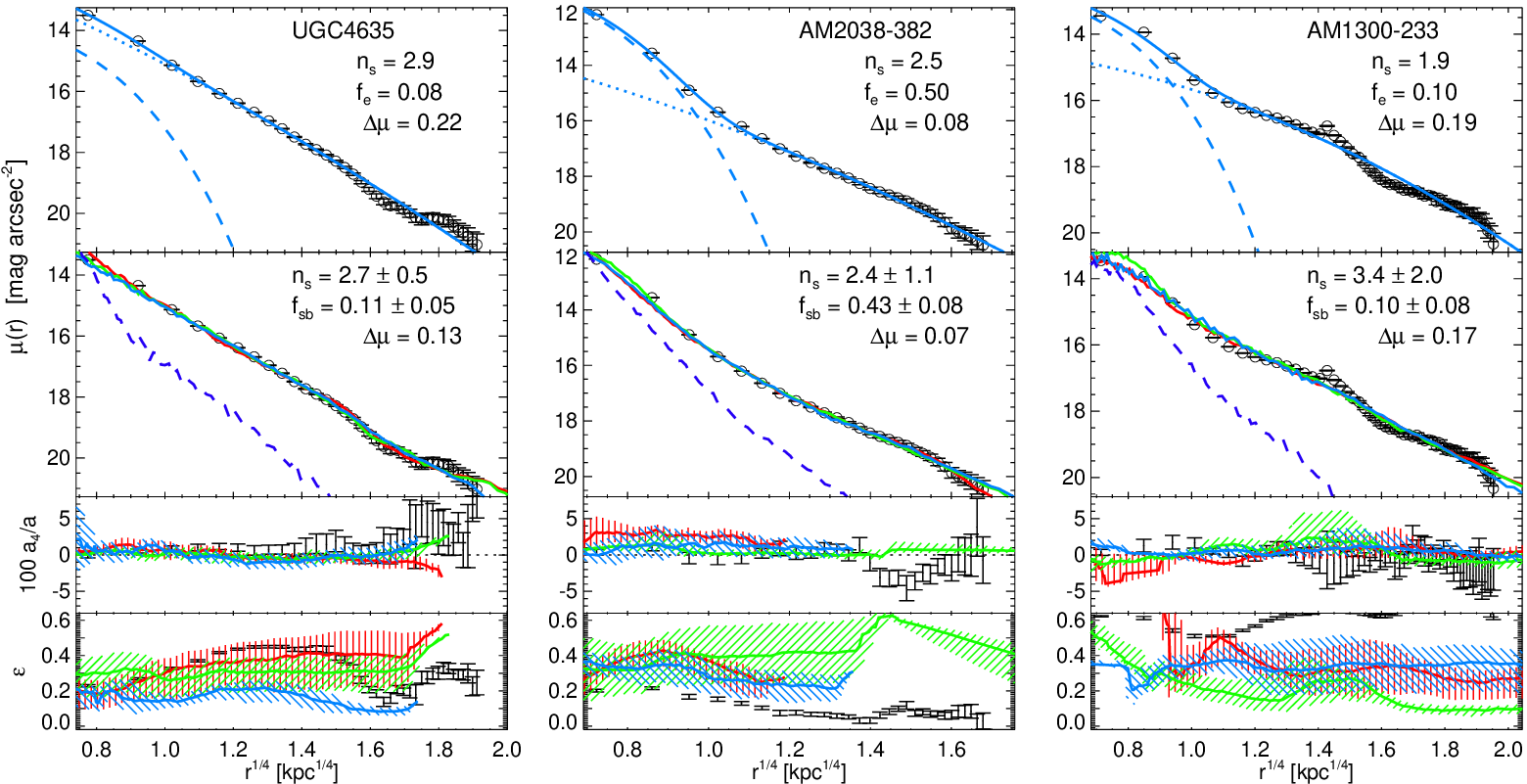}
    \caption{Figure~\ref{fig:rj.all.1.ps}, continued. 
    \label{fig:rj.all.9.ps}}
\end{figure*}

\begin{figure*}
    \centering
    \ascaleup
    \plotone{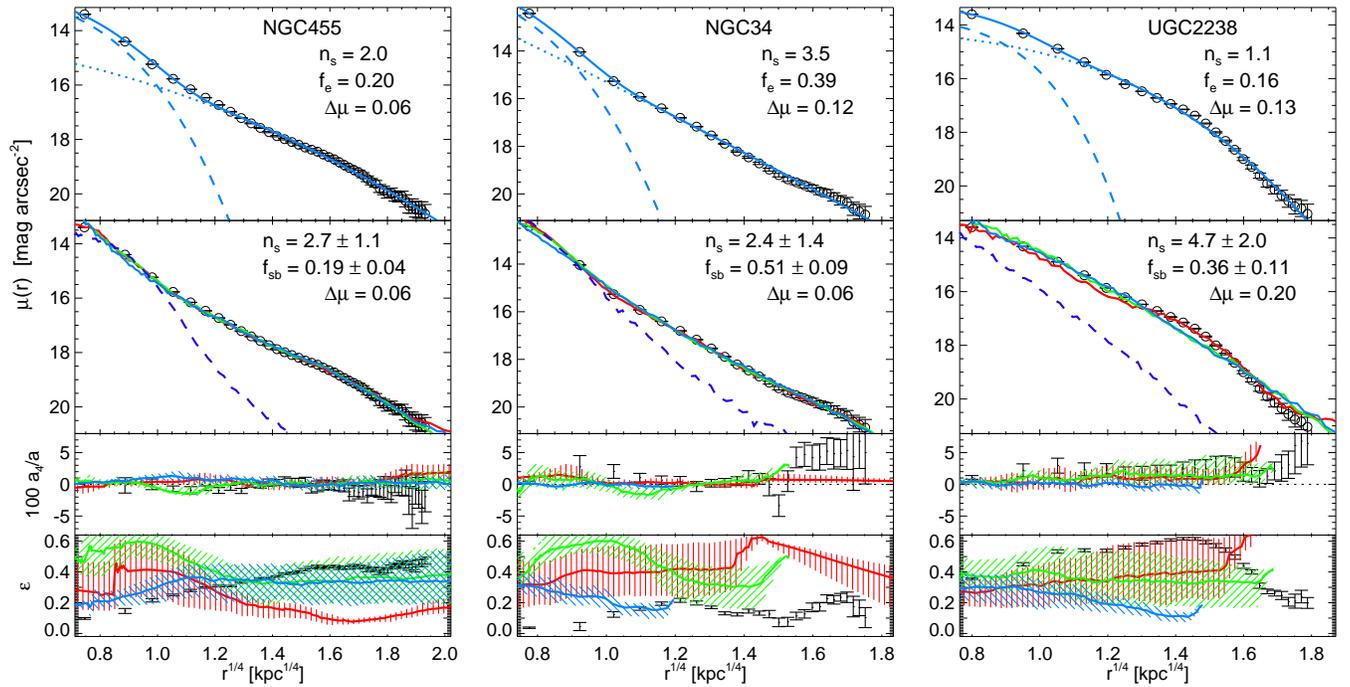}
    \caption{Figure~\ref{fig:rj.all.1.ps}, continued. UGC2238 contains a prominent 
    edge-on central disk.
    \label{fig:rj.all.10.ps}}
\end{figure*}

\begin{figure*}
    \centering
    \ascaleup
    \plotone{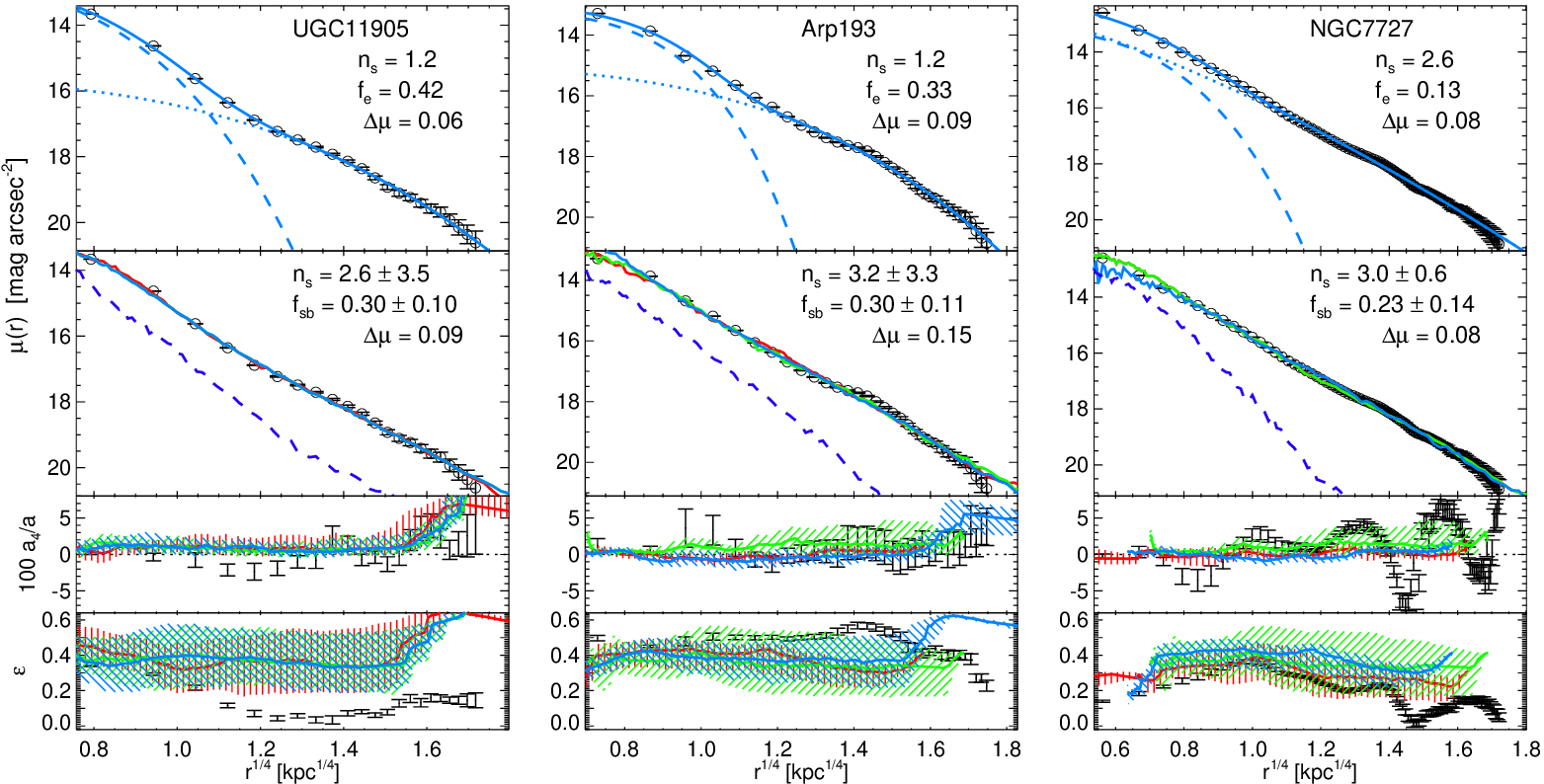}
    \caption{Figure~\ref{fig:rj.all.1.ps}, continued. 
    \label{fig:rj.all.11.ps}}
\end{figure*}

\begin{figure*}
    \centering
    \ascaleup
    \plotone{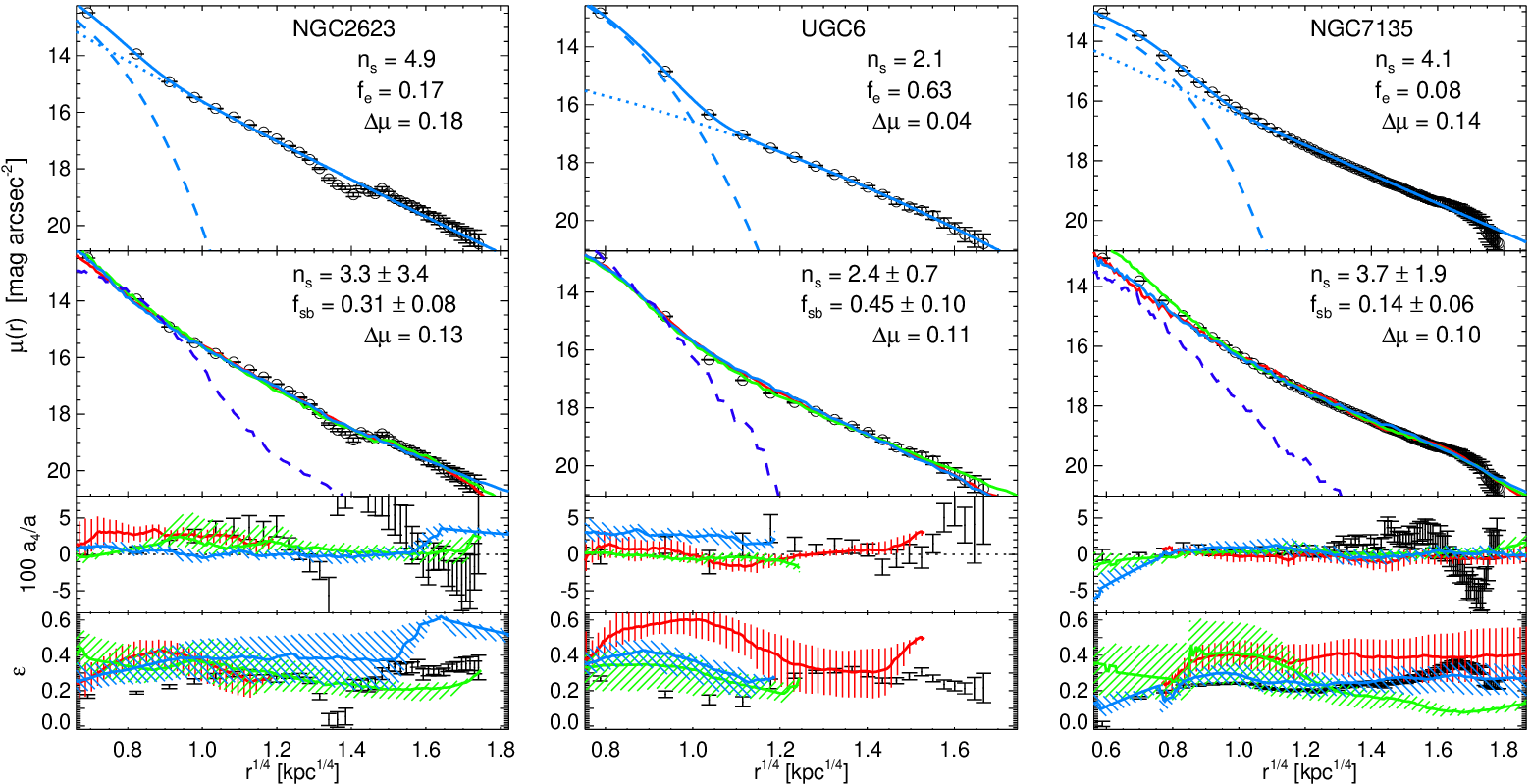}
    \caption{Figure~\ref{fig:rj.all.1.ps}, continued. 
    \label{fig:rj.all.12.ps}}
\end{figure*}

\begin{figure*}
    \centering
    \ascaleup
    \plotone{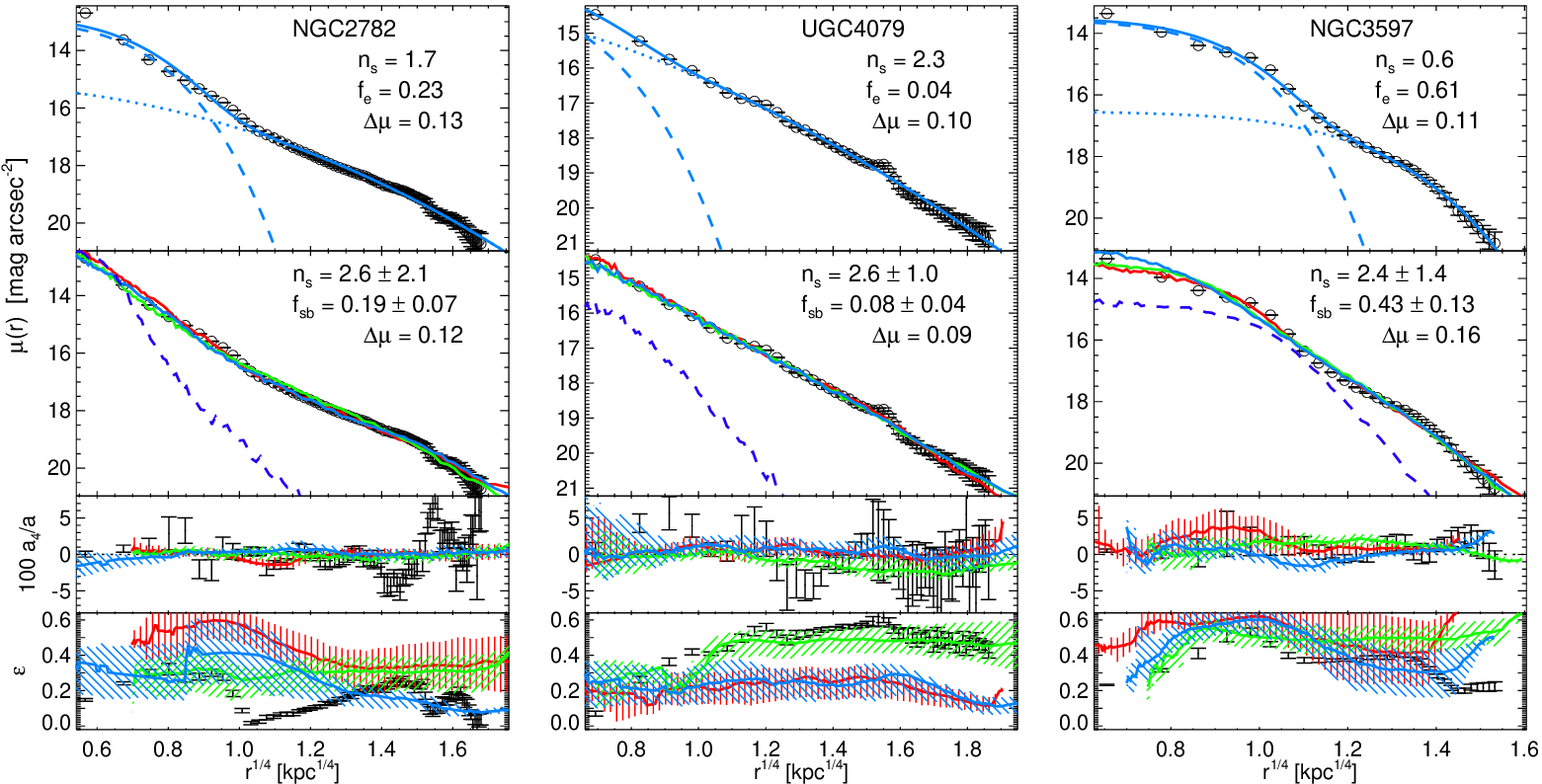}
    \caption{Figure~\ref{fig:rj.all.1.ps}, continued. 
    \label{fig:rj.all.13.ps}}
\end{figure*}

\begin{figure*}
    \centering
    \ascaleup
    \plotone{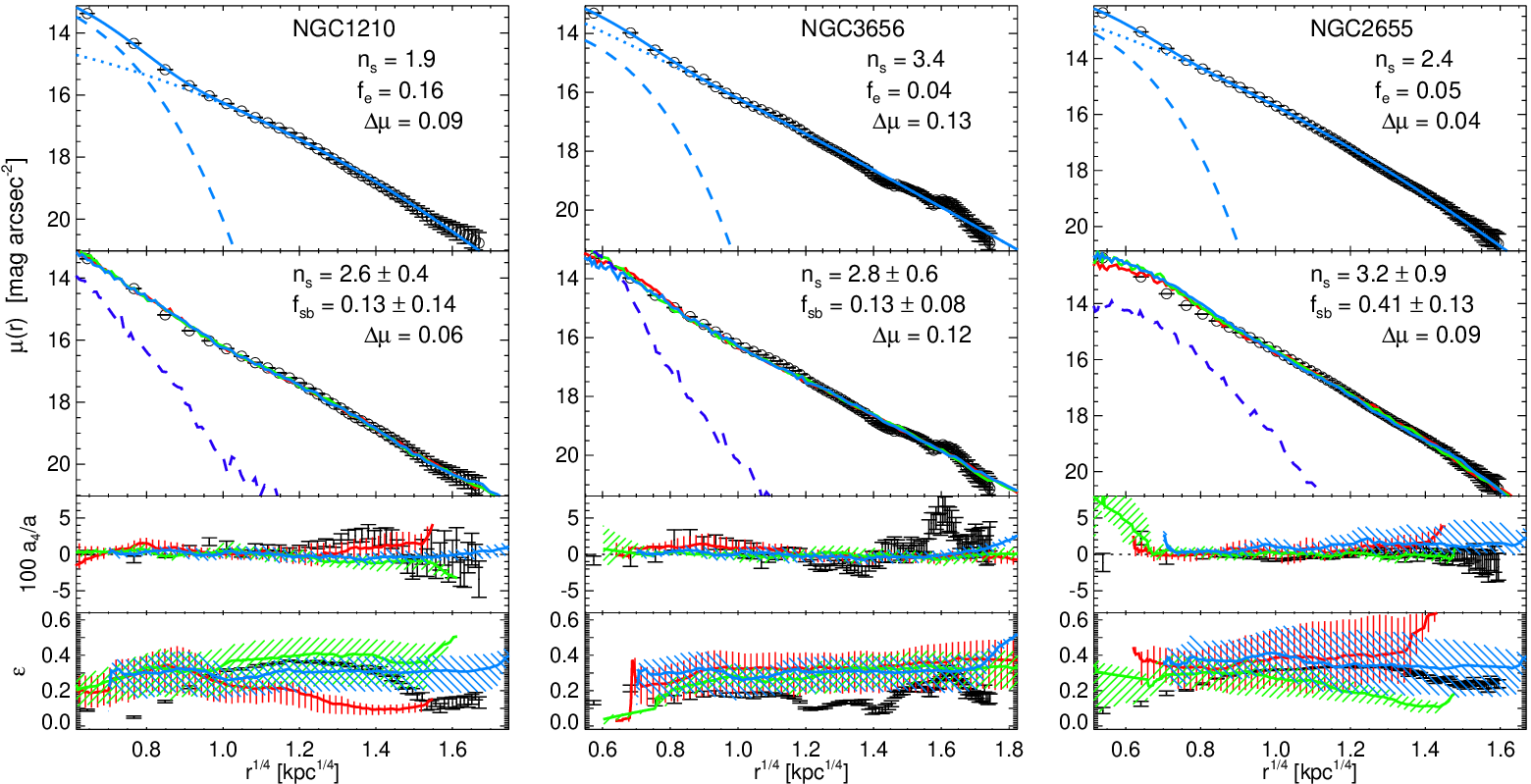}
    \caption{Figure~\ref{fig:rj.all.1.ps}, continued. 
    \label{fig:rj.all.14.ps}}
\end{figure*}

\begin{figure*}
    \centering
    \ascaleup
    \plotone{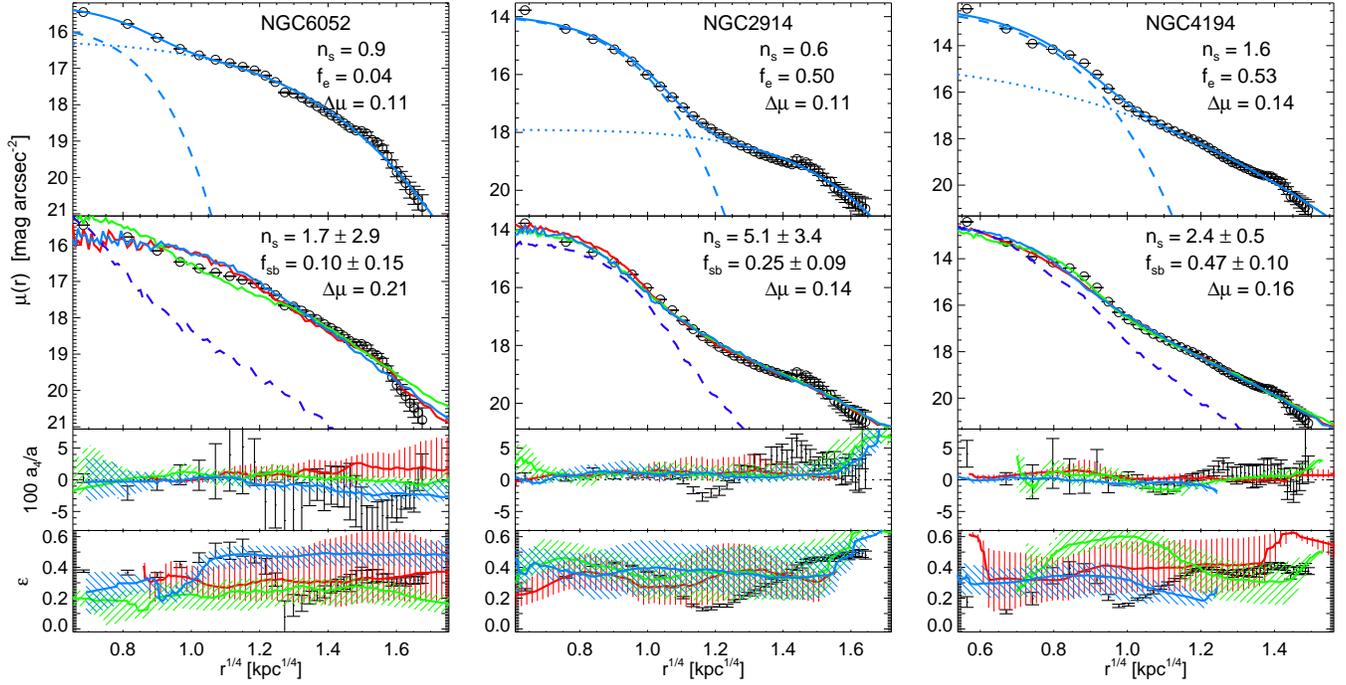}
    \caption{Figure~\ref{fig:rj.all.1.ps}, continued. NGC6052 appears to be still unrelaxed 
    at $\sim$a few kpc. 
    \label{fig:rj.all.15.ps}}
\end{figure*}

\begin{figure*}
    \centering
    \ascaleup
    \plotone{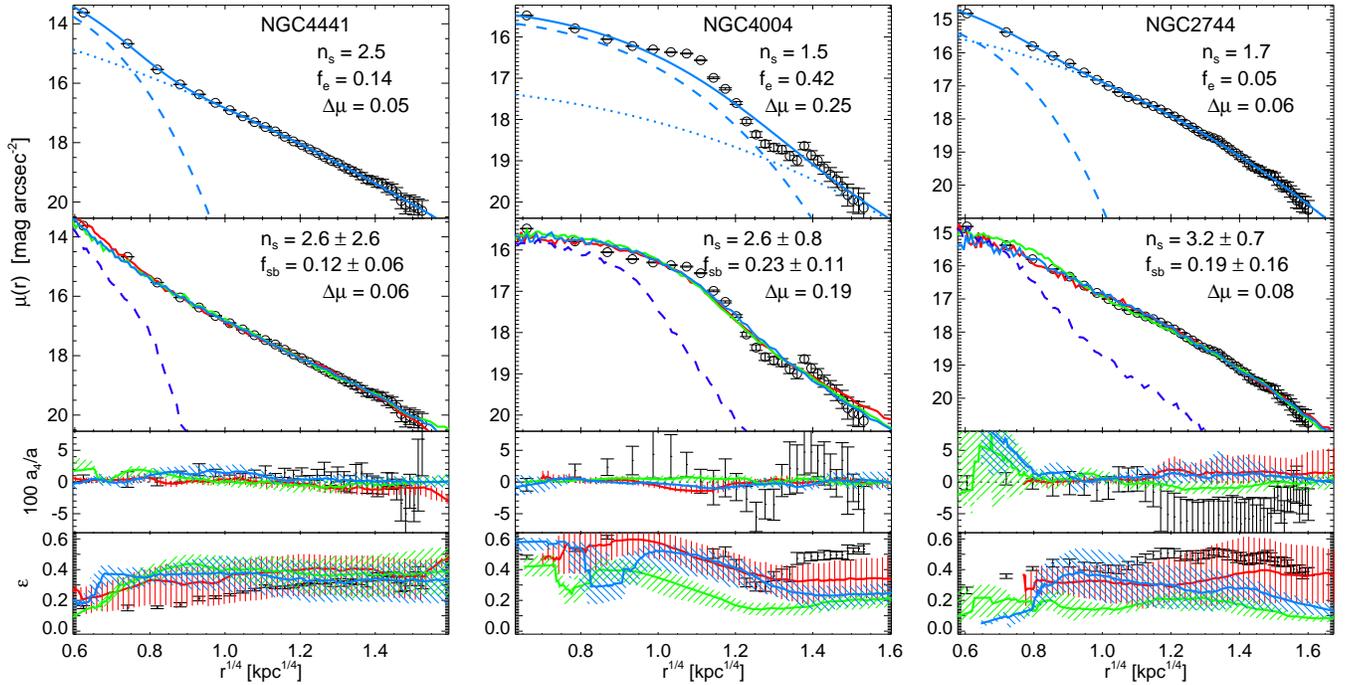}
    \caption{Figure~\ref{fig:rj.all.1.ps}, continued. There is a 
    thin, bar-like feature at $\sim1-2$\,kpc in NGC4004, not reproduced in our simulations 
    at the observed snapshots (although similar features do generically occur in earlier 
    merger stages). 
    \label{fig:rj.all.16.ps}}
\end{figure*}

\begin{figure*}
    \centering
    \ascaleup
    \plotone{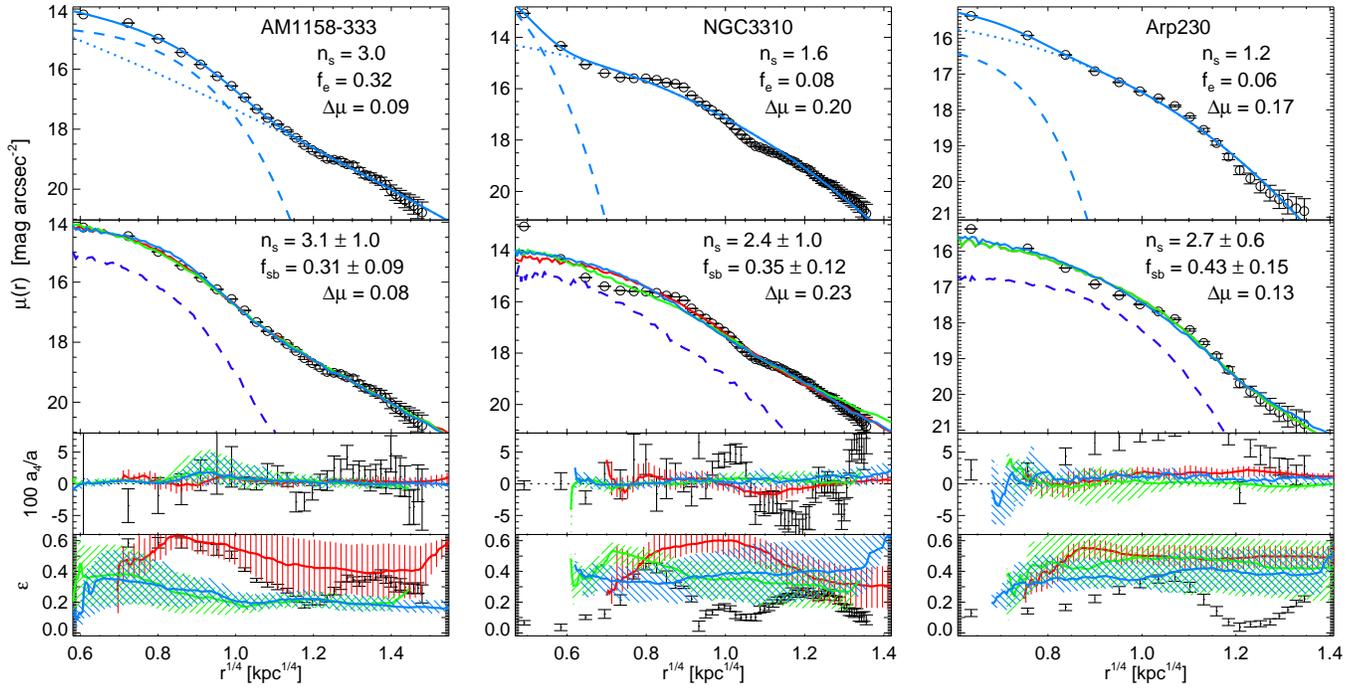}
    \caption{Figure~\ref{fig:rj.all.1.ps}, continued. NGC3310 shows a face-on 
    ring and arm structure at $\sim1$\,kpc, pushing the simulated extra light component to larger radii. 
    \label{fig:rj.all.17.ps}}
\end{figure*}

\begin{figure*}
    \centering
    \ascaleup
    \plotone{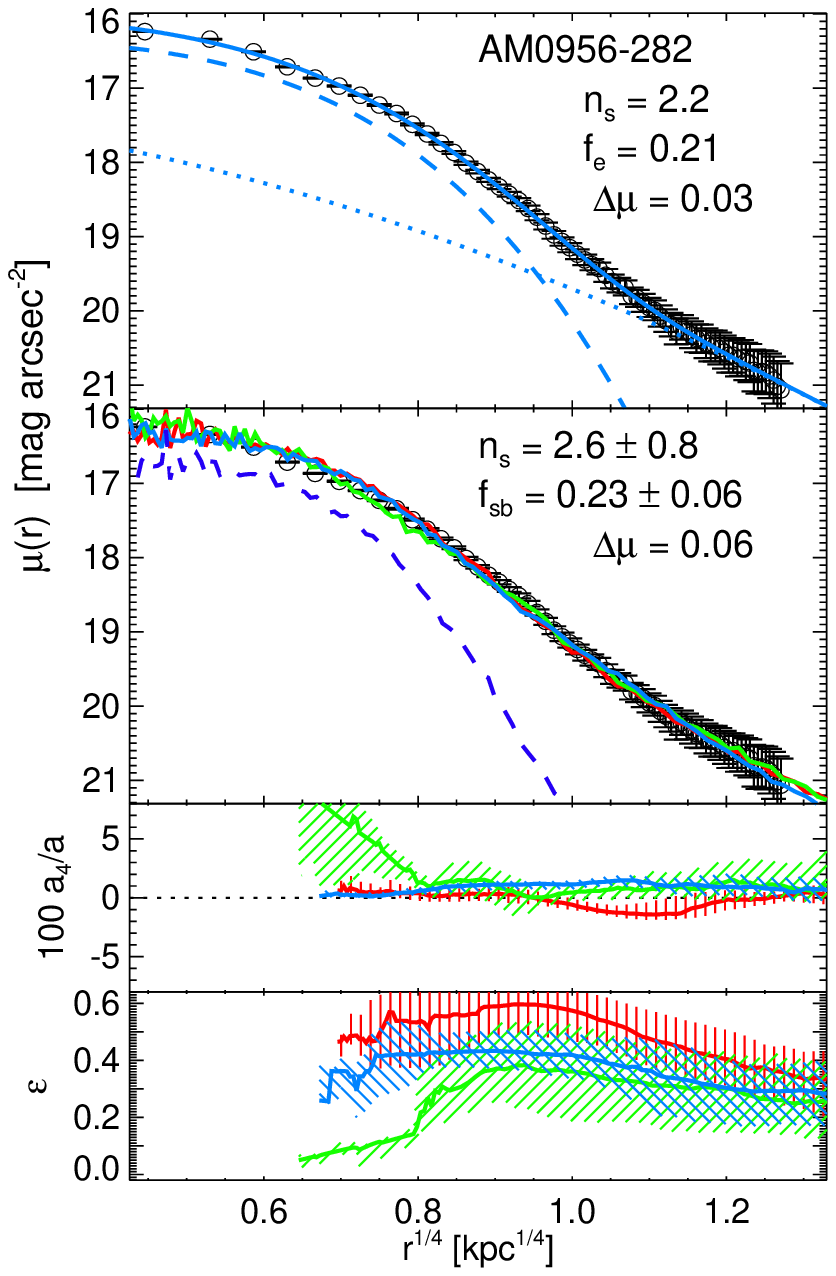}
    \caption{Figure~\ref{fig:rj.all.1.ps}, continued. 
    \label{fig:rj.all.18.ps}}
\end{figure*}

In Figures~\ref{fig:rj.all.1.ps}-\ref{fig:rj.all.18.ps} we explicitly show the results of 
our fitting and simulation comparison with each of the merger remnants 
in the sample of \rj. Table~\ref{tbl:rj.fits} summarizes the results, 
including the estimated integrated properties from simulations corresponding 
to each observed system. 

\clearpage
\clearpage
\clearpage
\clearpage
\clearpage
\clearpage
\clearpage
\clearpage
\clearpage
\clearpage

\tableclear
\begin{\tableset}{lccccccccccc}
\tabletypesize{\scriptsize}
\tablecaption{Fits to \rj\ Merger Remnants\label{tbl:rj.fits}}
\tablewidth{0pt}
\tablehead{
\colhead{Name} &
\colhead{R.A.\ (J2000)} &
\colhead{Decl.\ (J2000)} &
\colhead{$M_{K}$} &
\colhead{$n_{s}$ (fit)\tablenotemark{1}} &
\colhead{$n_{s}$ (sim)\tablenotemark{2}} &
\colhead{$n_{s}$ (RJ04)\tablenotemark{3}} &
\colhead{$f_{e}$ (fit)\tablenotemark{4}} & 
\colhead{$f_{sb}$ (sim)\tablenotemark{5}} &
\colhead{$f_{\rm gas}$ (sim)\tablenotemark{6}} & 
\colhead{$\Delta\mu$ (fit)} &
\colhead{$\Delta\mu$ (sim)} 
}
\startdata
      Mrk1014\tablenotemark{7} & $01$ $59$ $50$ & $\ \ \ 00$ $23$ $41$ & $-28.16$ & $3.9$ & $4.7\pm3.2$ & $10.0$ & $0.67$ & $0.24\pm0.15$ & $0.40\pm0.21$ & $0.06$ & $0.24$
\\
      UGC8058\tablenotemark{7} & $12$ $56$ $14$ & $\ \ \ 56$ $52$ $25$ & $-27.55$ & $4.6$ & $1.6\pm0.9$ & $10.0$ & $0.77$ & $0.37\pm0.14$ & $0.49\pm0.25$ & $0.14$ & $0.42$
\\
       Arp156 & $10$ $42$ $38$ & $\ \ \ 77$ $29$ $41$ & $-25.81$ & $3.1$ & $3.0\pm1.0$ & $9.64$ & $0.12$ & $0.17\pm0.06$ & $0.24\pm0.19$ & $0.11$ & $0.06$
\\
   AM0612-373 & $06$ $13$ $47$ & $-37$ $40$ $37$ & $-25.65$ & $1.6$ & $2.5\pm0.9$ & $3.44$ & $0.16$ & $0.08\pm0.04$ & $0.05\pm0.05$ & $0.07$ & $0.06$
\\
   AM2246-490 & $22$ $49$ $39$ & $-48$ $50$ $58$ & $-25.52$ & $2.8$ & $3.1\pm2.0$ & $10.0$ & $0.32$ & $0.19\pm0.07$ & $0.49\pm0.16$ & $0.12$ & $0.10$
\\
      NGC6598 & $18$ $08$ $56$ & $\ \ \ 69$ $04$ $04$ & $-25.51$ & $3.4$ & $3.2\pm1.1$ & $3.65$ & $0.05$ & $0.10\pm0.06$ & $0.05\pm0.06$ & $0.05$ & $0.04$
\\
      UGC5101 & $09$ $35$ $51$ & $\ \ \ 61$ $21$ $11$ & $-25.50$ & $4.6$ & $2.4\pm0.7$ & $10.0$ & $0.32$ & $0.45\pm0.14$ & $0.54\pm0.19$ & $0.04$ & $0.04$
\\
       NGC828 & $02$ $10$ $09$ & $\ \ \ 39$ $11$ $25$ & $-25.36$ & $3.6$ & $3.2\pm0.7$ & $2.99$ & $0.19$ & $0.24\pm0.16$ & $0.40\pm0.23$ & $0.07$ & $0.08$
\\
      NGC2418 & $07$ $36$ $37$ & $\ \ \ 17$ $53$ $02$ & $-25.31$ & $2.1$ & $2.6\pm0.8$ & $2.85$ & $0.13$ & $0.11\pm0.11$ & $0.16\pm0.23$ & $0.05$ & $0.09$
\\
       Arp187 & $05$ $04$ $53$ & $-10$ $14$ $51$ & $-25.25$ & $3.9$ & $2.5\pm1.1$ & $4.10$ & $0.06$ & $0.12\pm0.08$ & $0.14\pm0.21$ & $0.10$ & $0.08$
\\
     UGC10607 & $16$ $55$ $09$ & $\ \ \ 26$ $39$ $46$ & $-25.20$ & $1.7$ & $2.9\pm1.2$ & $2.94$ & $0.33$ & $0.30\pm0.17$ & $0.48\pm0.20$ & $0.10$ & $0.09$
\\
      NGC5018 & $13$ $13$ $00$ & $-19$ $31$ $05$ & $-25.15$ & $3.1$ & $3.7\pm2.0$ & $4.39$ & $0.06$ & $0.10\pm0.06$ & $0.34\pm0.20$ & $0.07$ & $0.07$
\\
      NGC3921 & $11$ $51$ $06$ & $\ \ \ 55$ $04$ $43$ & $-25.13$ & $2.5$ & $2.6\pm0.4$ & $5.03$ & $0.26$ & $0.27\pm0.13$ & $0.50\pm0.15$ & $0.09$ & $0.06$
\\
   AM0318-230 & $03$ $20$ $40$ & $-22$ $55$ $53$ & $-25.09$ & $2.3$ & $4.6\pm2.0$ & $5.01$ & $0.28$ & $0.25\pm0.11$ & $0.49\pm0.14$ & $0.13$ & $0.06$
\\
   AM2055-425 & $20$ $58$ $26$ & $-42$ $39$ $00$ & $-25.08$ & $0.8$ & $3.3\pm3.5$ & $2.34$ & $0.34$ & $0.25\pm0.13$ & $0.42\pm0.20$ & $0.16$ & $0.11$
\\
      NGC7585 & $23$ $18$ $01$ & $-04$ $39$ $01$ & $-24.98$ & $2.4$ & $3.3\pm0.8$ & $3.53$ & $0.11$ & $0.10\pm0.06$ & $0.18\pm0.20$ & $0.05$ & $0.06$
\\
      UGC9829\tablenotemark{8} & $15$ $23$ $01$ & $-01$ $20$ $50$ & $-24.96$ & $1.4$ & $3.2\pm1.1$ & $2.34$ & $0.02$ & $0.09\pm0.08$ & $0.13\pm0.20$ & $0.32$ & $0.22$
\\
   AM1419-263 & $14$ $22$ $06$ & $-26$ $51$ $27$ & $-24.94$ & $3.2$ & $3.7\pm3.5$ & $4.12$ & $0.06$ & $0.10\pm0.03$ & $0.12\pm0.18$ & $0.05$ & $0.05$
\\
   AM1255-430 & $12$ $58$ $08$ & $-43$ $19$ $47$ & $-24.93$ & $0.9$ & $4.7\pm2.2$ & $1.87$ & $0.26$ & $0.46\pm0.11$ & $0.40\pm0.24$ & $0.09$ & $0.07$
\\
       IC5298 & $23$ $16$ $00$ & $\ \ \ 25$ $33$ $24$ & $-24.92$ & $1.6$ & $2.6\pm3.5$ & $3.93$ & $0.34$ & $0.30\pm0.12$ & $0.49\pm0.18$ & $0.14$ & $0.14$
\\
      NGC7252 & $22$ $20$ $44$ & $-24$ $40$ $41$ & $-24.84$ & $1.3$ & $3.0\pm0.5$ & $3.32$ & $0.32$ & $0.23\pm0.14$ & $0.49\pm0.21$ & $0.06$ & $0.09$
\\
     UGC10675 & $17$ $03$ $15$ & $\ \ \ 31$ $27$ $29$ & $-24.80$ & $2.2$ & $4.6\pm2.8$ & $10.0$ & $0.64$ & $0.51\pm0.08$ & $0.53\pm0.18$ & $0.06$ & $0.10$
\\
      NGC1614 & $04$ $33$ $59$ & $-08$ $34$ $44$ & $-24.74$ & $2.3$ & $3.1\pm0.9$ & $10.0$ & $0.35$ & $0.34\pm0.11$ & $0.50\pm0.12$ & $0.14$ & $0.14$
\\
      NGC3256 & $10$ $27$ $51$ & $-43$ $54$ $14$ & $-24.72$ & $1.8$ & $2.9\pm0.8$ & $2.17$ & $0.09$ & $0.35\pm0.12$ & $0.49\pm0.24$ & $0.11$ & $0.15$
\\
      UGC4635 & $08$ $51$ $54$ & $\ \ \ 40$ $50$ $09$ & $-24.71$ & $2.9$ & $2.7\pm0.5$ & $5.07$ & $0.08$ & $0.11\pm0.05$ & $0.49\pm0.21$ & $0.22$ & $0.13$
\\
   AM2038-382 & $20$ $41$ $13$ & $-38$ $11$ $36$ & $-24.70$ & $2.5$ & $2.4\pm1.1$ & $10.0$ & $0.50$ & $0.43\pm0.08$ & $0.51\pm0.15$ & $0.08$ & $0.07$
\\
   AM1300-233 & $13$ $02$ $52$ & $-23$ $55$ $18$ & $-24.65$ & $1.9$ & $3.4\pm2.0$ & $4.99$ & $0.10$ & $0.10\pm0.08$ & $0.12\pm0.18$ & $0.19$ & $0.17$
\\
       NGC455 & $01$ $15$ $57$ & $\ \ \ 05$ $10$ $43$ & $-24.64$ & $2.0$ & $2.7\pm1.1$ & $6.21$ & $0.20$ & $0.19\pm0.04$ & $0.34\pm0.20$ & $0.06$ & $0.06$
\\
        NGC34 & $00$ $11$ $06$ & $-12$ $06$ $26$ & $-24.61$ & $3.5$ & $2.4\pm1.4$ & $10.0$ & $0.39$ & $0.51\pm0.09$ & $0.51\pm0.17$ & $0.12$ & $0.06$
\\
      UGC2238\tablenotemark{8} & $02$ $46$ $17$ & $\ \ \ 13$ $05$ $44$ & $-24.58$ & $1.1$ & $4.7\pm2.0$ & $1.46$ & $0.16$ & $0.36\pm0.11$ & $0.47\pm0.24$ & $0.13$ & $0.20$
\\
     UGC11905 & $22$ $05$ $54$ & $\ \ \ 20$ $38$ $22$ & $-24.51$ & $1.2$ & $2.6\pm3.5$ & $5.24$ & $0.42$ & $0.30\pm0.10$ & $0.51\pm0.15$ & $0.06$ & $0.09$
\\
       Arp193 & $13$ $20$ $35$ & $\ \ \ 34$ $08$ $22$ & $-24.40$ & $1.2$ & $3.2\pm3.3$ & $2.74$ & $0.33$ & $0.30\pm0.11$ & $0.49\pm0.20$ & $0.09$ & $0.15$
\\
      NGC7727 & $23$ $39$ $53$ & $-12$ $17$ $35$ & $-24.23$ & $2.6$ & $3.0\pm0.6$ & $3.41$ & $0.13$ & $0.23\pm0.14$ & $0.48\pm0.23$ & $0.08$ & $0.08$
\\
      NGC2623 & $08$ $38$ $24$ & $\ \ \ 25$ $45$ $17$ & $-24.22$ & $4.9$ & $3.3\pm3.4$ & $10.0$ & $0.17$ & $0.31\pm0.08$ & $0.51\pm0.09$ & $0.18$ & $0.13$
\\
         UGC6 & $00$ $03$ $09$ & $\ \ \ 21$ $57$ $37$ & $-24.01$ & $2.1$ & $2.4\pm0.7$ & $10.0$ & $0.63$ & $0.45\pm0.10$ & $0.51\pm0.15$ & $0.04$ & $0.11$
\\
      NGC7135 & $21$ $49$ $46$ & $-34$ $52$ $35$ & $-23.95$ & $4.1$ & $3.7\pm1.9$ & $10.0$ & $0.08$ & $0.14\pm0.06$ & $0.22\pm0.20$ & $0.14$ & $0.10$
\\
      NGC2782 & $09$ $14$ $05$ & $\ \ \ 40$ $06$ $49$ & $-23.83$ & $1.7$ & $2.6\pm2.1$ & $6.68$ & $0.23$ & $0.19\pm0.07$ & $0.37\pm0.19$ & $0.13$ & $0.12$
\\
      UGC4079 & $07$ $55$ $06$ & $\ \ \ 55$ $42$ $13$ & $-23.78$ & $2.3$ & $2.6\pm1.0$ & $3.08$ & $0.04$ & $0.08\pm0.04$ & $0.05\pm0.07$ & $0.10$ & $0.09$
\\
      NGC3597 & $11$ $14$ $41$ & $-23$ $43$ $39$ & $-23.72$ & $0.6$ & $2.4\pm1.4$ & $1.87$ & $0.61$ & $0.43\pm0.13$ & $0.50\pm0.18$ & $0.11$ & $0.16$
\\
      NGC1210 & $03$ $06$ $45$ & $-25$ $42$ $59$ & $-23.72$ & $1.9$ & $2.6\pm0.4$ & $4.08$ & $0.16$ & $0.13\pm0.14$ & $0.49\pm0.22$ & $0.09$ & $0.06$
\\
      NGC3656 & $11$ $23$ $38$ & $\ \ \ 53$ $50$ $30$ & $-23.70$ & $3.4$ & $2.8\pm0.6$ & $4.47$ & $0.04$ & $0.13\pm0.08$ & $0.34\pm0.20$ & $0.13$ & $0.12$
\\
      NGC2655 & $08$ $55$ $37$ & $\ \ \ 78$ $13$ $23$ & $-23.70$ & $2.4$ & $3.2\pm0.9$ & $3.01$ & $0.05$ & $0.41\pm0.13$ & $0.50\pm0.17$ & $0.04$ & $0.09$
\\
      NGC6052\tablenotemark{8} & $16$ $05$ $12$ & $\ \ \ 20$ $32$ $32$ & $-23.55$ & $0.9$ & $1.7\pm2.9$ & $1.13$ & $0.04$ & $0.10\pm0.15$ & $0.13\pm0.22$ & $0.11$ & $0.21$
\\
      NGC2914 & $09$ $34$ $02$ & $\ \ \ 10$ $06$ $31$ & $-23.51$ & $0.6$ & $5.1\pm3.4$ & $6.97$ & $0.50$ & $0.25\pm0.09$ & $0.49\pm0.13$ & $0.11$ & $0.14$
\\
      NGC4194 & $12$ $14$ $09$ & $\ \ \ 54$ $31$ $36$ & $-23.21$ & $1.6$ & $2.4\pm0.5$ & $4.59$ & $0.53$ & $0.47\pm0.10$ & $0.51\pm0.15$ & $0.14$ & $0.16$
\\
      NGC4441 & $12$ $27$ $20$ & $\ \ \ 64$ $48$ $06$ & $-22.98$ & $2.5$ & $2.6\pm2.6$ & $6.83$ & $0.14$ & $0.12\pm0.06$ & $0.36\pm0.19$ & $0.05$ & $0.06$
\\
      NGC4004\tablenotemark{8} & $11$ $58$ $05$ & $\ \ \ 27$ $52$ $44$ & $-22.89$ & $1.5$ & $2.6\pm0.8$ & $1.53$ & $0.42$ & $0.23\pm0.11$ & $0.32\pm0.24$ & $0.25$ & $0.19$
\\
      NGC2744 & $09$ $04$ $38$ & $\ \ \ 18$ $27$ $37$ & $-22.83$ & $1.7$ & $3.2\pm0.7$ & $2.35$ & $0.05$ & $0.19\pm0.16$ & $0.12\pm0.22$ & $0.06$ & $0.08$
\\
   AM1158-333 & $12$ $01$ $20$ & $-33$ $52$ $36$ & $-22.61$ & $3.0$ & $3.1\pm1.0$ & $3.75$ & $0.32$ & $0.31\pm0.09$ & $0.50\pm0.13$ & $0.09$ & $0.08$
\\
      NGC3310\tablenotemark{8} & $10$ $38$ $45$ & $\ \ \ 53$ $30$ $05$ & $-22.07$ & $1.6$ & $2.4\pm1.0$ & $2.41$ & $0.08$ & $0.35\pm0.12$ & $0.49\pm0.14$ & $0.20$ & $0.23$
\\
       Arp230 & $00$ $46$ $24$ & $-13$ $26$ $32$ & $-21.75$ & $1.2$ & $2.7\pm0.6$ & $1.56$ & $0.06$ & $0.43\pm0.15$ & $0.50\pm0.17$ & $0.17$ & $0.13$
\\
   AM0956-282 & $09$ $58$ $46$ & $-28$ $37$ $19$ & $-20.50$ & $2.2$ & $2.6\pm0.8$ & $2.45$ & $0.21$ & $0.23\pm0.06$ & $0.47\pm0.14$ & $0.03$ & $0.06$
\\
\enddata
\tablenotetext{1}{Outer Sersic index $n_{s}$ of the two-component best-fit profile.}
\tablenotetext{2}{Range of outer Sersic indices fit in the same manner to the best-fit simulations, 
at $t\approx1-3$\,Gyr after the merger when the system has relaxed.}
\tablenotetext{3}{Sersic index fit to the {\em entire} profile (i.e.\ not decomposed into 
an outer Sersic and inner extra light component) in \rj. Note the authors impose a 
maximum $n_{s}=10.0$.}
\tablenotetext{4}{Fraction of light in the inner or ``extra light'' component of the fits.}
\tablenotetext{5}{Fraction of light from stars produced in the central, merger-induced starburst 
in the best-fit simulations ($\pm$ the approximate interquartile range allowed). Note that this 
may be biased to high values for actively star-forming (especially LIRG and ULIRG) systems.}
\tablenotetext{6}{Initial gas fraction in the best-fitting simulations -- this is less robust
(note the large interquartile range), representing 
a rough gas fraction of the systems $\sim$a few Gyr before the final merger, if they 
evolved in isolation.}
\tablenotetext{7}{AGN contamination affects the central regions (i.e.\ estimated extra light) and 
prevents a good simulation fit (we only model the stellar profile here).}
\tablenotetext{8}{Unrelaxed or prominent disk/bar features make the comparison with these 
objects uncertain.}
\end{\tableset}

\end{appendix}

\end{document}